\documentclass{jfm} 

\usepackage[hidelinks]{hyperref}
\usepackage{natbib}
\usepackage{multicol}
\usepackage{mathrsfs}
\usepackage{fancybox}
\usepackage{textcomp}
\usepackage{gensymb}
\usepackage{mathrsfs}
\usepackage{amsmath,amsfonts,amssymb}
\usepackage{accents}
\usepackage{multirow}
\usepackage{graphicx}
\usepackage{booktabs}
\usepackage{adjustbox}
\usepackage{overpic}

\newcommand{\dimensional}[1]{#1}
\newcommand{\dimensionless}[1]{\hat{#1}}
\renewcommand{\vector}[1]{\boldsymbol{#1}}
\newcommand{\grad}{\boldsymbol \nabla}
\newcommand{\cross}{\times}

\def\dimensionalScaling#1{\dimensional{\mathcal{#1}}}%
\def\region#1{\mathfrak{#1}}

\def\regionLithium{\region L}%
\def\regionSubstrate{\region S}%
\def\inTheLiquid#1{#1^{\regionLithium}}%
\def\inTheSolid#1{#1^{\regionSubstrate}}%
\def\scaleWidth{\dimensionalScaling W}%
\def\scaleDepth{\dimensionalScaling H}%
\def\numberHartmann{\mathit{Ha}}%

\newcommand{\order}{\mathrm{O}}

\shorttitle{TEMHD flow in a trench}
\shortauthor{O.~G.~Bond and P.~D.~Howell
}

\title{Thermoelectric magnetohydrodynamic flow in a liquid metal-infused trench}

\author{O.~G.~Bond
  \corresp{\email{ogb3142@gmail.com}} and
  P.~D.~Howell\aff{1} 
  }

\affiliation{\aff{1}Mathematical Institute, University of Oxford, Oxford, OX2 6GG, UK
}

\begin{document}

\maketitle

\begin{abstract}
We derive a mathematical model for steady, unidirectional, thermoelectric magnetohydrodynamic (TEMHD) flow of liquid lithium along a solid metal trench, subject to an imposed heat flux.
%
We use a finite-element method implemented in COMSOL Multiphysics\textregistered\ to solve the problem numerically, demonstrating how the fluid velocity, induced magnetic field and temperature change depending on the key physical and geometrical parameters.
The observed flow structures are elucidated by using the method of matched asymptotic expansions to obtain approximate solutions in the limit where the Hartmann number is large and the trench walls are thin.
\end{abstract}

\begin{keywords}
Authors should not enter keywords on the manuscript, as these must be chosen by the author during the online submission process and will then be added during the typesetting process (see http://journals.cambridge.org/data/\linebreak[3]relatedlink/jfm-\linebreak[3]keywords.pdf for the full list)
\end{keywords}
\section{Introduction}\label{sec:intro}

The divertor is a vital component designed to absorb heat and waste products exhausted from a tokamak nuclear fusion reactor.
The divertor must survive continuous extreme heat loads of around $10\,\mathrm{MW/m^2}$, and
a promising proposal is to coat it with a layer of flowing liquid lithium which can be constantly recycled.
The Liquid Metal-Infused Trenches (LiMIT) concept has been devised to exploit the large heat flux experienced by the divertor to drive the lithium flow via
thermoelectric effects.
In this paper we derive and solve a model for the resulting flow along a single lithium-filled trench. Our aim is to determine how the flow properties depend on the applied magnetic field and heat flux, and how they can be beneficially influenced by varying the trench geometry.

\begin{figure}
\begin{minipage}{0.65\textwidth}\centering
\textbf{(a)}\\
\includegraphics[width=0.8\textwidth]{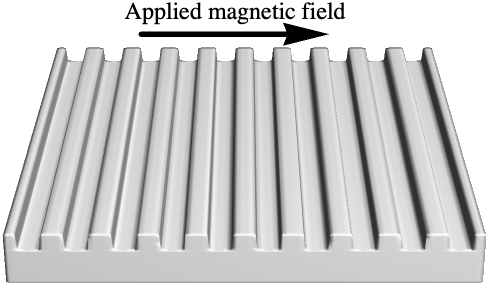}
\end{minipage}%
\begin{minipage}{0.35\textwidth}\centering
\textbf{(b)}\\
\includegraphics[width=0.9\textwidth]{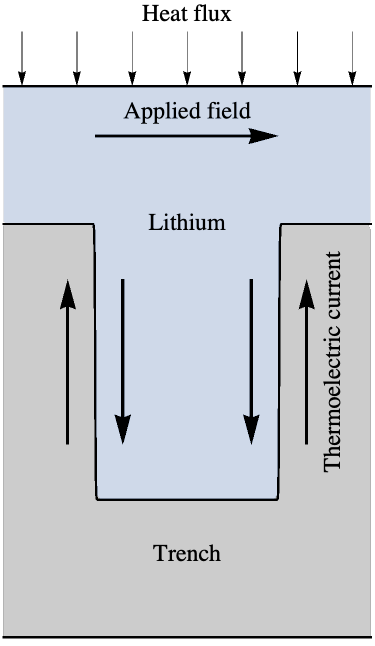}
\end{minipage}
\caption{(a)~Schematic of an array of trenches, indicating the orientation of the applied magnetic field. (b)~Schematic of the heat flux, magnetic field and induced thermoelectric current in a single trench.\label{fig:LiMIT}}
\end{figure}

The mathematical foundations of thermoelectric magnetohydrodynamic (TEMHD) duct flow were established by \citet{Shercliff1979,Shercliff1979con}.
The proposal to use TEMHD effects to drive liquid lithium flow in fusion applications was pioneered at the University of Illinois at Urbana-Champaign \citep{Jaworski2009}. The 
LiMIT concept \citep{Xu2014,Ren2014,Fiflis2015} comprises
an array of parallel solid metal trenches that are filled with liquid lithium and cooled from below.
As illustrated in figure~\ref{fig:LiMIT}, the combination of an applied magnetic field orthogonal to the trench walls and a temperature gradient parallel to the walls generates a thermoelectric current which, in turn, gives rise to a Lorentz force that drives the flow of lithium along the trench.
The LiMIT concept has been tested experimentally in fusion devices
\citep[for example][]{Ren2014,Fiflis2015,Xu2015} and simulated numerically \citep{Xu2014}.
In this paper, we formulate a two-dimensional mathematical model for unidirectional liquid lithium flow in LiMIT which permits large parameter sweeps to be performed numerically and the trends to be understood using asymptotic analysis.


Our model resembles classical models for
magnetohydrodynamic (MHD) flow of liquid metal in ducts and channels, which have been of great theoretical interest ever since the 1950s \citep{Shercliff1953}.
Much of the early progress in MHD duct flow was based on asymptotic analysis in the limit of large Hartmann number $\numberHartmann$
\citep[see, for example,][]{Hunt1976,Walker1971}.
The basic structure in a rectangular duct consists of a core plug flow, with \emph{Hartmann layers} of width $\mathrm{O}\bigl(\numberHartmann^{-1}\bigr)$ on walls orthogonal to the applied magnetic field and \emph{side layers} of width $\mathrm{O}\bigl(\numberHartmann^{-1/2}\bigr)$ on walls parallel to the applied field \citep{Temperley1971}.
It is also well known that conductivity of the duct walls can have a significant influence and give rise to velocity jets in the side layers \citep{Hunt1965}.
%
%
%
More recently, advances in computational methods have made it possible to simulate MHD duct flow with novel geometries, including a fan-shaped insert \citep{Kim1997}, sudden expansions \citep{Mistrangelo2007} and other obstacles \citep{Dousset2009}.

In \S\ref{sec:modelling_principles}, we state the leading-order dimensionless governing equations and boundary conditions for steady, unidirectional flow of liquid lithium down a LiMIT-type trench.
We also provide some preliminary numerical solutions for parameter regimes relevant to both theoretical and experimental setups. In \S\ref{sec:asymptotics}, we use asymptotic analysis in the limit of large Hartmann number to obtain approximate solutions to the problem when the trench walls are thin. In \S\ref{sec:parameter_sweeps}, we perform several parameter sweeps to demonstrate how the flow properties depend on the geometrical and physical variables of interest. Finally, we provide a summary and discussion of our findings in \S\ref{sec:discussion}.



\section{Mathematical model}\label{sec:modelling_principles}

\begin{table}
\centering
\begin{tabular}{cccccc}
\toprule 
\textbf{Name} & \textbf{Symbol} & \textbf{SI Unit} & \textbf{Lithium} & \multirow{1}{*}{\textbf{SS 316}} & \multirow{1}{*}{\textbf{Tungsten}}\tabularnewline
\midrule 
$\begin{matrix}\text{Electrical}\\
\text{conductivity}
\end{matrix}$ & $\dimensional{\sigma}$ & ${\rm S\,m^{-1}}$ & $3.46\times10^{6}$ & $1.04\times10^{6}$ & $8.23\times10^{6}$\tabularnewline
\midrule 
$\begin{matrix}\text{Thermal}\\
\text{conductivity}
\end{matrix}$ & $\dimensional k$ & ${\rm W\,m^{-1}\,K^{-1}}$ & $4.80\times10^{1}$ & $1.79\times10^{1}$ & $1.40\times10^{2}$\tabularnewline
\midrule 
$\begin{matrix}\text{Magnetic}\\
\text{permeability}
\end{matrix}$ & $\mu$ & $\mathrm{N\,A}^{-2}$ & $1.26\times10^{-6}$ & $1.28\times10^{-6}$ & $1.26\times10^{-6}$ \tabularnewline
\midrule 
$\text{\ensuremath{\begin{matrix}\text{Seebeck}\\
 \text{coefficient} 
\end{matrix}}}$ & $\dimensional S$ & ${\rm V\,K^{-1}}$ & $2.4\times10^{-5}$ & $-2.69\times10^{-6}$ & $5.58\times10^{-6}$\tabularnewline
\midrule 
$\text{\ensuremath{\begin{matrix}\text{Mass}\\
 \text{density} 
\end{matrix}}}$ & $\dimensional{\rho}$ & ${\rm kg\,m^{-3}}$
& $5.05\times10^{2}$&
--- & ---
\tabularnewline
\midrule 
$\text{\ensuremath{\begin{matrix}\text{Kinematic}\\
 \text{viscosity} 
\end{matrix}}}$ & $\dimensional{\nu}$ & ${\rm m^{2}\,s^{-1}}$ & $1.06\times10^{-6}$ & --- & ---\tabularnewline
\midrule 
$\text{\ensuremath{\begin{matrix}\text{Surface}\\
\text{tension}
\end{matrix}}}$ & $\dimensional{\gamma}$ & ${\rm {\rm N\,m^{-1}}}$ & $3.81\times10^{-1}$ & --- & ---\tabularnewline
\bottomrule
\end{tabular}
\caption{Typical values of relevant thermophysical properties for lithium, stainless steel and tungsten over temperatures in the range
$[200^{\circ}{\rm C},400^{\circ}{\rm C}]$
\citep{Ho1977,Davison1968,Tolias2017,Choong1975,Fiflis2013,White1984,Marel1988,Itami1988,Kreissman1953}.
\label{tab:thermophysical_properties}}
\end{table}

\begin{figure}
\begin{minipage}{0.4\textwidth}\centering
\textbf{(a)}\\
\includegraphics[width=0.8\textwidth]{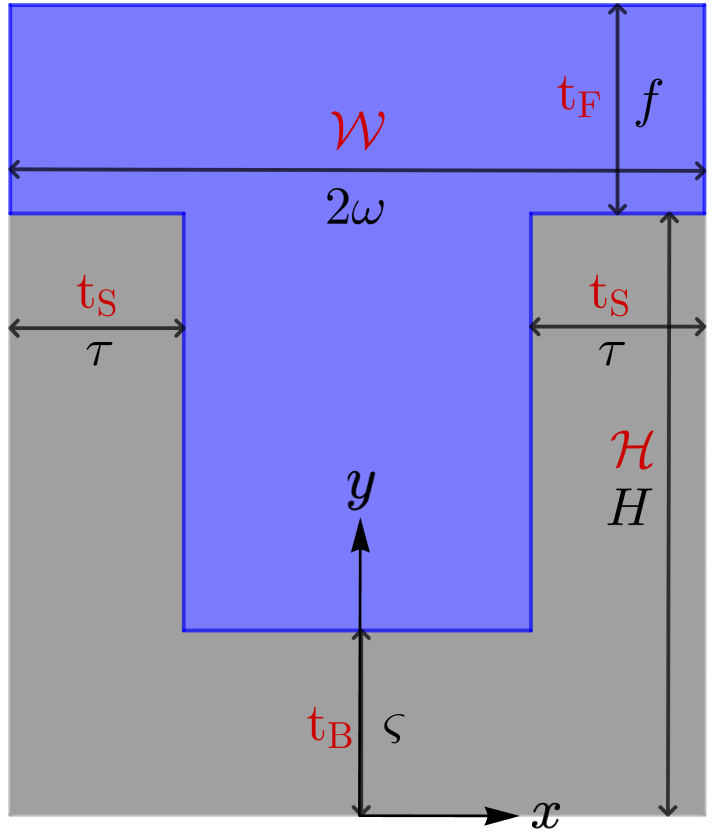}
\end{minipage}%
\begin{minipage}{0.6\textwidth}\centering
\textbf{(b)}\\
\includegraphics[width=0.85\textwidth]{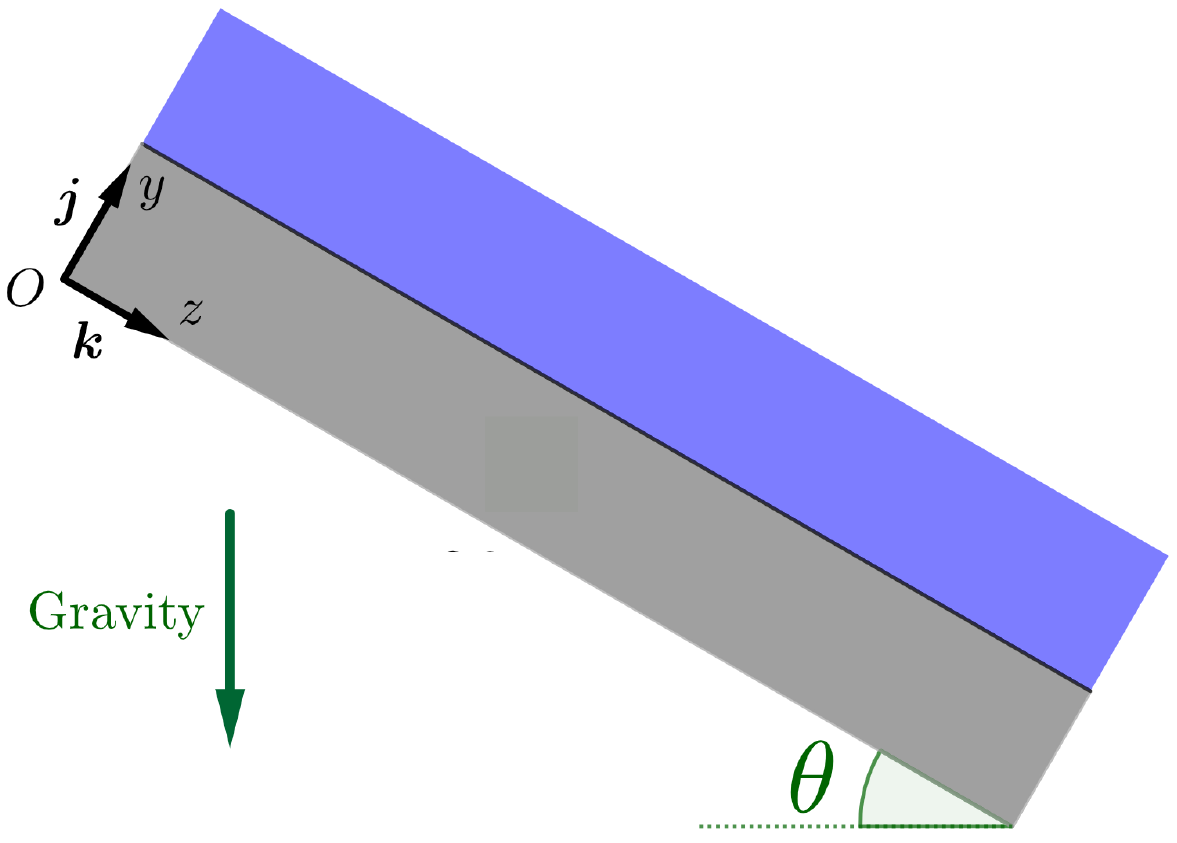}
\end{minipage}
\caption{Diagram summarising the coordinate system and the dimensional geometrical parameters associated with a rectangular trench (in red) with their dimensionless analogues (in black).\label{fig:trench geometry}}
\end{figure}

\begin{table}
\begin{centering}
\begin{tabular}{cccccc}
\toprule 
\multirow{2}{*}{\textbf{Name}} & \multirow{2}{*}{\textbf{Symbol}} & \multirow{2}{*}{\textbf{Formula}} & \multicolumn{3}{c}{\textbf{Typical values}}\tabularnewline
\cmidrule{4-6} \cmidrule{5-6} \cmidrule{6-6} 
 &  &  & \textbf{Default} & \textbf{SLiDE} & \textbf{ST40}\tabularnewline
\midrule
\midrule
$\begin{matrix}\text{Dimensional}\\
\text{trench width (mm)}
\end{matrix}$ & $\scaleWidth$ & --- & 1 & 3 & 0.5\tabularnewline
\midrule 
$\begin{matrix}\text{Dimensionless}\\
\text{trench half-width}
\end{matrix}$ & $\omega$ & --- & $\dfrac{1}{2}$ & $\dfrac{1}{2}$ & $\dfrac{1}{2}$\tabularnewline
\midrule 
$\begin{matrix}\text{Dimensionless}\\
\text{side-wall thickness}
\end{matrix}$ & $\tau$ & $\dfrac{\dimensional{\textrm{t}}_{{\rm S}}}{\dimensionalScaling W}$ & $0.15$ & $\dfrac{1}{6}$ & 0.2\tabularnewline
\midrule 
$\begin{matrix}\text{Dimensionless}\\
\text{base thickness}
\end{matrix}$ & $\varsigma$ & $\dfrac{\dimensional{\textrm{t}}_{{\rm B}}}{\dimensionalScaling W}$ & $0.15$ & $\dfrac{1}{3}$ & 0.2\tabularnewline
\midrule 
$\begin{matrix}\text{Dimensionless}\\
\text{wall height}
\end{matrix}$ & $H$ & $\dfrac{\scaleDepth}{\dimensionalScaling W}$ & $1$ & 1 & 2\tabularnewline
\midrule 
$\begin{matrix}\text{Dimensionless}\\
\text{film thickness}
\end{matrix}$ & $f$ & $\dfrac{\dimensionalScaling F}{\dimensionalScaling W}$ & $0.1$ & $\dfrac{1}{12}$ & 0.2\tabularnewline
\midrule
$\begin{matrix}\text{Downward slope}\\ \text{angle (rad)}\end{matrix}$ & $\theta$ & --- & 0.1 & 0 & 0.316\tabularnewline
\midrule 
$\begin{matrix}\text{Applied magnetic}\\ \text{field angle (rad)}\end{matrix}$ & $\psi$ & --- & 0 & 0 & 0\tabularnewline
\midrule 
$\begin{matrix}\text{Solid-to-fluid electrical}\\
\text{conductivity ratio}
\end{matrix}$ & $\Sigma$ & $\dfrac{\inTheSolid{\dimensional{\sigma}}}{\inTheLiquid{\dimensional{\sigma}}}$ & 2.3786 & 0.3006 & 2.3786\tabularnewline
\midrule 
$\begin{matrix}\text{Solid-to-fluid thermal}\\
\text{conductivity ratio}
\end{matrix}$ & $K$ & $\dfrac{\inTheSolid{\dimensional k}}{\inTheLiquid{\dimensional k}}$ & 2.9167 & 0.3729 & 2.9167\tabularnewline
\midrule
$\begin{matrix}\text{Applied magnetic}\\ \text{field (T)}\end{matrix}$ & $\dimensionalScaling B^{{\rm a}}$ & --- & 1 & 0.07 & 3\tabularnewline
\midrule
$\begin{matrix}\text{Applied heat}\\ \text{flux } {\rm (MW\,m^{-2})}\end{matrix}$
& $\dimensionalScaling Q$ & --- & $1$ & 10 & 10\tabularnewline
\midrule
$\begin{matrix}\text{Reciprocal of}\\
\text{Hartmann number}
\end{matrix}$ & $\epsilon$ & $\dfrac{1}{\dimensionalScaling B^{{\rm a}}\dimensionalScaling W}\sqrt{\dfrac{\inTheLiquid{\dimensional{\rho}}{\dimensional{\nu}}}{\inTheLiquid{\dimensional{\sigma}}}}$ & $0.012$ & $0.058$ & 0.0081\tabularnewline
\midrule
$\begin{matrix}\text{Gravitational}\\
\text{parameter}
\end{matrix}$ & $\varGamma$ & $\dfrac{\inTheLiquid{\dimensional{k}}g\dimensionalScaling W\sin\theta}{\dimensionalScaling Q(\inTheLiquid{S}-\inTheSolid{S})}\sqrt{\dfrac{\inTheLiquid{\dimensional{\rho}}}{{\dimensional{\nu}}\inTheLiquid{\dimensional{\sigma}}}}$ & $0.03$ & $0$ & $0.0047$\tabularnewline
\bottomrule
\end{tabular}
\par\end{centering}
\caption{Default model parameter values, as well as typical values seen in the experimental setup SLiDE, and proposed values for the ST40 tokamak \citep{Buxton2023}. The superscripts $\mathfrak{L}$ and $\mathfrak{S}$ refer to values for the liquid lithium and the solid metal, respectively.}
\label{tab:model_dimensionless_parameters}
\end{table}

\subsection{Parameter values}\label{ss:params}

Values and definitions of typical relevant thermophysical properties are given in table~\ref{tab:thermophysical_properties}. The liquid metal inside the trench is taken to be lithium, and two candidate materials are considered for the trench itself: Stainless Steel~316 and tungsten.

The setup of a single periodic cell in an array of rectangular trenches is illustrated in figure~\ref{fig:trench geometry}. The geometry is characterised by the trench width $\mathcal{W}$, the heights $\mathrm{t}_\text{B}$ and $\mathcal{H}$ of the base and the wide walls, the wall half-thickness $\mathrm{t}_\text{S}$ and the height $\mathrm{t}_\text{F}$ of the liquid layer above the trench. The dimensionless analogues of these quantities are scaled with $\mathcal{W}$, i.e.,
$(\mathcal{W},\mathrm{t}_\text{B},\mathcal{H},\mathrm{t}_\text{S},\mathrm{t}_\text{F})=\mathcal{W}(2\omega,\varsigma,H,\tau,f)$. Following this scaling, we retain the parameter $\omega$ for the dimensionless trench half-width to allow ourselves the freedom to vary the width easily in the calculations below (thus taking $\mathcal{W}$ to be a typical but not necessarily exact value for the trench width).

Typical trench geometrical parameters are shown in table~\ref{tab:model_dimensionless_parameters}, along with typical values of the applied magnetic field and heat flux, and the relevant dimensionless parameters. Three different cases are considered. The ``default'' parameter regime has a typical geometrical setup, with thermophysical properties relevant to a tungsten divertor, and is used as a base case in the computations performed below. The ``SLiDE'' parameter regime is one relevant to the Solid-Liquid Divertor Experiment setup, which involves a stainless steel divertor \citep{Ruzic2011,Xu2013,Ren2014}. The ``ST40'' parameter regime is named after Tokamak Energy's spherical tokamak device, which uses a tungsten divertor, with impinging heat fluxes and applied magnetic fields relevant to a fusion reactor
\citep{McNamara_2023}.
In all three cases the applied magnetic field is effectively horizontal, but the effects of the field being applied at a nonzero angle $\psi$ to the horizontal
will also be examined in \S\ref{ss:gps}.

\subsection{Governing equations}\label{sec:governing_equations}

Here we briefly state the basic equations governing  liquid-metal TEMHD flow.
In the non-relativistic limit where the displacement current is negligible, the magnetic flux density $\vector{B}$ satisfies Gauss' and Amp\`{e}re's Laws, namely
\refstepcounter{equation}\label{eq:maxwell}
\begin{align*}
\vector{\nabla}\cdot\vector{B}&=0,
&
\vector{\nabla}\times\vector{B}&=\mu\vector{J},
\tag{\textrm{\theequation}\textit{a,b}}
\end{align*}
where $\vector{J}$ is the electrical current density and $\mu$ is the magnetic permeability. Since the magnetic susceptibilities of lithium, SS~316 and tungsten are all very small (see table~\ref{tab:thermophysical_properties}), we treat $\mu$ as a constant throughout.
The incompressible Navier--Stokes equations, including the Lorentz body force, may thus be expressed in the form
\begin{subequations}
\begin{gather}\label{eq:divu}
\vector{\nabla}\cdot\vector{u}=0,
\\\label{eq:momentum}
\rho\left(\frac{\partial\vector{u}}{\partial t}+(\vector{u}\cdot\vector{\nabla})\vector{u}\right)=-\vector{\nabla}p
+\rho\nu\nabla^2\vector{u}+\rho\vector{g}+\frac{1}{\mu}(\vector{\nabla}\times\vector{B})\times\vector{B},
\end{gather}
\end{subequations}
where $\vector{u}$ and $p$ denote the liquid velocity and pressure, while $\rho$, $\nu$ and $\vector{g}$ denote the density, kinematic viscosity and gravitational acceleration (all assumed constant).
Finally, we assume that internal viscous and Ohmic heating are negligible compared with large externally applied heat flux $\mathcal{Q}$, so the temperature $T$ satisfies the heat equation
\begin{equation}
\rho c\left(\frac{\partial T}{\partial t}+\vector{u}\cdot\vector{\nabla} T\right)=
-\vector{\nabla}\cdot\vector{q},
\end{equation}
where $c$ is the heat capacity and $\vector{q}$ the heat flux.

Thermoelectric effects enter through Ohm's and Fourier's constitutive laws
\citep{Shercliff1979}, which are modified to
\refstepcounter{equation}\label{eq:OhmFourier}
\begin{align*}
\dimensional{\vector J}&=\dimensional{\sigma}\left(\dimensional{\vector E}+\dimensional{\vector u}\cross\dimensional{\vector B}-\dimensional S\dimensional{\grad}\dimensional T\right),
&
\dimensional{\vector q}&=-\dimensional k\dimensional{\grad}\dimensional T+\dimensional S\dimensional T\dimensional{\vector J},
\tag{\textrm{\theequation}\textit{a,b}}
\end{align*}
respectively, where $\vector{E}$ is the electric field,  $\dimensional{\sigma}$ is the electrical conductivity, $\dimensional{k}$ is the thermal conductivity, and $S$ is the Seebeck coefficient of the medium. The additional thermoelectric terms (proportional to $S$) can be derived from the Onsager reciprocal relations in thermodynamics \citep{Callen1948}, and physically they arise as a result of the Seebeck, Peltier and Thomson effects \citep{Shercliff1979}.

The relative importance of the final term in (\ref{eq:OhmFourier}\textit{b}) is measured by the dimensionless grouping $\sigma S^2\mathcal{W}\mathcal{Q}/k^2$, which is always small (of order $10^{-2}$), so this term is neglected henceforth.
Moreover, taking the curl of (\ref{eq:OhmFourier}\textit{a}) to eliminate the electric field also eliminates the Seebeck term, so the induction equation
\begin{equation}
\frac{\partial\vector{B}}{\partial t}=\vector{\nabla}\times(\vector{u}\times\vector{B})+\eta\nabla^2\vector{B}
\end{equation}
is unaffected (where $\eta=1/(\sigma\mu)$ is the magnetic diffusivity).
However,  thermoelectric effects enter the model through the boundary conditions at the interface between the liquid lithium and the solid trench, where the Seebeck coefficient is discontinuous. Since the liquid velocity is zero at this interface, continuity of the tangential electric field leads to the boundary condition
\begin{equation}
\left[\vector{n}\times\left(\eta\vector{\nabla}\times\vector{B}+S\vector{\nabla}T\right)\right]_-^+=\vector{0},
\end{equation}
where $[\cdot]_-^+$ denotes the jump in a quantity across the solid--liquid interface, whose unit normal is $\vector{n}$. Thus a magnetic field can be generated by a temperature gradient parallel to the interface.

\subsection{Modelling assumptions}

We proceed to set out a model of steady unidirectional flow of liquid lithium along a single trench within a periodic array.
We assume that the geometry and the flow are all uniform along the trench, with velocity given by
\begin{equation}
\vector{u}(\vector{x})=w(x,y)\vector{k},
\end{equation}
in the coordinate system depicted in figure~\ref{fig:trench geometry}.
We consider a uniform applied magnetic field $\mathcal{B}^\text{a}$ inclined at an angle $\psi$ to the $x$-axis, which induces a field $b$ along the trench, i.e.
\begin{equation}
\vector{B}(\vector{x})=\mathcal{B}^{\text{a}}\left(\vector{i}\cos\psi+\vector{j}\sin\psi\right)+b(x,y)\vector{k}.
\end{equation}
As usual in unidirectional flow, the pressure $p$ in general varies linearly with distance along the trench. However, since we assume the free surface is uniform in the $z$-direction, it follows that the down-trench pressure gradient must be zero.
The variations in pressure due to the induced magnetic field are measured by the modified Weber number \mbox{$\mathit{We}=(\nu/\eta)(\rho\mathcal{W}\mathcal{U}^2/\gamma)$}, where $\gamma$ is the surface tension and $\mathcal{U}$ is a typical scale for the velocity, which will be chosen below. Since this parameter is small in all of the cases considered here, we can take the pressure to be purely hydrostatic and, hence, the free surface to be flat
(as depicted in figure~\ref{fig:trench geometry}).


The plasma above the trench is assumed to provide a uniform heat source $\mathcal{Q}$ at the free surface,
while the cooling pipes below the trench are effectively treated as a perfect heatsink, so the base of the trench is held at a fixed temperature $T_0$. We also assume that all material properties (such as density, viscosity and conductivity) are constants, i.e., that any temperature dependencies may be neglected in the first instance.

\subsection{Nondimensionalisation}
The problem is nondimensionalised as follows, with dimensionless variables denoted with hats. We scale all lengths with the width of the trench $\dimensionalScaling{W}$, that is:
\begin{align}\label{eq:2dTrenchSteadyNondimensionalisationCoordinates}
\left(\dimensional x,\dimensional y\right) =\dimensionalScaling W\left(\dimensionless x,\dimensionless y\right).
\end{align}
As noted in \S\ref{ss:params}, although
the dimensionless trench width may be set to unity without loss of generality, it is convenient to allow it to vary, and we denote it by $2\omega$. 
%
Since
the pressure gradient along the trench must be zero, given our modelling assumptions,
the only remaining unknowns are the temperature $T$ and the velocity $w$ and induced magnetic field $b$ along the trench. The temperature is scaled using the imposed heat flux $\dimensionalScaling Q$, while $w$ and $b$ are scaled using a typical velocity $\mathcal{U}$ which will be selected below. We thus set
\refstepcounter{equation}\label{eq:NDTwb}
\begin{align*}
\dimensional T & =\dimensional T_{0}+\left(\dfrac{\dimensionalScaling Q\dimensionalScaling W}{\inTheLiquid{\dimensional k}}\right)\dimensionless T,
&
\dimensional w & =\mathcal{U}\dimensionless w,
&
\dimensional b & =\epsilon \mathit{Rm}\, \dimensionalScaling B^{{\rm a}}\dimensionless b,
\tag{\textrm{\theequation}\textit{a--c}}
\end{align*}
%
where $\epsilon\ll1$ and $\mathit{Rm}\ll1$ are the reciprocal Hartmann number and magnetic Reynolds number, respectively, given by
\refstepcounter{equation}\label{eq:defeRm}
\begin{align*}
\epsilon & =\dfrac{1}{{\rm Ha}}=\dfrac{1}{{\cal B}^{{\rm a}}{\cal W}}\sqrt{\dfrac{\rho\nu}{\sigma^{\mathfrak{L}}}},
&
\mathit{Rm} & =\mathcal{UW}\mu\sigma^{\mathfrak{L}}.
\tag{\textrm{\theequation}\textit{a,b}}
\end{align*}

The appropriate velocity scale is found to be proportional to the difference in Seebeck coefficient between liquid and solid, namely
\begin{align}\label{eq:nondimensionalisationVelocity-Steady}
\mathcal{U} & =\dfrac{\dimensionalScaling Q(\inTheLiquid{S}-\inTheSolid{S})}{\inTheLiquid{\dimensional k}\dimensionalScaling B^{{\rm a}}}.
\end{align}
This scaling reflects the way the flow is driven by the heat flux $\dimensionalScaling Q$ applied to the free surface, through the Seebeck effect, while increasing the applied magnetic field too much
slows
down the flow, as observed qualitatively by \citet{Xu2014}.

\begin{figure}
    \centering
    \includegraphics[width=\textwidth]{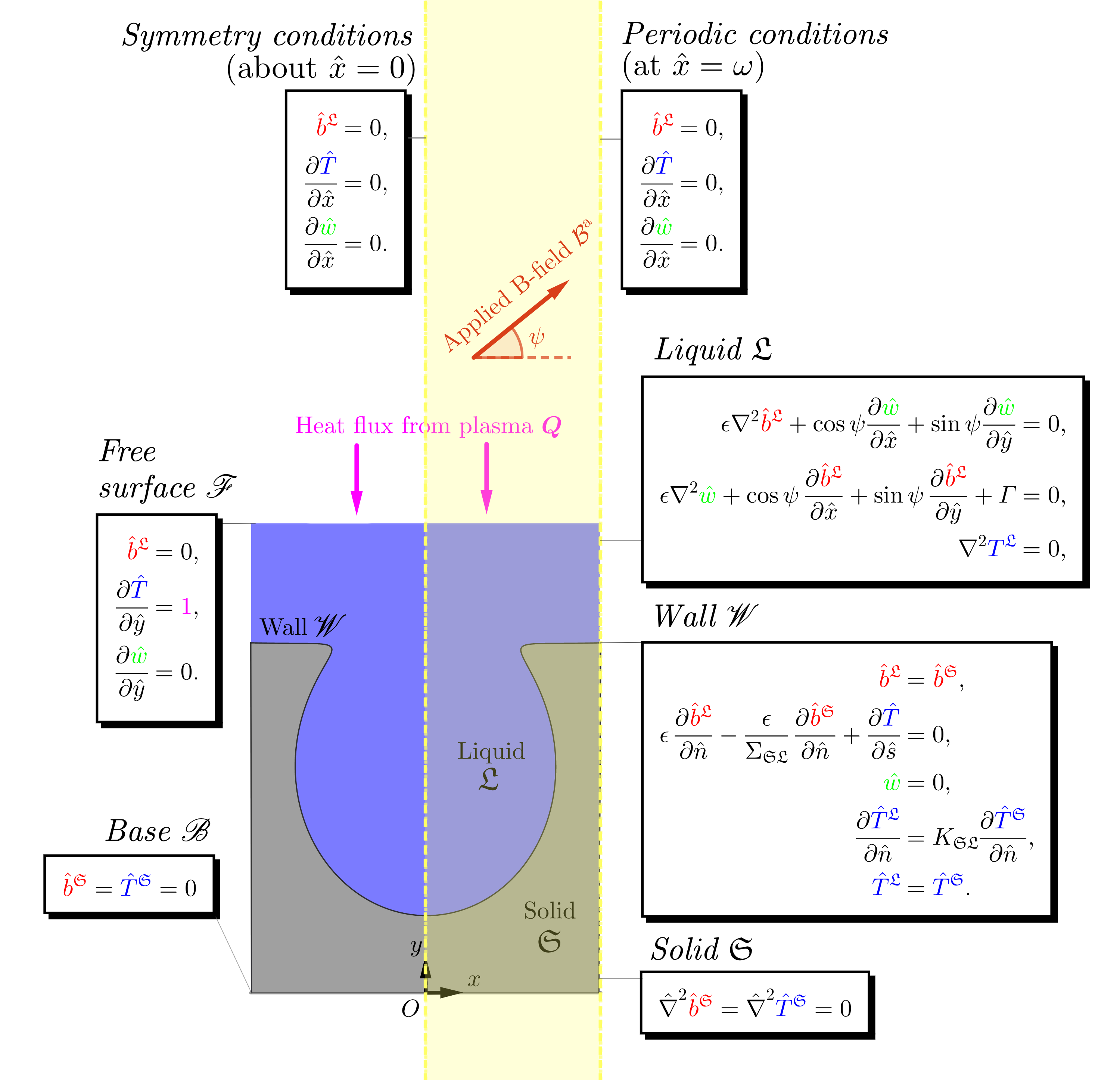}
    \caption{A summary of the steady, unidirectional TEMHD trench problem, assuming symmetry about $\dimensionless{x}$ = 0.}
    \label{fig:leading_order}
\end{figure}

The dimensionless versions of the governing equations and boundary conditions, evaluated in each domain and on each boundary, are summarised in figure~\ref{fig:leading_order}. 
For simplicity, we restrict attention to trenches with reflectional symmetry about $\hat{x}=0$.
The problem contains several more dimensionless parameters, namely
\refstepcounter{equation}\label{eq:ndparams}
\begin{align*}
\varGamma &=
\frac{\epsilon \mathcal{W}^2g\sin\theta}{\nu\mathcal{U}},
&
\Sigma_{\mathfrak{SL}}&=\dfrac{\sigma_{\mathfrak S}}{\sigma_{\mathfrak L}},
&
K_{\mathfrak{SL}}&=\dfrac{k_{\mathfrak S}}{k_{\mathfrak L}}. 
\tag{\textrm{\theequation}\textit{a--c}}
\end{align*}
%
The solid-to-liquid electrical conductivity ratio is denoted by
$\Sigma_{\mathfrak{SL}}$, so the regimes \mbox{$\Sigma_{\mathfrak{SL}}\ll1$} and $\Sigma_{\mathfrak{SL}}\gg1$ correspond to the limits of perfectly electrically insulating or conducting trench walls, respectively. A similar analogy holds for the thermal conductivity ratio $K_{\mathfrak{SL}}$. Both of these constants are $\order(1)$ in practice. The dimensionless constant $\varGamma$ meaures the importance of gravity in driving the flow along the channel.
It is generally small (see table~\ref{tab:model_dimensionless_parameters}) but not completely negligible, so we retain this term for the time being.

Because we have neglected the thermoelectric heat flux and viscous dissipation, the thermal problem decouples from the rest of the equations. The temperature $T$ satisfies Laplace's equation in the lithium and the trench wall, with standard continuity conditions at the interface between them, and is driven by a dimensionless unit heat flux at the free surface.
Indeed, the whole system is driven by this heat flux, which is
the only inhomogeneous term in the problem,
The resulting temperature gradient induces a magnetic field through the Seebeck boundary condition at the solid-liquid interface, and the induced field in turn generates the flow along the trench.

Due to the ease of design and manufacture, we focus on rectangular trenches in this paper. The geometry of such a trench can be characterised by several geometrical parameters, which are shown visually in figure~\ref{fig:trench geometry}
and whose typical values are listed in table~\ref{tab:model_dimensionless_parameters}.
\subsection{Numerical solutions}

\begin{figure}
    \centering
    \includegraphics[width=0.9\textheight,angle=90]{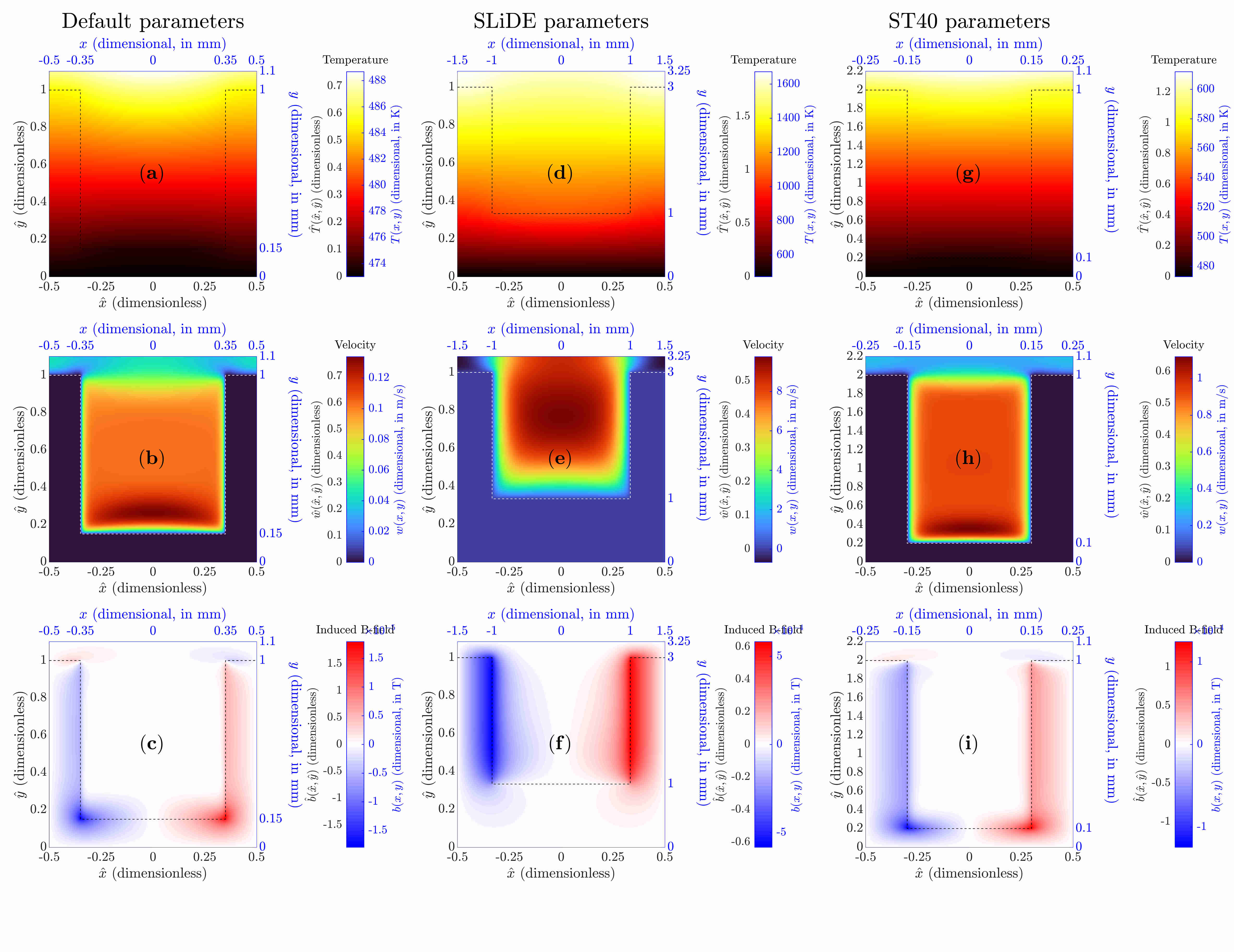}
     \caption{Plots of the \textbf{(a,d,g)} temperatures, \textbf{(b,e,h)} velocities and \textbf{(c,f,i)} induced magnetic fields for the ``default'', ``SLiDE'' and ``ST40''  parameter cases. 
     \label{fig:global_trench}}
 \end{figure}

We solve the problem shown in figure~\ref{fig:leading_order} numerically using finite element methods implemented
in
COMSOL Multiphysics\textregistered\ simulation software. Plots of the temperature, velocity and magnetic field in the three different parameter regimes described in \S\ref{ss:params} are shown in figure~\ref{fig:global_trench}. 
%
In all three cases, the temperature appears to be approximately a linear function of $\dimensionless{y}$, reaching its peak at the free surface where the external heat flux is applied, although with
some variation in the $\dimensionless{x}$-direction near the walls due to the change in thermal conductivity.

Meanwhile, the velocity plots \textbf{(b,e,h)} show an obvious boundary layer structure which arises because the MHD problem (shown in figure~\ref{fig:leading_order}) becomes singular as $\epsilon \to 0$. There are Hartmann boundary layers of width $\order(\epsilon)$ at the vertical walls, as well as a weaker ``side'' layer of width $\order(\epsilon^{1/2})$ near the horizontal boundaries at the base and the tops of the walls.
Outside these layers, it appears that the velocity inside the trench approaches a constant as $\epsilon \to 0$.
%
The boundary layers can also be seen in the induced magnetic field plots \textbf{(c,f,i)}, which show a magnetic field dipole.
The boundary layer structure will be further elucidated below in \S\ref{sec:asymptotics}.

Another important feature particularly evident in plots~\textbf{(b)} and~\textbf{(h)} is the appearance of a conducting-base velocity jet near the bottom of the trench. This jet arises due to the fact that the electrically conducting base is in direct contact with a vacuum, which forces the electrical current loops to close partially in the fluid. The velocity jet appears near the wall that is parallel to the applied magnetic field, as observed by \citet{Hunt1965}.

Regarding the SLiDE setup, we note that the maximum temperature shown in plot~\textbf{(d)} is close to the boiling point of lithium (1615\,K), and that the velocities shown in plot~\textbf{(e)} are much faster than observed in practice \citep{Xu2013}. This behaviour occurs because the magnetic field in this case is relatively weak (0.07\,T) while the impinging heat flux is fusion-relevant ($10\,{\rm MW\,m}^{-2}$), both of which increase the velocity scaling in equation \eqref{eq:nondimensionalisationVelocity-Steady}.
However, we note that the heat flux in the SLiDE experiments is confined to a narrow electron beam rather than being applied uniformly along the trench, as assumed in our model.

\subsection{Thin wall approximation}

\begin{figure}
\centering
\includegraphics[width=0.7\textwidth]{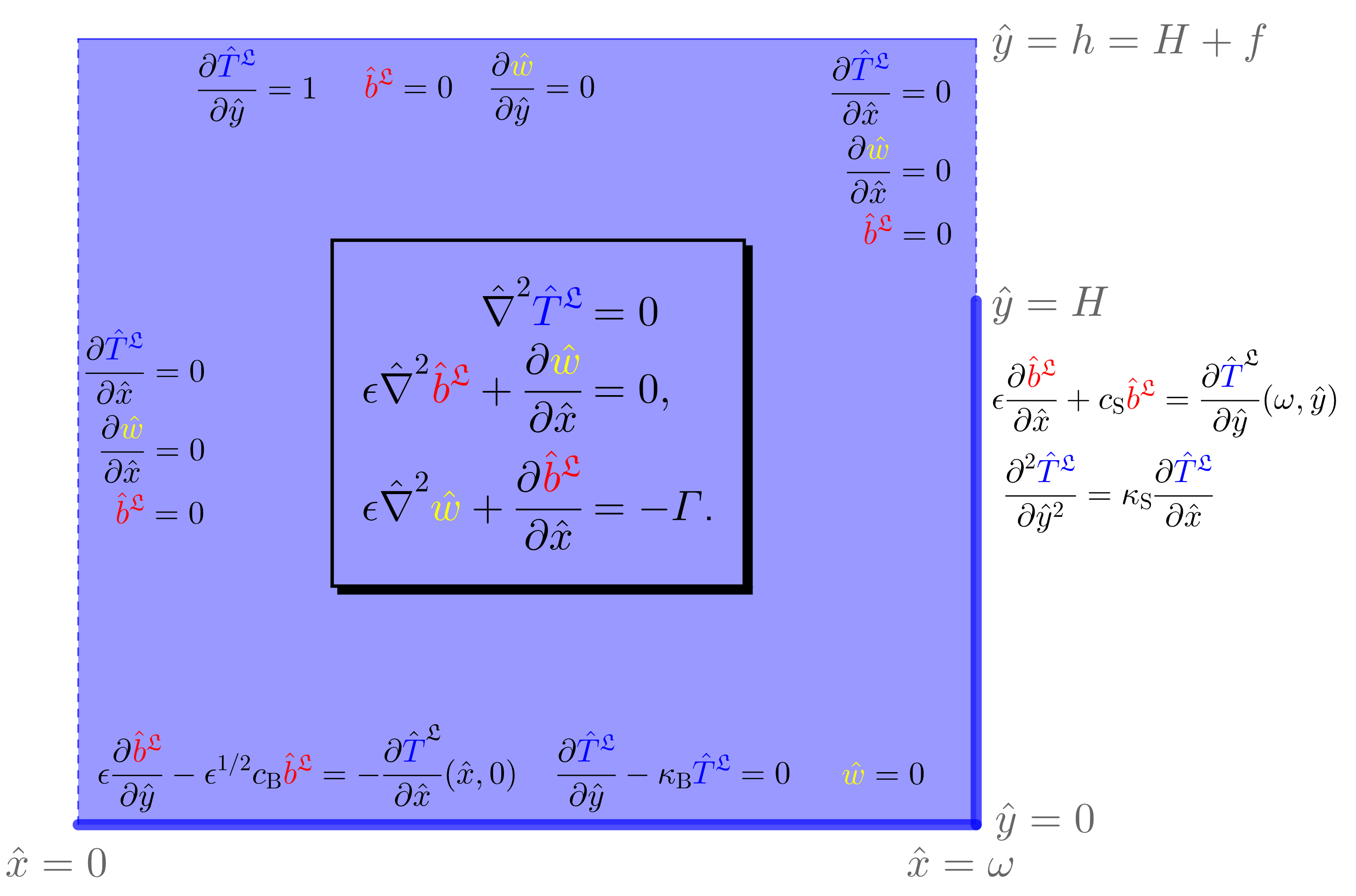}
\caption{Summary of the thin-wall problem for a rectangular trench.}
\label{fig:thin_wall_problem}
\end{figure}

In general, analytical progress can be made with the steady problem summarised in figure~\ref{fig:leading_order} only when the trench side walls and base are thin. 
In the limit where the wall thickness $\tau$ and base thickness $\varsigma$ tend to zero, the governing equations need to be solved only in the liquid domain, with the walls and base represented by effective boundary conditions, as
illustrated in figure~\ref{fig:thin_wall_problem}. Here we introduce several further dimensionless parameters, namely
\refstepcounter{equation}\label{eq:twparams}
\begin{align*}
c_{{\rm S}}&=\dfrac{\epsilon}{\Sigma_{\mathfrak{SL}}^{\mathrm{S}}\tau},
&
c_{{\rm B}}&=\dfrac{\epsilon^{1/2}}{\Sigma_{\mathfrak{SL}}^{\mathrm{B}}\varsigma},
&
\kappa_{{\rm S}}&=\dfrac{1}{K_{\mathfrak{SL}}^{{\rm S}}\tau},
&
\kappa_{{\rm B}}&=\dfrac{K_{\mathfrak{SL}}^{{\rm B}}}{\varsigma},
\tag{\textrm{\theequation}\textit{a--d}}
\end{align*}
%
in which the superscripts ``S'' and ``B'' denote the parameter values on the side walls and the base, respectively. 
At present, the side walls and base of the trench are made of the same material; however, the most interesting distinguished limit occurs when the above constants are all $\order(1)$ while $\epsilon$, $\tau$ and $\varsigma$ tend to zero.

We note that the temperature in the thin-wall problem is still decoupled from the magnetohydrodynamic problem.
In the regime where $K_\mathfrak{SL}^{\rm S}$ is large, the solution to the temperature problem in figure~\ref{fig:thin_wall_problem} is given by
\begin{equation}\label{eq:2dTrenchThinWallProblemTemperature}
     \inTheLiquid{\hat{T}}(\hat{x},\hat{y})=\hat{y}+\dfrac{1}{\kappa_{{\rm B}}},
\end{equation}
which is a linear function of $\hat{y}$, in approximate agreement with the full numerical solutions shown in figure~\ref{fig:global_trench}.
In general, however, the temperature must still be determined numerically, so in the subsequent analysis we solve the remaining magnetohydrodynamic problem, treating the liquid temperature as an in principle known function.
 
\section{Asymptotic analysis}\label{sec:asymptotics}

\subsection{Outer regions and boundary layers}
\label{ss:regions}

We use matched asymptotic expansions to analyse the thin-wall 2D TEMHD trench problem set out in figure~\ref{fig:thin_wall_problem} in the limit as $\epsilon\rightarrow0$.
To simplify matters, we take \mbox{$\psi=0$}, so that the applied magnetic field is strictly in the toroidal direction in a tokamak configuration.
The bulk flow inside the trench ($0<\dimensionless{y}<H$) is primarily driven by thermoelectric effects in the
boundary layer of thickness $\order(\epsilon)$ at the trench wall $\dimensionless{x}=\omega$.
The velocity in the outer region above the trench ($\dimensionless{y}>H$) is driven by gravity, and is an order of magnitude larger. 
These two regions are matched through a cross-over region near $\dimensionless{y}=H$ of thickness $\order(\epsilon^{1/2})$, and there is also a
``side'' layer near the base $\dimensionless{y}=0$ of thickness $\order(\epsilon^{1/2})$.
We will state the leading-order problems satisfied in these two layers, although they are not in general amenable to analytical solution.
We do not consider the further inner regions near the corners of the domain, first noted by \citet{Todd1967}, as they do not affect the leading-order outer solution.



\subsection{Outer problem above the trench}
Above the trench walls, byintegrating the Navier-Stokes equation in figure \ref{fig:thin_wall_problem} from $\dimensionless x=0$ to $\dimensionless x=\omega$, we find
\begin{equation}\label{eq:2dTrenchThinWallProblemNavierStokesEquationOuterIntegrateWRTx-3}
\dfrac{\rm d}{{\rm d}\dimensionless y}\int_{0}^{\omega}\hat{{w}}(\dimensionless x,\dimensionless y)\,{\rm d}\dimensionless x=\dfrac{\varGamma\omega}{\epsilon}\left(H+f-\dimensionless y\right)
\quad\text{in }H<\dimensionless y<H+f.
\end{equation}
From this solvability condition, we deduce that $\hat{{w}}$ must be order $1/\epsilon$
above the trench.
To match with an $\mathrm{O}(1)$ velocity field $\hat{w}$ inside the trench, we must have
\begin{equation}
\int_{0}^{\omega}\hat{{w}}(\dimensionless x,\dimensionless y)\,{\rm d}\dimensionless x=
\dfrac{\varGamma}{2\epsilon}(\dimensionless y-H)\left(H+2f-\dimensionless y\right)+C
\quad\text{in }H<\dimensionless y<H+f,
\end{equation}
where $C$ is an order-one constant.

We also have that $\hat{w}$ is independent of $\hat{x}$ to leading order, and thus
\begin{equation}\label{eq:2dTrenchThinWallProblemNavierStokesEquationAboveTrenchRescaled-AE-LO-Sol}
\hat{w}(\hat{x},\hat{y})\sim
\dfrac{\varGamma}{2\epsilon}(\dimensionless y-H)\left(H+2f-\dimensionless y\right)+\mathrm{O}(1)
\quad\text{in }H<\dimensionless y<H+f.
\end{equation}
By seeking further terms in an asymptotic expansion, one can show that the induced magnetic field $\inTheLiquid{\hat{b}}$ above the trench is zero to all algebraic orders in $\epsilon$, whilst the corrections to the velocity are all constants.

\subsection{Outer problem in the trench}


The governing equations in the bulk of the trench are shown inside the shadowed box in figure \ref{fig:thin_wall_problem}. 
At leading order, solutions to this problem inside the trench are given by
\begin{equation}
\hat{w}(\hat{x},\hat{y})\sim\hat{w}_0(\hat{y}),
\quad
\hat{b}(\hat{x},\hat{y})\sim-\Gamma\hat{x}
\quad\text{in }0<\hat{y}<H.
\end{equation}
The outer problem on its own does not completely determine the form of $\hat{w}_{0}(\dimensionless y)$, which must therefore be found by matching with the Hartmann layer at the trench wall.


\subsection{Inner problem near the right wall}

Letting $\dimensionless x=\omega-\epsilon\hat{X}_{\rm R}$, with ${\hat{w}(\dimensionless x,\dimensionless y)}={\hat{w}_{\rm R}(\dimensionless X_{\rm R},\dimensionless y)}$ and ${\hat{b}^{\mathfrak{L}}(\dimensionless x,\dimensionless y)}={\hat{b}_{\rm R}^{\mathfrak{L}}(\dimensionless X_{\rm R},\dimensionless y)}$, one obtains the leading-order problem%
\begin{subequations}
\label{eq:2dTrenchThinWallProblem-RightWall-Order0}
\begin{align}
\dfrac{\partial^{2}{\inTheLiquid{\dimensionless b}_{{\rm R},0}}}{\partial\dimensionless X_{\rm R}^{2}}-\dfrac{\partial{\dimensionless w_{{\rm R},0}}}{\partial\dimensionless X_{\rm R}} & =0\text{\ensuremath{\qquad}in }\dimensionless X_{\rm R}>0,\quad0<\hat{y}<H,\label{eq:2dTrenchThinWallProblemInductionEquationInnerRightOrder0}\\
\dfrac{\partial^{2}{{\dimensionless w_{{\rm R},0}}}}{\partial\dimensionless X_{\rm R}^{2}}-\frac{\partial{\inTheLiquid{\dimensionless b}_{{\rm R},0}}}{\partial\dimensionless X_{\rm R}} & =0\text{\ensuremath{\qquad}in }\dimensionless X_{\rm R}>0,\quad0<\hat{y}<H,\label{eq:2dTrenchThinWallProblemNavierStokesEquationInnerRightOrder0}\\
{\hat{w}_{{\rm R},0}} & =0\text{\ensuremath{\qquad}on }\dimensionless X_{\rm R}=0,\quad0<\hat{y}<H,\label{eq:2dTrenchThinWallProblemVelocityRightBC1InnerRightOrder0}\\
-\dfrac{\partial{\inTheLiquid{\dimensionless b}_{{\rm R},0}}}{\partial\dimensionless X_{\rm R}}+c_{{\rm S}}{\inTheLiquid{\dimensionless b}_{{\rm R},0}}&=\dfrac{\partial{\inTheLiquid{\dimensionless T}}}{\partial\dimensionless y}\left(\omega,\dimensionless y\right) \text{\ensuremath{\qquad}on }\dimensionless X_{\rm R}=0,\quad0<\hat{y}<H,\label{eq:2dTrenchThinWallProblemMagneticFieldRightBC1InnerRightOrder0}
\end{align}
along with leading-order matching conditions
\begin{align}
{\hat{w}_{{\rm R},0}(\dimensionless X_{\rm R},\dimensionless y)} & \to\hat{w}_{0}(\dimensionless y)\text{ as }\dimensionless X_{\rm R}\to\infty,\label{eq:2dTrenchThinWallProblemVelocityInnerRightMatch-Order0}\\
{{\hat{b}_{{\rm R},0}^{\mathfrak{L}}(\dimensionless X_{\rm R},\dimensionless y)}} & \to-\varGamma\omega\text{ as }\dimensionless X_{\rm R}\to\infty.\label{eq:2dTrenchThinWallProblemMagneticFieldInnerRightMatch-Order0}
\end{align}
\end{subequations}

The problem \eqref{eq:2dTrenchThinWallProblem-RightWall-Order0} has solution
\begin{subequations}
\begin{align}
{\inTheLiquid{\dimensionless b}_{{\rm R},0}}(\dimensionless X_{\rm R},\dimensionless y) & =\dfrac{1}{1+c_{{\rm S}}}\left(\dfrac{\partial{\inTheLiquid{\dimensionless T}}}{\partial\dimensionless y}\left(\omega,0\right)+\omega\varGamma c_{{\rm S}}\right){\rm e}^{-\dimensionless X_{\rm R}}-\omega\varGamma,\label{eq:2dTrenchThinWallProblemInductionEquationInnerRightOrder0Sol}\\
{\dimensionless w_{{\rm R},0}}(\dimensionless X_{\rm R},\dimensionless y) & =\dfrac{1}{1+c_{{\rm S}}}\left(\dfrac{\partial{\inTheLiquid{\dimensionless T}}}{\partial\dimensionless y}\left(\omega,0\right)+\omega\varGamma c_{{\rm S}}\right)\left(1-{\rm e}^{-\dimensionless{X}_{\rm R}}\right),\label{eq:2dTrenchThinWallProblemNavierStokesEquationInnerRightOrder0Sol}
\end{align}
which also determines the leading-order outer velocity in the trench, namely
\begin{align}
\hat{w}_{0}(\dimensionless y)&=\dfrac{1}{1+c_{{\rm S}}}\left(\dfrac{\partial{\inTheLiquid{\dimensionless T}}}{\partial\dimensionless y}\left(\omega,0\right)+\omega\varGamma c_{{\rm S}}\right).\label{eq:2dTrenchThinWallProblemInductionEquationOuterAEOrder0Sol-Velocity}
\end{align}
\end{subequations}
This simple formula for the bulk velocity demonstrates the importance of the temperature gradient in driving TEMHD flow. As will be demonstrated below, the expression \eqref{eq:2dTrenchThinWallProblemInductionEquationOuterAEOrder0Sol-Velocity} shows good agreement with numerical results for the bulk velocity, even when the side-wall thickness $\tau$ is not especially small. The formula \eqref{eq:2dTrenchThinWallProblemInductionEquationOuterAEOrder0Sol-Velocity} also resembles that obtained by in \citep{Xu2015} in the limit of large $\numberHartmann$.

\subsection{Composite approximations}

The solutions we have obtained thus far allow us to create composite expressions which provide good approximations to the velocity, magnetic field and temperature profiles inside the trench, even when the walls are no longer thin,
provided we are not close to any horizontal boundary layers near the top or bottom of the trench.
In $0<\hat{y}<H$ and $0<\hat{x}<\omega-\tau$ we have
\begin{subequations}\label{pdh:comp1}
\begin{align}
{{{\dimensionless w}}}(\dimensionless x,\dimensionless y) & \approx\dfrac{1}{1+c_{{\rm S}}}\left(\dfrac{\partial{\inTheLiquid{\dimensionless T}}}{\partial\dimensionless y}\left(\omega-\tau,\dimensionless y\right)+(\omega-\tau)\varGamma c_{{\rm S}}\right)\left(1-2{\rm e}^{-\left(\omega-\tau\right)/\epsilon}\cosh\left(\dfrac{\dimensionless x}{\epsilon}\right)\right),\label{eq:2dTrenchThinWallProblem-ApproxHorizontalProfile-NonThin-Velocity}\\
{\inTheLiquid{\dimensionless b}}(\dimensionless x,\dimensionless y) & \approx\dfrac{2}{1+c_{{\rm S}}}\left(\dfrac{\partial\inTheLiquid{\dimensionless T}}{\partial\dimensionless y}(\omega-\tau,\dimensionless y)+(\omega-\tau)\varGamma c_{{\rm S}}\right){\rm e}^{-(\omega-\tau)/\epsilon}\sinh\left(\dfrac{\dimensionless x}{\epsilon}\right)-\varGamma\dimensionless x.\label{eq:2dTrenchThinWallProblem-ApproxHorizontalProfile-NonThin-BField}
\end{align}
\end{subequations}
From equation \eqref{eq:2dTrenchThinWallProblem-ApproxHorizontalProfile-NonThin-BField}, one can construct an approximate expression for the overall induced magnetic field, including that inside the solid, assuming that it is approximately linear inside the trench wall, i.e.,
\begin{equation}
{\dimensionless b}(\dimensionless x,\dimensionless y)\approx\begin{cases}
{\inTheLiquid{\dimensionless b}}(\dimensionless x,\dimensionless y) & \text{for }0\le\dimensionless x\le\omega-\tau,\\
{\inTheLiquid{\dimensionless b}}(\omega-\tau,\dimensionless y)\left(\dfrac{\omega-\dimensionless x}{\tau}\right) & \text{for }\omega-\tau\le\dimensionless x\le\omega.
\end{cases}\label{eq:2dTrenchThinWallProblem-ApproxHorizontalProfile-NonThin-BField-Overall}
\end{equation}
Furthermore, we can use the solution for the temperature, given by equation \eqref{eq:2dTrenchThinWallProblemTemperature}, to construct an approximate vertical temperature profile in the middle of the trench. This approximation is given by 
\begin{equation}
T(0,y)\approx\begin{cases}\displaystyle
\frac{\hat{y}}{K_{\mathfrak{SL}}^{{\rm B}}} & \text{if }0<\hat{y}\le\varsigma,\\[4mm]\displaystyle
\hat{y} + \frac{\varsigma}{K_{\mathfrak{SL}}^{{\rm B}}}-\varsigma & \text{if }\varsigma<\hat{y}\le H+f,\label{eq:2DTrenchApproximateVerticalTemperature}
\end{cases}
\end{equation}
and applies when $K_{\mathfrak{SL}}^{{\rm B}}$ is small. 


\subsection{Inner problem near the base}
The external magnetic field applied tangentially to the base of the trench causes a so-called side layer of thickness $\mathrm{O}(\epsilon^{1/2})$.
Upon rescaling $\dimensionless y=\epsilon^{1/2}\dimensionless Y_{\mathrm B}$, with ${\hat{w}(\dimensionless x,\dimensionless y)}={\hat{w}_{\mathrm B}(\hat{x},\dimensionless Y_{\mathrm B})}$ and ${\hat{b}^{\mathfrak{L}}(\dimensionless x,\dimensionless y)}={\hat{b}_{\mathrm B}^{\mathfrak{L}}(\dimensionless x,\dimensionless Y_{\mathrm B})}$, we obtain leading-order problem
\begin{subequations}
\label{eq:2DTrenchThinWallProblemBottomBL-Order0}
\begin{align}
\dfrac{\partial^{2}{\inTheLiquid{\dimensionless b}_{{\rm B},0}}}{\partial\hat{Y}_{\mathrm B}^{2}}+\dfrac{\partial{\dimensionless w_{{\rm B},0}}}{\partial\dimensionless x} & =0\text{\ensuremath{\qquad}in }\dimensionless Y_{\mathrm B}>0,\quad0<\hat{x}<\omega,\label{eq:2dTrenchThinWallProblemInductionEquationBottomOrder0}\\
\dfrac{\partial^{2}{{\hat{w}_{{\rm B},0}}}}{\partial\hat{Y}_{\mathrm B}^{2}}+\frac{\partial{\inTheLiquid{\dimensionless b}_{{\rm B},0}}}{\partial\dimensionless x}+\varGamma & =0\text{\ensuremath{\qquad}in }\dimensionless Y_{\mathrm B}>0,\quad0<\hat{x}<\omega,\label{eq:2dTrenchThinWallProblemNavierStokesEquationBottomOrder0}
\end{align}
with boundary conditions
\begin{align}
&&~\hat{w}_{{\rm B},0}=
\dfrac{\partial{\inTheLiquid{\dimensionless b}_{{\rm B},0}}}{\partial\hat{Y}_{\mathrm B}}-c_{{\rm B}}{\inTheLiquid{\dimensionless b}_{{\rm B},0}} & =0
&&\text{on}\quad\hat{Y}_{\mathrm B}=0,
&&~\\&&~
\dfrac{\partial{\hat{w}_{{\rm B},0}}}
{\partial\dimensionless x}=
{\inTheLiquid{\dimensionless b}_{{\rm B},0}}
& =0
&&\text{on}\quad\hat{x}=0,
&&~
\end{align}
%
and far-field conditions
\begin{align}
{\hat{w}_{{\rm B},0}}
\to\dfrac{1}{1+c_{{\rm S}}}\left(\dfrac{\partial{\inTheLiquid{\dimensionless T}}}{\partial\dimensionless y}\left(\omega,0\right)+\omega\varGamma c_{{\rm S}}\right),
\quad
{\inTheLiquid{\dimensionless b}_{{\rm B},0}}
\to-\varGamma\dimensionless x\
&&&
\text{as}\quad\hat{Y}_{\mathrm B}\to\infty,
\label{eq:2dTrenchThinWallProblemMagneticFieldInnerBottomMatchOrder0}
\end{align}
as well as the matching condition
\begin{equation}
\left(1+c_{{\rm S}}\right){\dimensionless w_{{\rm B},0}}+c_{{\rm S}}{{\inTheLiquid{\dimensionless b}_{{\rm B},0}}}=\dfrac{\partial{\inTheLiquid{\dimensionless T}}}{\partial\dimensionless y}\left(\omega,0\right)\qquad\text{at }\dimensionless x=\omega.\label{eq:2dTrenchThinWallProblemMagneticFieldInnerBottomRightMatchOrder0}
\end{equation}
\end{subequations}
The effective boundary condition \eqref{eq:2dTrenchThinWallProblemMagneticFieldInnerBottomRightMatchOrder0} 
is derived by analysing
%
the ``inner-bottom-right corner'' problem
below in \S\ref{ss:brc}.

It is possible to decompose the problem
\eqref{eq:2DTrenchThinWallProblemBottomBL-Order0}
into two independent problems, each of which has only one inhomogeneous boundary condition.
Let us write the liquid velocity and induced magnetic field in this region in the forms
\begin{subequations}
\begin{align}
{\dimensionless w_{{\rm B},0}}(\dimensionless x,\dimensionless Y_{\mathrm B}) & =
\dfrac{1}{1+c_{{\rm S}}}\left(\dfrac{\partial{\inTheLiquid{\dimensionless T}}}{\partial\dimensionless y}\left(\omega,0\right)+\omega\varGamma c_{{\rm S}}\right)
\left(1+{\hat{w}}_{{\rm B},0}^{[1]}(\dimensionless x,\dimensionless Y_{\mathrm B})\right)
+\varGamma{\hat{w}}_{{\rm B},0}^{[2]}(\dimensionless x,\dimensionless Y_{\mathrm B}),
\\
{{\inTheLiquid{\dimensionless b}_{{\rm B},0}}}(\dimensionless x,\dimensionless Y_{\mathrm B}) & =
\dfrac{1}{1+c_{{\rm S}}}\left(\dfrac{\partial{\inTheLiquid{\dimensionless T}}}{\partial\dimensionless y}\left(\omega,0\right)+\omega\varGamma c_{{\rm S}}\right){\hat{b}_{{\rm B},0}^{\regionLithium[1]}}(\dimensionless x,\dimensionless Y_{\mathrm B})
+\varGamma\left(-\dimensionless x
+\hat{b}_{{\rm B},0}^{\regionLithium[2]}(\dimensionless x,\dimensionless Y_{\mathrm B})
\right).
\end{align}
\end{subequations}
We then need to solve the homogeneous PDEs ($j=1,2$)
\begin{subequations}
\begin{align}
\dfrac{\partial^{2}\hat{b}_{\text{B},0}^{\regionLithium[j]}}
{\partial\hat{Y}_{\text{B}}^{2}}
+\dfrac{\partial{\dimensionless w}_{\text{B},0}^{[j]}}{\partial\dimensionless x} =
\dfrac{\partial^{2}{{\hat{w}}_{\text{B},0}^{[j]}}}{\partial\hat{Y}_{\text{B}}^{2}}
+\frac{\partial\hat{b}_{\text{B},0}^{\regionLithium[j]}}{\partial\dimensionless x} & =0
\quad\text{in}\quad\dimensionless Y_{\text{B}}>0,\quad0<\hat{x}<\omega,
\end{align}
subject to the boundary conditions
\begin{align}&&~
\dfrac{\partial{\hat{w}_{\text{B},0}^{[j]}}}{\partial\dimensionless x} =
\hat{b}_{\text{B},0}^{\regionLithium[j]} & =0
&&\text{on}\quad\dimensionless x=0,
&&~
\\ &&~
(1+c_\text{S})\hat{w}_{\text{B},0}^{[j]} +c_\text{S}
\hat{b}_{\text{B},0}^{\regionLithium[j]} & =0
&&\text{on}\quad\dimensionless x=\omega,
&&~
\\ &&~
\hat{w}_{\text{B},0}^{[j]},~
\hat{b}_{\text{B},0}^{\regionLithium[j]} & \rightarrow0
&&\text{as}\quad\hat{Y}_\text{B}\rightarrow\infty,
&&~
\end{align}
and
\begin{align}\label{eq:wbjfinalbc}
{\hat{w}_{\text{B},0}^{[j]}} & =-\delta_{1j},
&
\dfrac{\partial\hat{b}_{\text{B},0}^{\regionLithium[j]}}{\partial\hat{Y}_{\text{B}}}-c_{{\rm B}}
\hat{b}_{\text{B},0}^{\regionLithium[j]} & =-\delta_{2j}c_{{\rm B}}\dimensionless x
&\text{on}\quad
\hat{Y}_{\text{B}}&=0.
\end{align}
%
\end{subequations}

In either of the two cases, the solution to the problem may be expressed in the form
\begin{subequations}\label{pdh:wbss}
\begin{align}
{\dimensionless w}_{{\rm B},0}^{[j]}(\dimensionless{x},\dimensionless{Y}_{\mathrm B}) & =\phantom{-}\operatorname{Re}\left[\sum_{n=0}^{\infty}{\cal U}_{n}^{[j]}{\rm e}^{-\mu_{n}\dimensionless Y_{\mathrm B}}\cosh\mu_{n}^{2}\dimensionless x\right]\\
\hat{b}_{\text{B},0}^{\regionLithium[j]}(\dimensionless{x},\dimensionless{Y}_{\mathrm B}) & =-\operatorname{Re}\left[\sum_{n=0}^{\infty}{\cal U}_{n}^{[j]}{\rm e}^{-\mu_{n}\dimensionless Y_{\mathrm B}}\sinh\mu_{n}^{2}\dimensionless x\right],
\end{align}
\end{subequations}
where
\begin{align}\mu_{n}=\dfrac{1}{\sqrt{\omega}}\sqrt{\phi+\left(n+\frac{1}{2}\right){\rm i}\pi}\quad\text{with }\quad\phi=\textrm{arctanh}\left(\dfrac{c_{\rm S}}{1 + c_{\rm S}} \right),\label{eq:2dTrenchThinWallProblemNavierStokesEquationBottom-Sol-mu}
\end{align}
The complex coefficients ${\cal U}_n$ are determined in principle from the final inhomogeneous boundary condition \eqref{eq:wbjfinalbc}, i.e.,
\begin{subequations}
    \begin{align}
\operatorname{Re}\left[\sum_{n=0}^{\infty}{\cal U}_{n}^{[j]}\cosh\mu_{n}^{2}\dimensionless x\right] & =\begin{cases}
-1 & \text{if }j=1,\\
0 & \text{if }j=2,
\end{cases}\\
\operatorname{Re}\left[\sum_{n=0}^{\infty}\left(\mu_{n}+c_{{\rm B}}\right){\cal U}_{n}^{[j]} \sinh\mu_{n}^{2}\dimensionless x\right] & =\begin{cases}
0 & \text{if }j=1,\\
-c_{{\rm B}}\dimensionless x & \text{if }j=2.
\end{cases}
\end{align}
\end{subequations}
It is not straightforward to evaluate the coefficients ${\cal U}_n$ in general. We present below two limiting cases in which they can be found analytically using Fourier series. 

\begin{figure}
\begin{minipage}{0.5\textwidth}\centering\textbf{(a)}\\
    \begin{overpic}[width=0.8\textwidth]{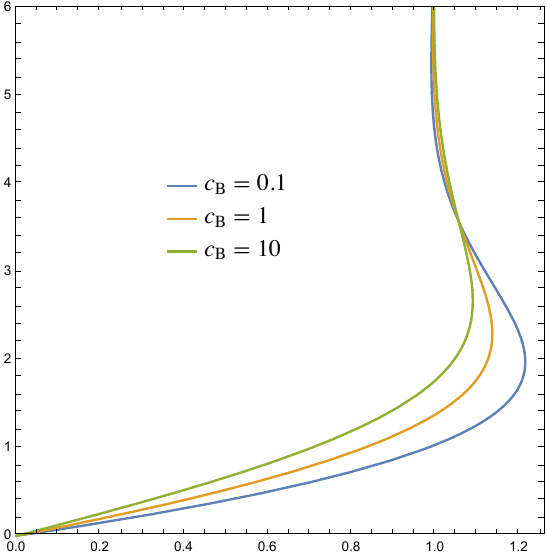}
\put(35,-7){$1+\hat{w}^{[1]}_{\text{B},0}(0,\hat{Y}_\text{B})$}
\put(-7,50){$\hat{Y}_\text{B}$}
\end{overpic}
\end{minipage}\hfill%
\begin{minipage}{0.5\textwidth}\centering\textbf{(b)}\\
    \begin{overpic}[width=0.8\textwidth]{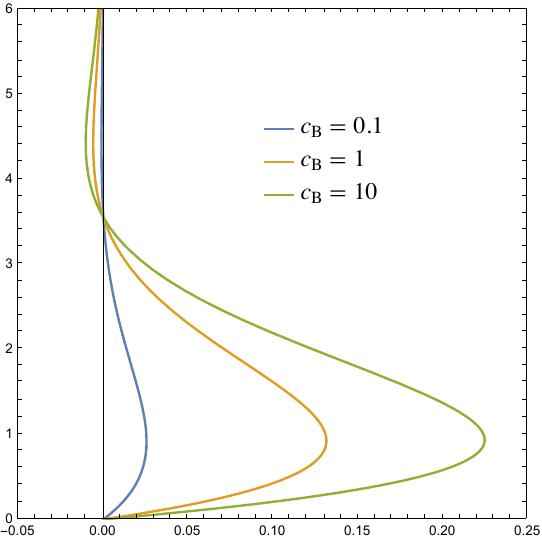}
\put(40,-7){$\hat{w}^{[2]}_{\text{B},0}(0,\hat{Y}_\text{B})$}
\put(-7,50){$\hat{Y}_\text{B}$}
\end{overpic}
\end{minipage}\\~\\~\\
\caption{The contributions $1+\hat{w}^{[1]}_{\text{B},0}$ and $\hat{w}^{[2]}_{\text{B},0}$ to the fluid velocity plotted versus $\hat{Y}_\text{B}$ at the centre of the trench in Case~\textbf{(I)}: $c_\text{S}\rightarrow0$.}
    \label{fig:w12case1}
\end{figure}
\begin{figure}
\begin{minipage}{0.5\textwidth}\centering \textbf{(a)}\\
    \begin{overpic}[width=0.8\textwidth]{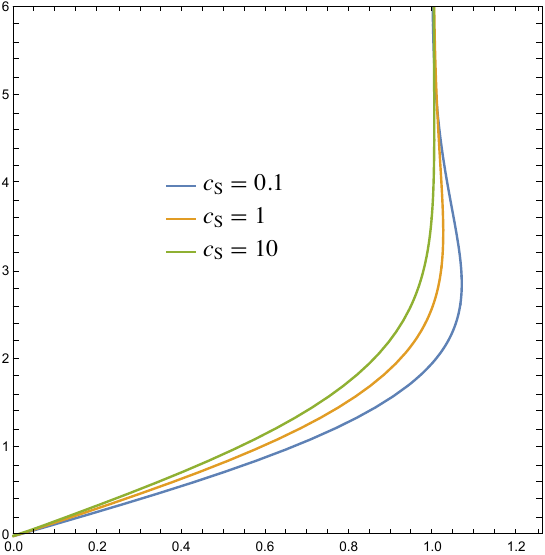}
\put(35,-7){$1+\hat{w}^{[1]}_{\text{B},0}(0,\hat{Y}_\text{B})$}
\put(-7,50){$\hat{Y}_\text{B}$}
\end{overpic}
\end{minipage}\hfill%
\begin{minipage}{0.5\textwidth}\centering \textbf{(b)}\\
    \begin{overpic}[width=0.8\textwidth]{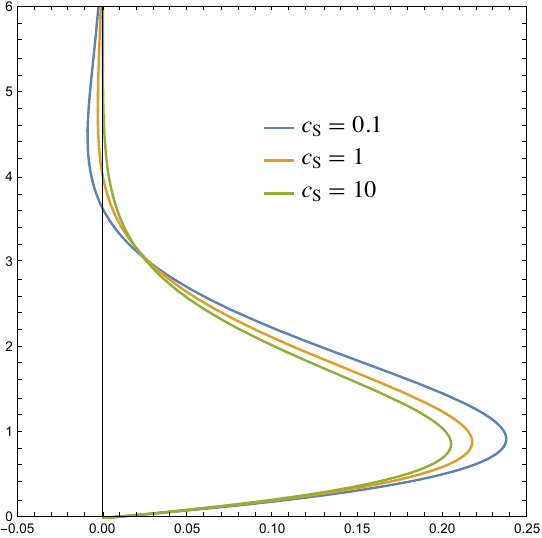}
\put(40,-7){$\hat{w}^{[2]}_{\text{B},0}(0,\hat{Y}_\text{B})$}
\put(-7,50){$\hat{Y}_\text{B}$}
\end{overpic}
\end{minipage}\\~\\~\\
\caption{The contributions $1+\hat{w}^{[1]}_{\text{B},0}$ and $\hat{w}^{[2]}_{\text{B},0}$ to the fluid velocity plotted versus $\hat{Y}_\text{B}$ at the centre of the trench in Case~\textbf{(II)}: $c_\text{B}\rightarrow0$.}
    \label{fig:w12case2}
\end{figure}

\subsubsection{Case \textbf{\textup{(I)}}: $c_{\rm S}\to0$}

For $\mathrm{O}(1)$ conductivity ratio, this first case corresponds to the industrially relevant regime where the Hartmann layer is much narrower than the trench side-wall.
In this limit, we have $\phi\to 0$, and the coefficients in the series solution \eqref{pdh:wbss} are given by
\begin{subequations}
\begin{align}
{\cal U}_{n}^{[1]} & =\dfrac{2(-1)^{n+1}}{\left(n+\frac{1}{2}\right)\pi}\left(1-\dfrac{{\rm i}\sqrt{\left(n+\frac{1}{2}\right)\pi}}{\sqrt{\left(n+\frac{1}{2}\right)\pi}+c_{{\rm B}}\sqrt{2\omega}}\right),\\
{\cal U}_{n}^{[2]} & =\dfrac{{\rm i}(-1)^{n}(2\omega)^{3/2}c_{{\rm B}}}{\left(\sqrt{\left(n+\frac{1}{2}\right)\pi}+c_{{\rm B}}\sqrt{2\omega}\right)\left(n+\frac{1}{2}\right)^{2}\pi^{2}}.
\end{align}
\end{subequations}
The contributions $1+\hat{w}^{[1]}_{\text{B},0}$ and $\hat{w}^{[2]}_{\text{B},0}$ to the fluid velocity in this limit are plotted in figure~\ref{fig:w12case1}. Here we set $\omega=1$, without loss of generality, and plot the velocities evaluated at the trench centre-line $\hat{x}=0$ versus the scaled vertical coordinate $\hat{Y}_B$.

Figure~\ref{fig:w12case1}\textbf{(a)} shows the contribution to the velocity generated by the temperature gradient through the Seebeck effect, while figure~\ref{fig:w12case1}\textbf{(b)}
shows the contribution due to the gravitational forcing along the trench. In both cases, we observe the the expected jet caused by conductivity of the trench base, with the velocity significantly over-shooting its far-field limit as well as a noticeable return flow.

\subsubsection{Case \textbf{\textup{(II)}}: $c_{\rm S
B}\to\infty$}\mbox{}\\ 
This case corresponds to the regime where the trench base is much thinner than the side boundary layer, which is of order $\epsilon^{1/2}$. Although it is less likely to be practically relevant, for completeness we note that the coefficients in this limiting case are given by
\begin{equation}
{\cal U}_{n}^{[1]}=\dfrac{2(-1)^{n+1}{\rm i}\cosh\phi}{\mu_{n}^{2}\omega},\qquad{\cal U}_{n}^{[2]}=2(-1)^{n+1}{\rm i}\left(\dfrac{\cosh\phi}{\mu_{n}^{4}\omega}-\dfrac{\sinh\phi}{\mu_{n}^{2}}\right).
\end{equation}

The resulting solutions for the velocity contributions are plotted in figure~\ref{fig:w12case2}. Again we observe wall jets near the trench base, with behaviour qualitatively similar to that in figure~\ref{fig:w12case1}.

\subsection{Inner problem in bottom-right corner}
\label{ss:brc}

After rescaling both $\dimensionless x=\omega-\epsilon\dimensionless X_{\rm R}$
and $\dimensionless y=\epsilon^{1/2}\dimensionless Y_{\mathrm B}$, with ${\hat{w}(\dimensionless x,\dimensionless y)}={\hat{w}_{\rm BR}(\hat{X}_{\rm R},\dimensionless Y_{\mathrm B})}$
and ${\hat{b}^{\mathfrak{L}}(\dimensionless x,\dimensionless y)}={\hat{b}_{\rm BR}^{\mathfrak{L}}(\dimensionless X_{\rm R},\dimensionless Y_{\mathrm B})}$, we obtain the leading-order problem
\begin{subequations}\label{eq:2dTrenchThinWallProblemBottomRightAEOrdero0}
\begin{align}
\dfrac{\partial^{2}{\inTheLiquid{\dimensionless b}_{{\rm BR},0}}}{\partial\dimensionless X_{\rm R}^{2}}-\dfrac{\partial{\dimensionless w_{{\rm BR},0}}}{\partial\dimensionless X_{\rm R}} =
\dfrac{\partial^{2}{\dimensionless w_{{\rm BR},0}}}{\partial\dimensionless X_{\rm R}^{2}}-\frac{\partial{\inTheLiquid{\dimensionless b}_{{\rm BR},0}}}{\partial\dimensionless X_{\rm R}} & =0\text{\ensuremath{\qquad}in }\hat{X}_{\rm R}>0,\quad\dimensionless Y_{\mathrm B}>0,\label{eq:2dTrenchThinWallProblemNavierStokesEquation-BottomRight-AE-Order0}
\end{align}
along with boundary and matching conditions
\begin{align}\label{Brprobb}
{\hat{w}_{{\rm BR},0}} =0,
\quad
-\dfrac{\partial{\inTheLiquid{\dimensionless b}_{{\rm BR},0}}}{\partial\hat{X}_{\rm R}}+c_{{\rm S}}{\inTheLiquid{\dimensionless b}_{{\rm BR},0}}&=\dfrac{\partial{\inTheLiquid{\dimensionless T}}}{\partial\dimensionless y}\left(\omega,0\right)
&&\text{on }\hat{X}_{\rm R}=0,
\\
{\hat{w}_{{\rm BR},0}}(\dimensionless X_{\rm R},\dimensionless Y_{\mathrm B}) \to{\hat{w}_{{\rm B},0}}(\omega,\dimensionless Y_{\mathrm B}),
\quad
{{\inTheLiquid{\dimensionless b}_{{\rm BR},0}}}(\dimensionless X_{\rm R},\dimensionless Y_{\mathrm B}) & \to{\inTheLiquid{\dimensionless b}_{{\rm BR},0}}(\omega,\dimensionless Y_{\mathrm B})
&&\text{as }\dimensionless X_{\rm R}\to\infty,
\end{align}
\end{subequations}
and
\begin{subequations}\label{BRprobY0}
\begin{align}
{\hat{w}_{{\rm BR},0}} & =0\quad\text{on }\hat{Y}_{\mathrm B}=0,
\\
\dfrac{\partial{\inTheLiquid{\dimensionless b}_{{\rm BR},0}}}{\partial\hat{Y}_{\mathrm B}}+c_{{\rm B}}{\inTheLiquid{\dimensionless b}_{{\rm BR},0}} & =0\text{\ensuremath{\quad}on }\hat{Y}_{\mathrm B}=0.
\end{align}
\end{subequations}

The solution to  problem \eqref{eq:2dTrenchThinWallProblemBottomRightAEOrdero0} is given by
\begin{subequations}\label{BRprobsol}
\begin{align}
{\dimensionless w_{{\rm BR},0}}(\dimensionless X_{\rm R},\dimensionless Y_{\mathrm B}) & ={\hat{w}_{{\rm B},0}}(\omega,\dimensionless Y_{\mathrm B})\left(1-{\rm e}^{-\dimensionless X_{\rm R}}\right),
\\
{{\inTheLiquid{\dimensionless b}_{{\rm BR},0}}}(\dimensionless X_{\rm R},\dimensionless Y_{\mathrm B}) & ={\inTheLiquid{\dimensionless b}_{{\rm B},0}}(\omega,\dimensionless Y_{\mathrm B})+{\hat{w}_{{\rm B},0}}(\omega,\dimensionless Y_{\mathrm B}){\rm e}^{-\dimensionless X_{\rm R}}.
\end{align}
\end{subequations}
where the boundary condition \eqref{Brprobb} is satisfied provided
\begin{equation}
\left(1+c_{{\rm S}}\right){\dimensionless w_{{\rm B},0}}(\omega,\dimensionless Y_{\mathrm B})+c_{{\rm S}}{{\inTheLiquid{\dimensionless b}_{{\rm B},0}}}(\omega,\dimensionless Y_{\mathrm B})=\dfrac{\partial{\inTheLiquid{\dimensionless T}}}{\partial\dimensionless y}\left(\omega,0\right).
\end{equation}
Thus, as promised, we obtain the matching condition \eqref{eq:2dTrenchThinWallProblemMagneticFieldInnerBottomRightMatchOrder0} required to solve the inner problem near the base.

The solution \eqref{BRprobsol} 
depends only parametrically on $\dimensionless{Y}_{\mathrm B}$, and therefore cannot in general satisfy the final boundary conditions \eqref{BRprobY0}.
This observation indicates the presence of a further inner layer inside the corner region. We do not present the governing equations in this inner-inner region, as they do not appear to be analytically tractable and do not affect the outer solution, at leading order at least
\citep{Cook1972}.

\section{Parameter sweeps}\label{sec:parameter_sweeps}

\subsection{Parameter values and geometrical ratios}

We now carry out some parameter sweeps on the full steady trench problem shown in figure~\ref{fig:leading_order}, keeping all parameter values fixed at the ``default'' values in table~\ref{tab:model_dimensionless_parameters}, unless specified otherwise. The numerical solutions shown (solid) include comparison with  predictions \eqref{pdh:comp1}--\eqref{eq:2DTrenchApproximateVerticalTemperature} provided by the asymptotics (dashed). To facilitate varying the trench geometry while avoiding unphysical parameter regimes,
we introduce various physically relevant geometrical ratios: the \textit{aspect ratio} $r_{{\rm A}} = {H}/{\omega}$, the \textit{overfilling ratio} $r_{{\rm F}} = {\omega f}/{(\omega-\tau)(H-\varsigma)}$, the \textit{side-wall fraction} $r_{{\rm S}}={\tau}/{\omega}$ and the \textit{base fraction} $r_{{\rm B}}={\varsigma}/{H}$.


In performing parameter sweeps, we show the effect of changing each parameter on the horizontal and vertical velocity profiles, as well as the horizontal magnetic field profiles. In this context, by ``horizontal'' and ``vertical'' profiles, we mean one-dimensional slices through the bisecting lines 
$\dimensionless{y}=\tfrac{1}{2}(H+\varsigma)$ and
$\dimensionless{x}=0$, respectively.
The vertical magnetic field profiles are ommitted because $\hat{b}(0,\hat{y})$ is identically zero.
The temperature is plotted only in cases where it is affected by the parameter being swept over.


\begin{figure}
   \begin{minipage}{0.5\textwidth}
    \centering
    \textbf{(a)}\\
    \includegraphics[width=\textwidth]{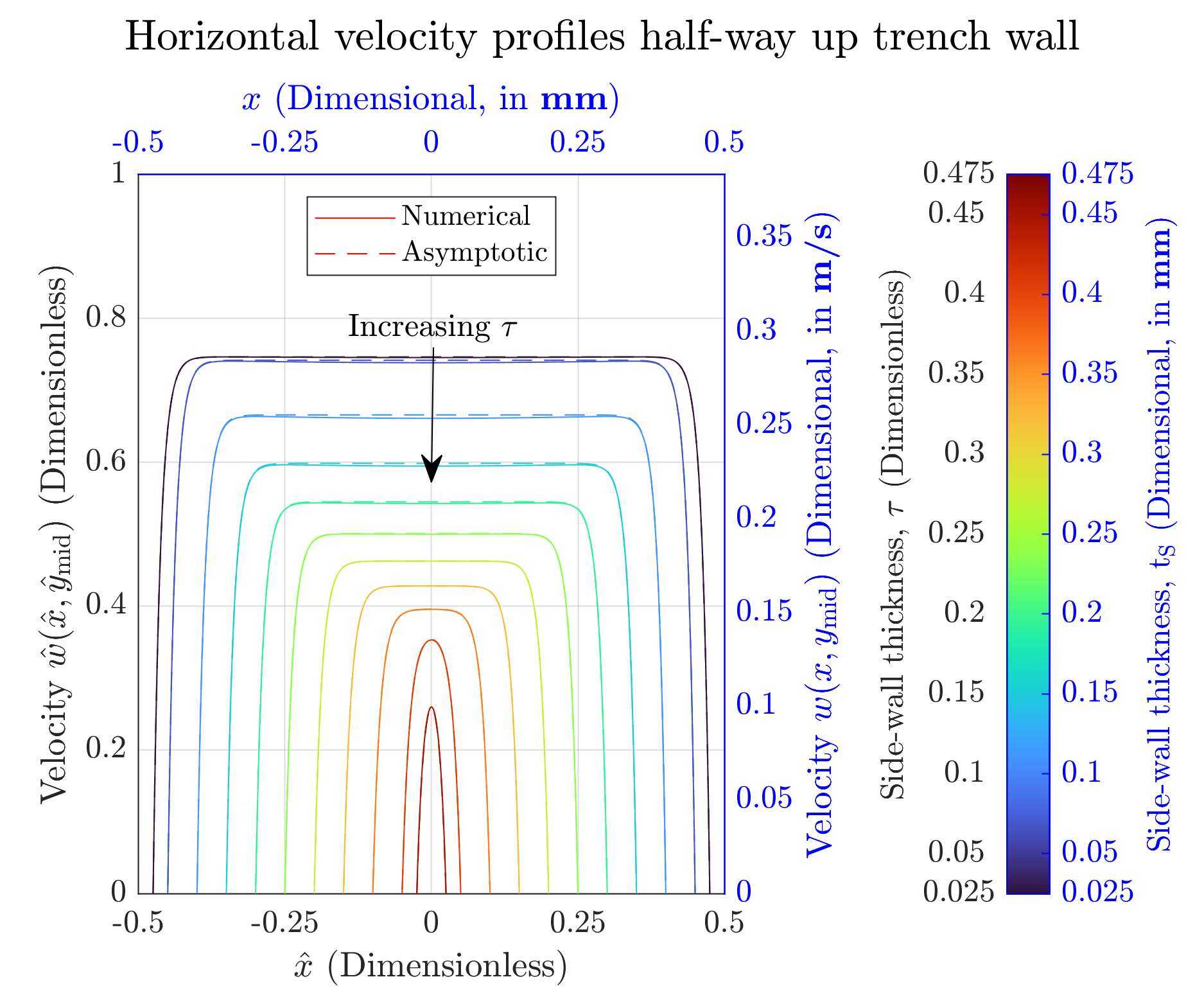}\\
    \textbf{(b)}\\
    \includegraphics[width=\textwidth]{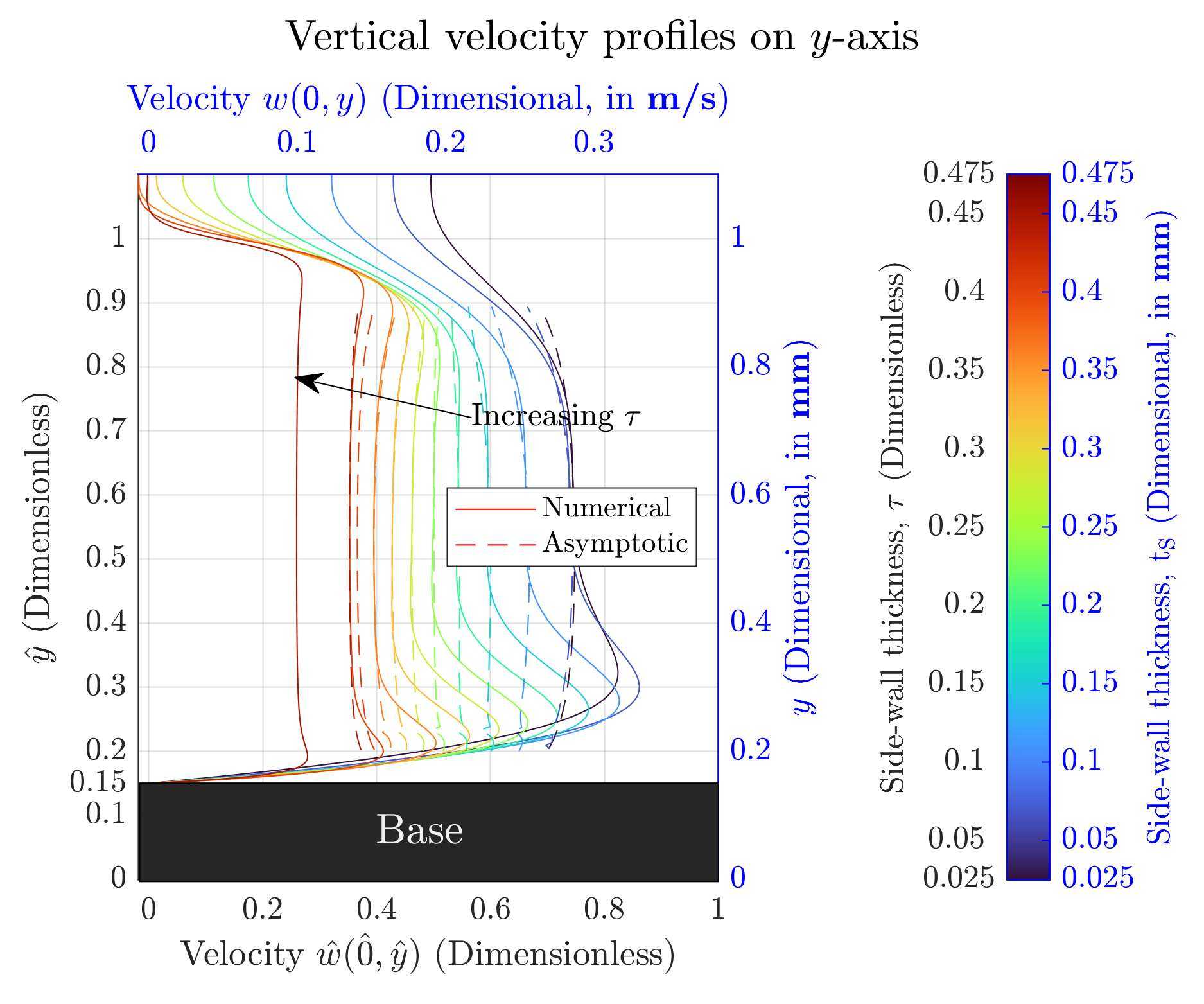} \\
    \textbf{(c)}\\
    \includegraphics[width=\textwidth]{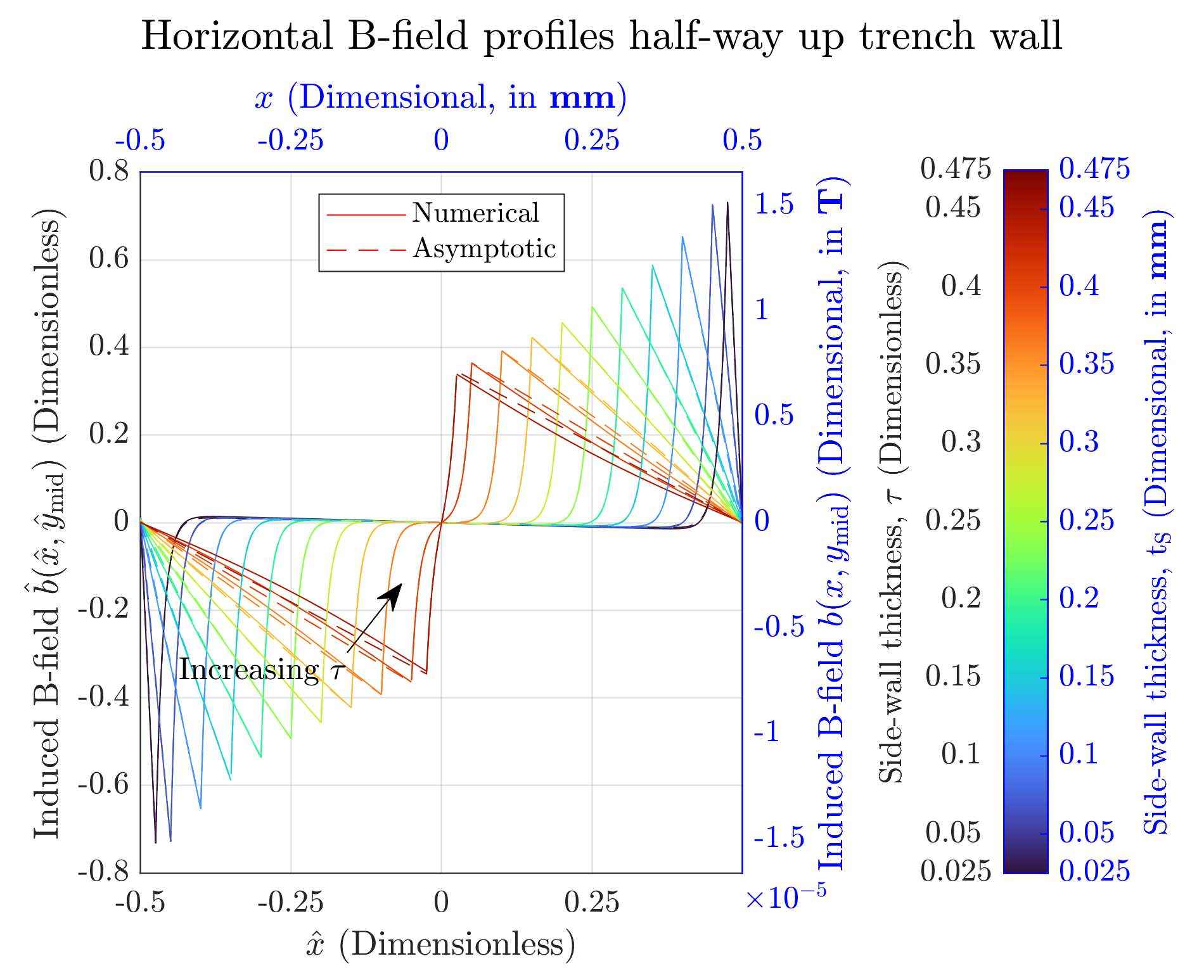}
    \end{minipage}
    \begin{minipage}{0.5\textwidth}
    \centering
    \textbf{(d)}\\
    \includegraphics[width=\textwidth]{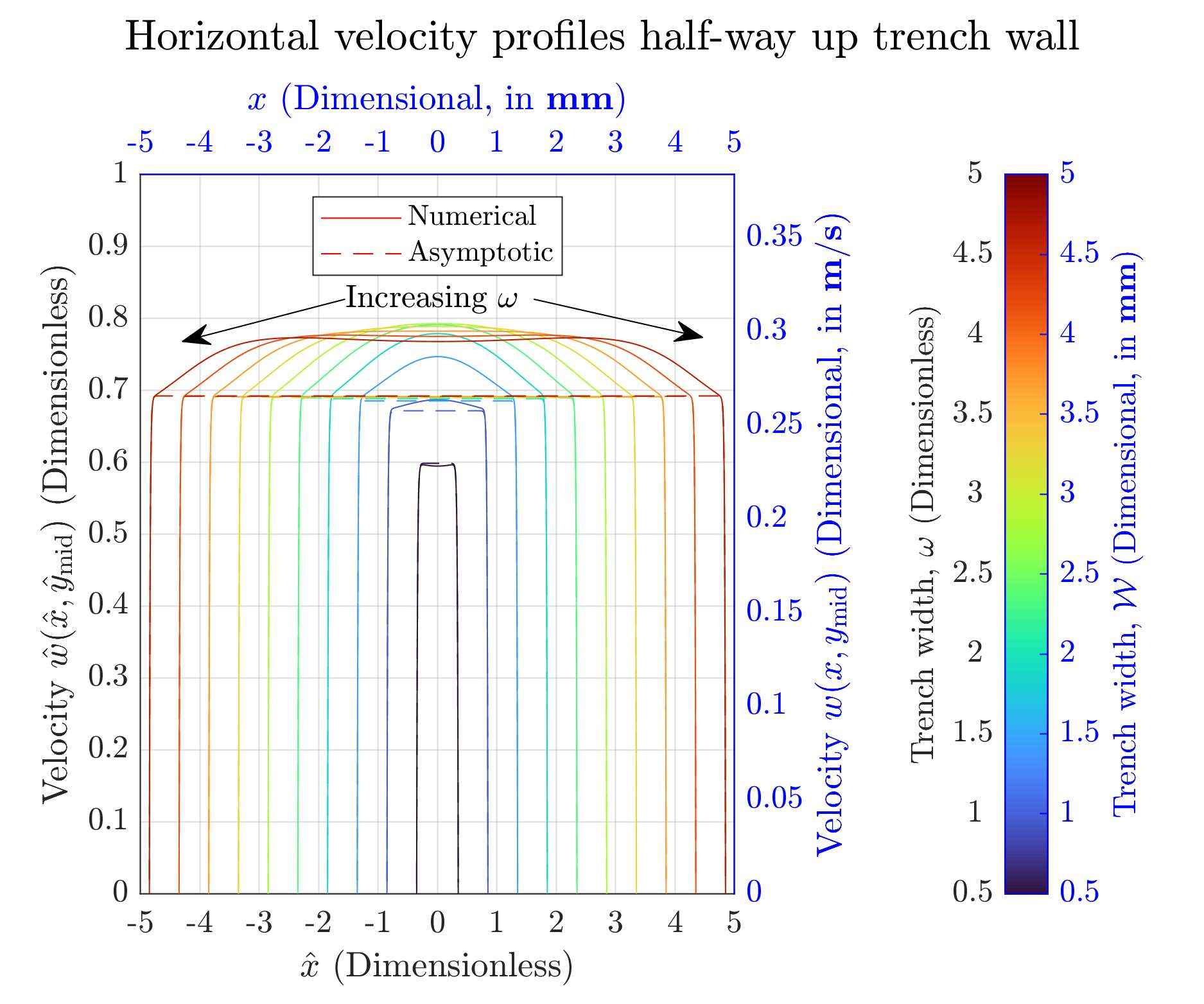}\\
    \textbf{(e)}\\
    \includegraphics[width=\textwidth]{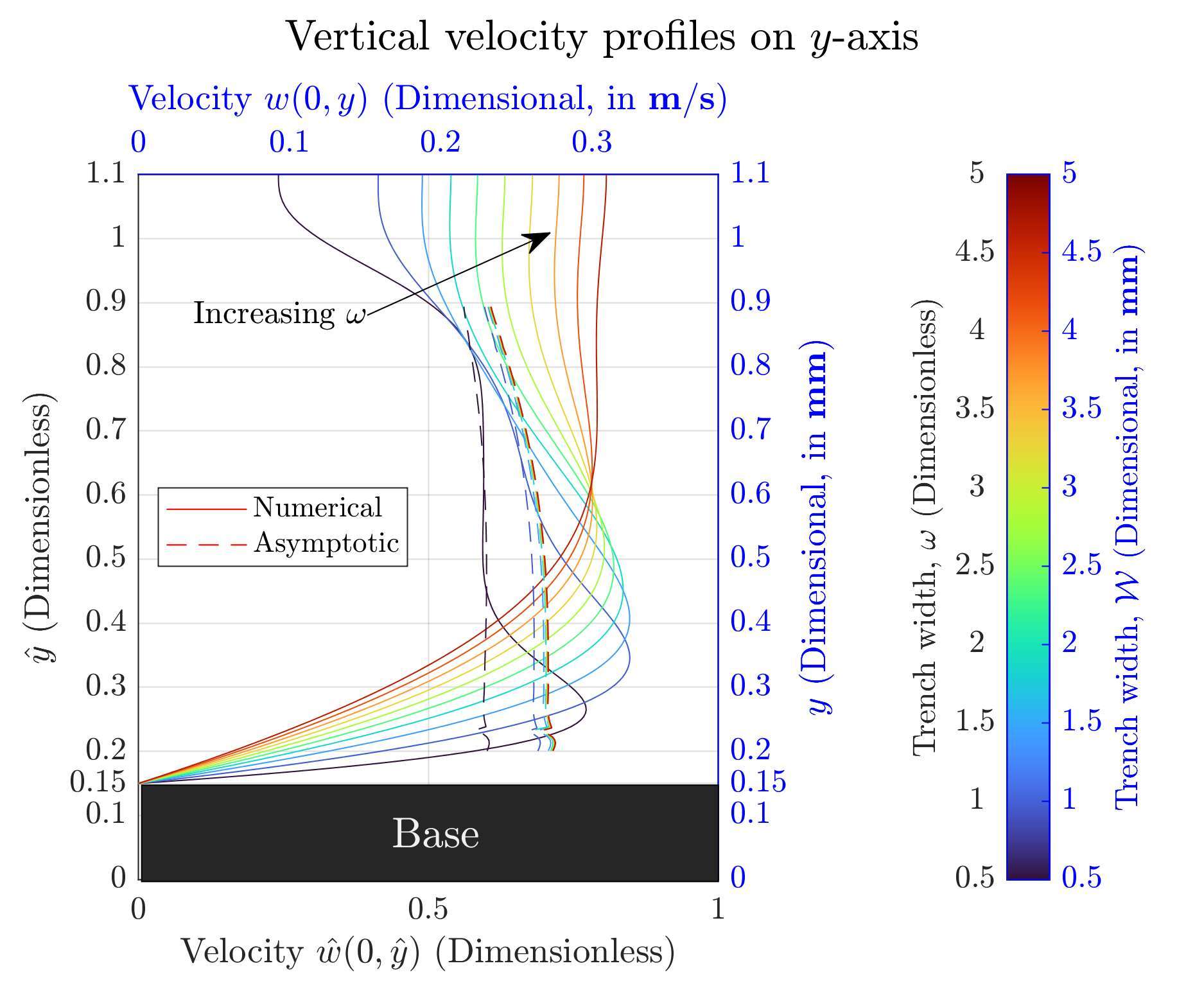}\\
    \textbf{(f)}\\
    \includegraphics[width=\textwidth]{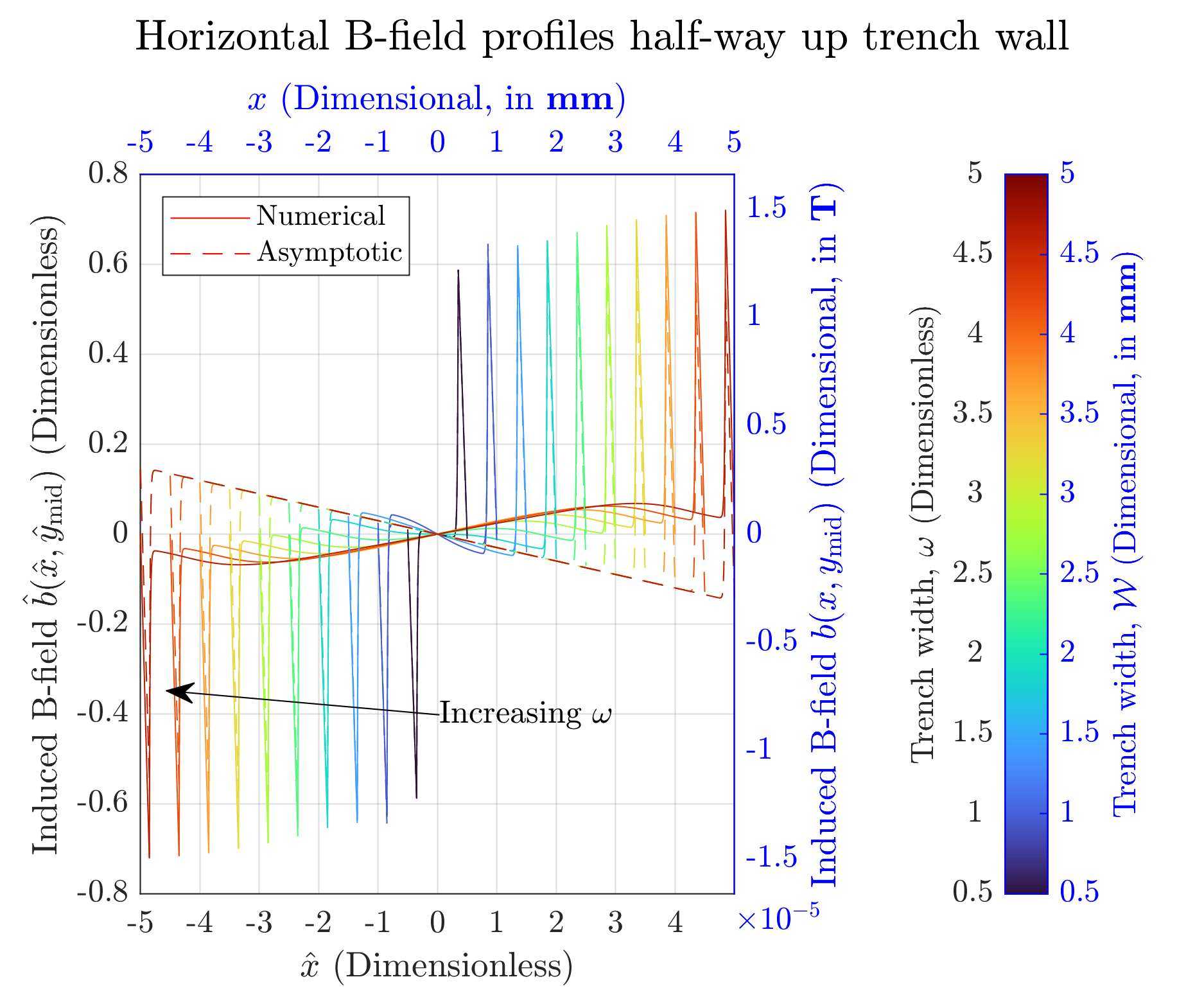}\\
    \end{minipage}
    \caption{Numerical plots of \textbf{(a,d)} horizontal velocity profiles half-way up the trench wall, \textbf{(b,e)} vertical velocity profiles along the $\dimensionless{y}$-axis, and \textbf{(c,f)} horizontal magnetic field profiles half-way up the trench wall, for dimensionless side-wall thickness $\tau \in \{ 0.005, 0.05, 0.1,\ldots,0.45, 0.495 \}$ (in \textbf{(a,b,c)}) and dimensionless trench width $\omega \in \{0.4,0.6,\ldots,1.8,2 \}$ (in \textbf{(d,e,f)}).} 
    
    \label{fig:increasing_tau_and_omega}
\end{figure}

\subsection{Changing the horizontal dimensions \texorpdfstring{$\tau$}{tau} and \texorpdfstring{$\omega$}{omega}}

The effects of changing the side wall thickness $\tau$ while keeping the total trench width $\omega$ fixed are shown in figures \ref{fig:increasing_tau_and_omega}\textbf{(a-c)}.
The horizontal velocity profiles \textbf{(a)} demonstrate
the anticipated flow structure, with the velocity approximately uniform outside Hartmann layers at the trench walls
Increasing $\tau$ has the effect of both increasing the net wall conductivity and squeezing the fluid into a narrower channel, and thus decreases the maximum velocity of the fluid enclosed between the walls.
For $\tau\gtrsim0.35$,
the Hartmann layers appear to merge and create an approximately parabolic velocity profile.
The dashed curves show that the numerical results are in excellent agreement with with the asymptotic prediction \eqref{eq:2dTrenchThinWallProblem-ApproxHorizontalProfile-NonThin-Velocity}. 

The horizontal magnetic field profiles in plot~\textbf{(c)} show approximate piecewise linear behaviour, which becomes steeper and more pronounced when the walls become thinner, resulting in a stronger electrical current density passing through them, since their electrical conductivity remains constant. Outside the Hartmann layers, the induced field is very small as expected, given the small value of $\varGamma$ here, namely 0.0299.

Plot~\textbf{(b)} shows that the boundary layer at the bottom of the trench becomes thinner and less curved as the wall thickness increases, suppressing the velocity jet. The asymptotic prediction \eqref{eq:2dTrenchThinWallProblem-ApproxHorizontalProfile-NonThin-Velocity} works well inside the trench and away from the bottom boundary layer. Above the trench, the velocity seems to approach a constant value somewhat lower than that inside the trench. Making the walls thicker causes the speed of the flow above the walls to decrease because of the increased drag due to the tops of the trench walls. Also, the ``cross-over region'' near the tops of the walls becomes narrower and less curved as the side walls are brought closer together.

Figure~\ref{fig:increasing_tau_and_omega}\textbf{(d-f)} shows the effects of varying the dimensionless trench width $\omega$ whilst keeping the side wall thickness $\tau$ fixed.
%
In
figure~\ref{fig:increasing_tau_and_omega}\textbf{(d)}, we see that the approximately uniform core velocity predicted by the asymptotics (dashed curves) stays almost constant, as now only the trench width is varying, but not the effective wall conductivity.
However, the numerical solutions (solid curves) show that the velocity inside the trench continues to increase slightly with increasing $\omega$, with a central bulge in the velocity profile.
Plot \textbf{(e)} shows the cause of this discrepancy: as the trench becomes wider, the conducting-wall jet near the base becomes elongated, thus being captured by the central horizontal line $\dimensionless{y} = \tfrac{1}{2}(H+\varsigma)$. The intensity of the magnetic field dipole shown in plot \textbf{(f)} does not vary as significantly as in plot \textbf{(c)}, because the effective conductivity of the side walls does not change as they move apart. However, the magnetic field gradient in the bulk changes sign as $\omega$ increases, again because of infiltration by the bottom boundary layer.

\begin{figure}
   \begin{minipage}{0.5\textwidth}
    \centering
    \textbf{(a)}\\
    \includegraphics[width=\textwidth]{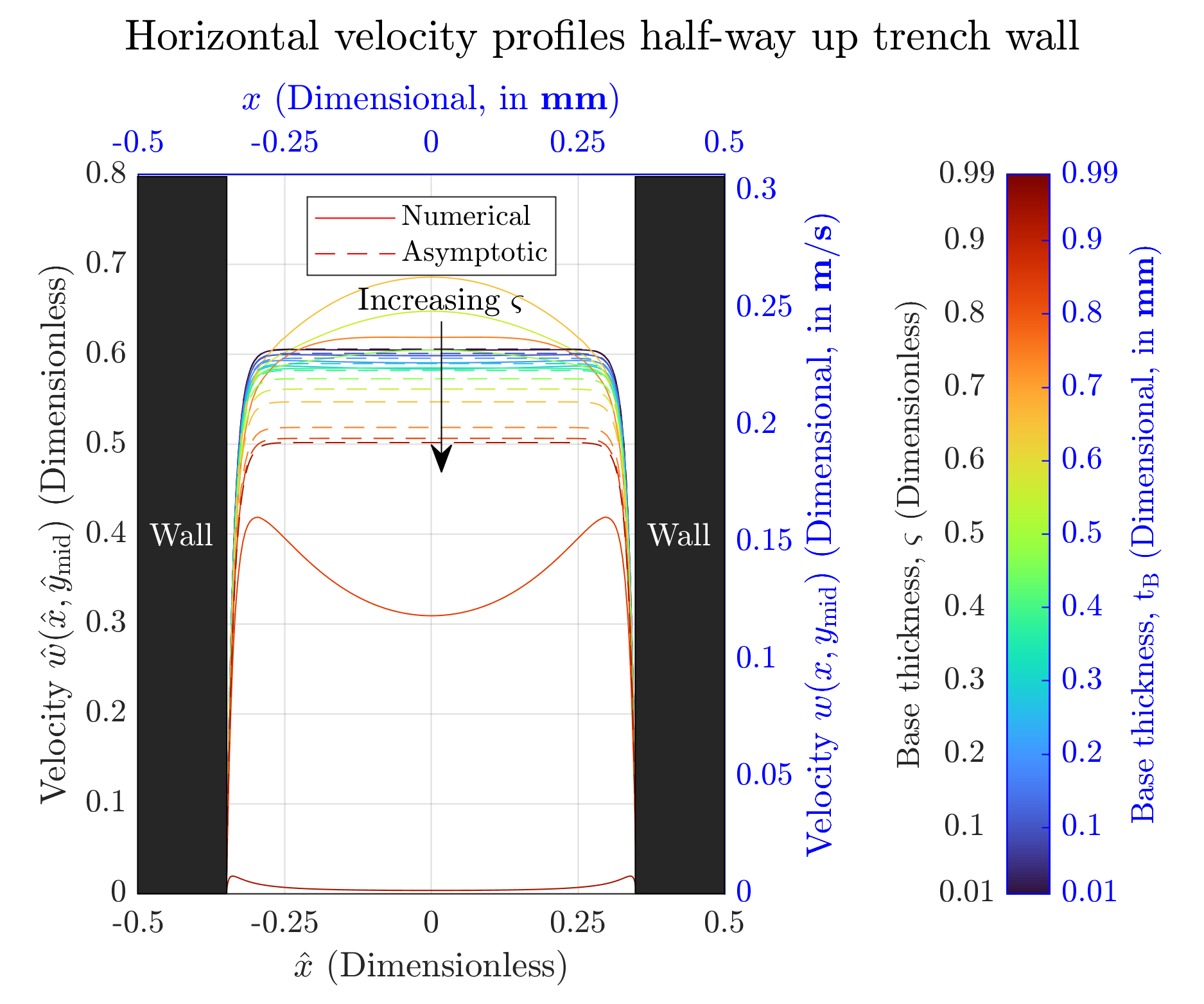}\\
    \textbf{(b)}\\
    \includegraphics[width=\textwidth]{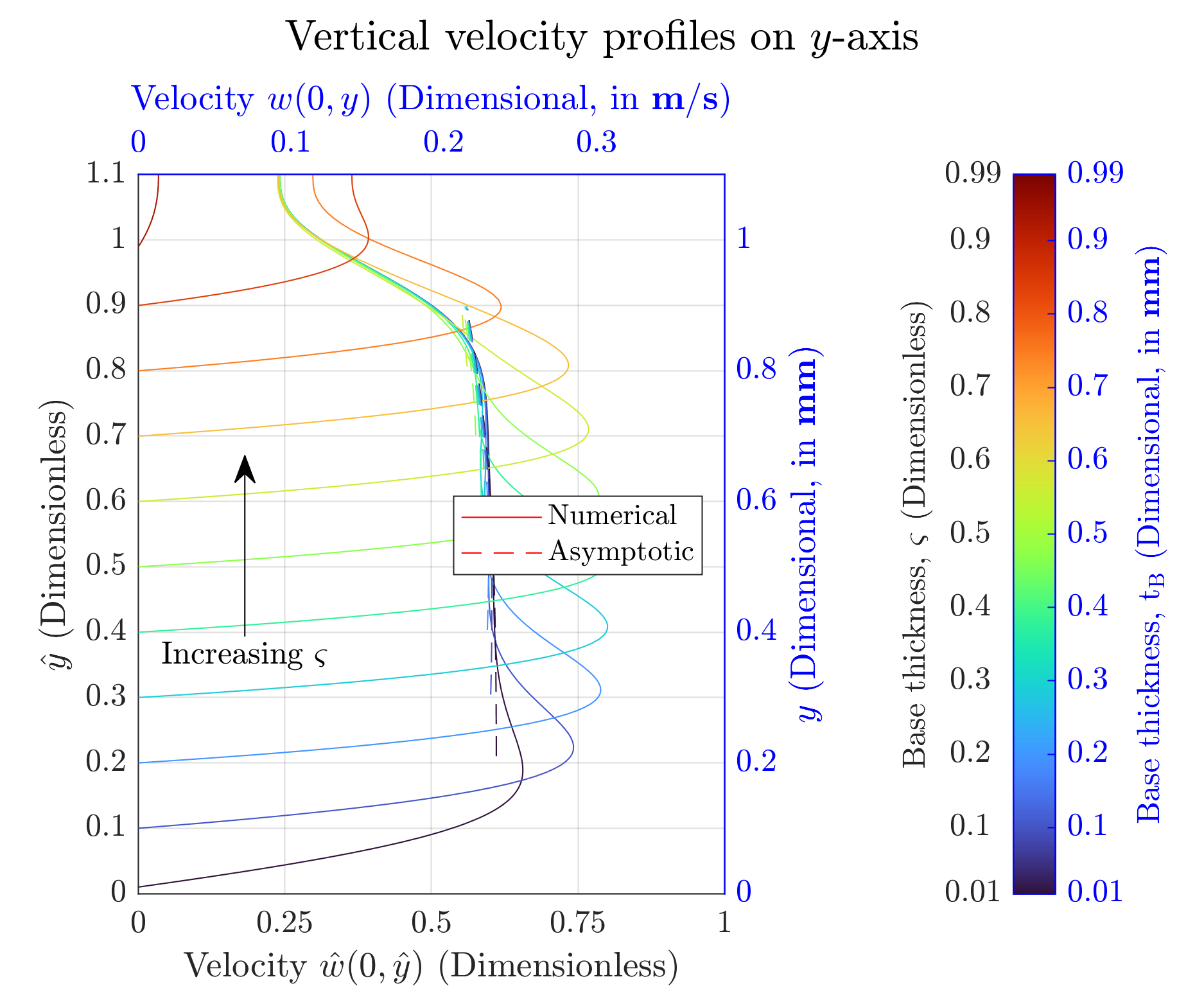} \\
    \textbf{(c)}\\
    \includegraphics[width=\textwidth]{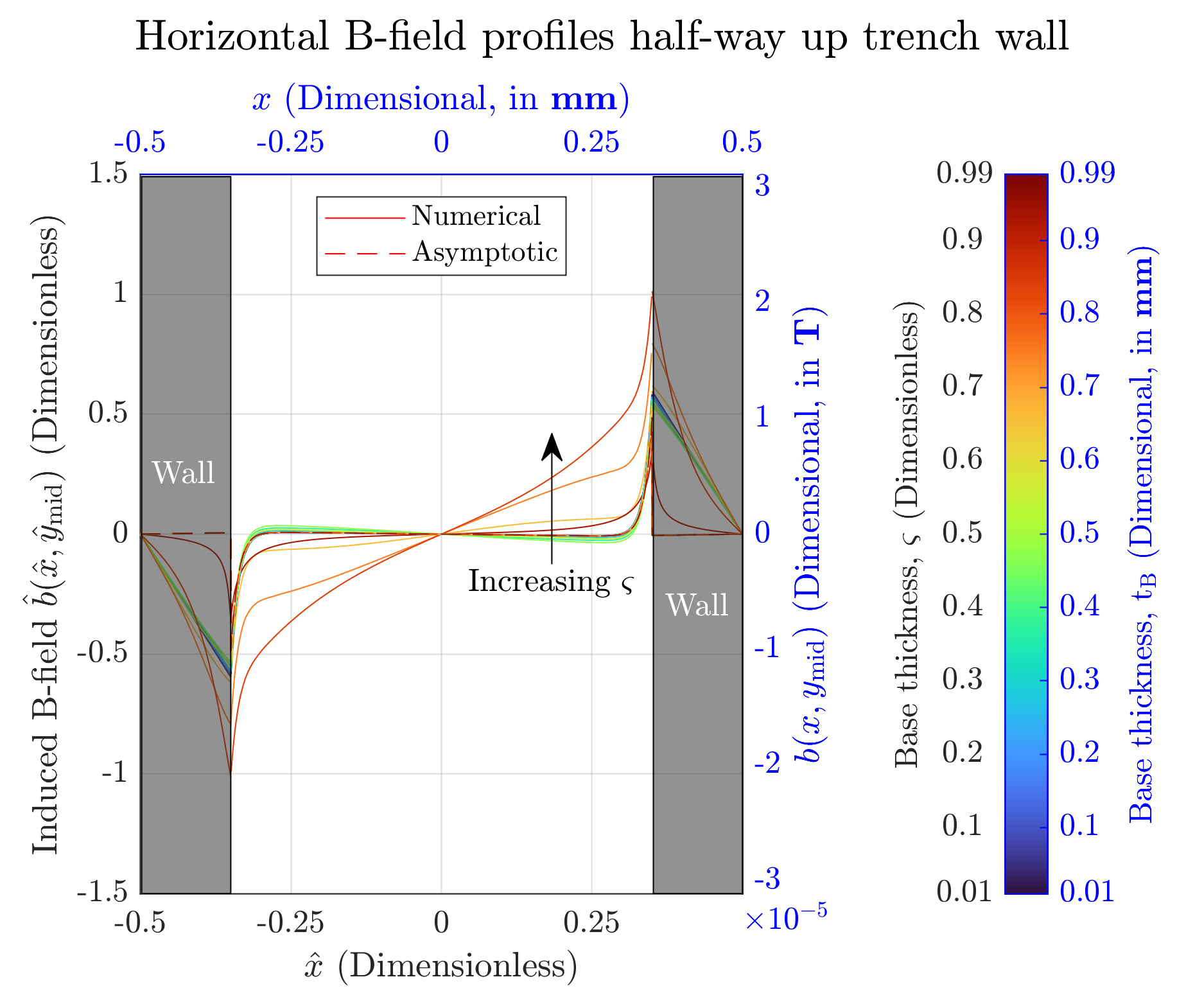}
    \end{minipage}
    \begin{minipage}{0.5\textwidth}
    \centering
    \textbf{(d)}\\
    \includegraphics[width=\textwidth]{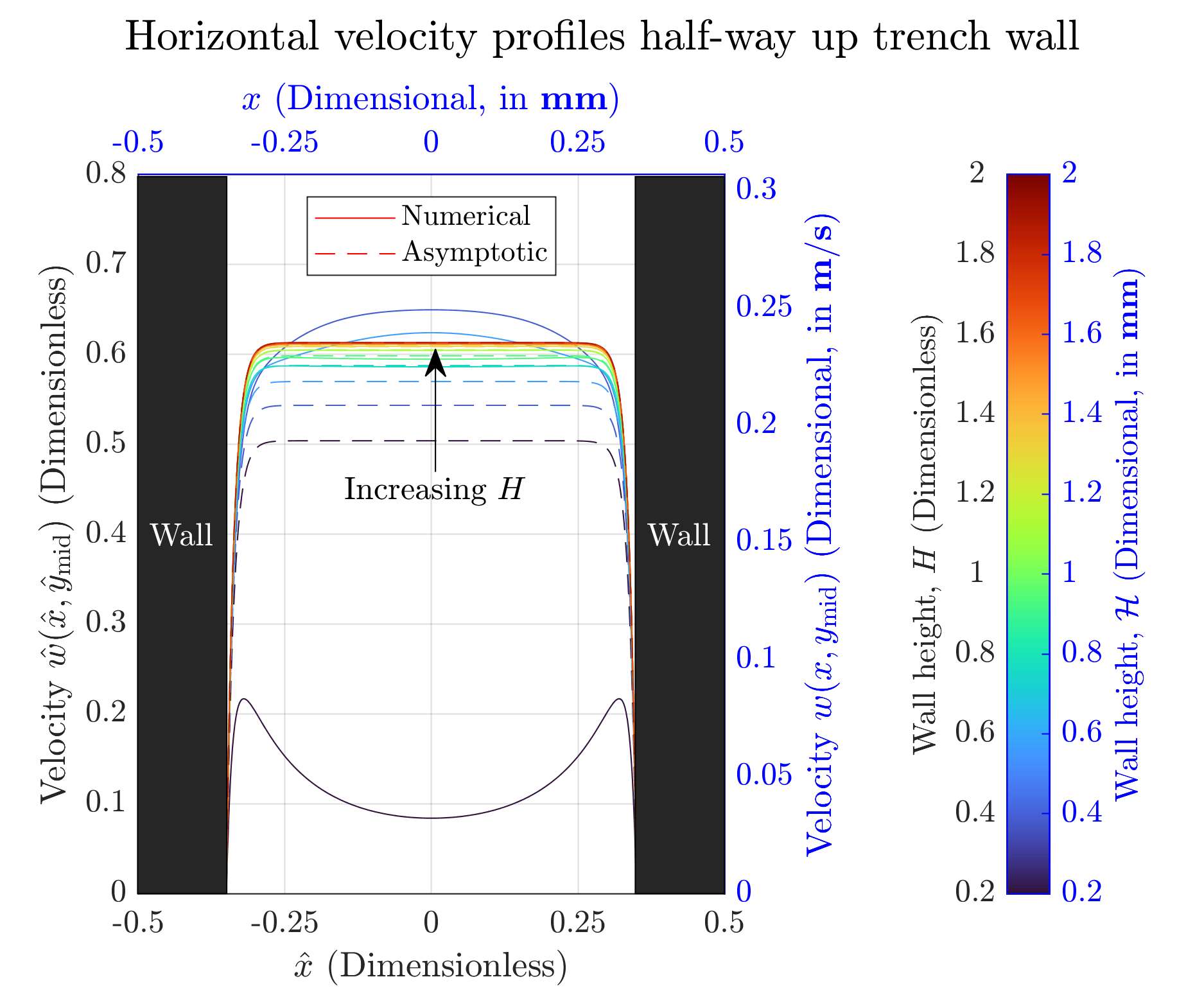}\\
    \textbf{(e)}\\
    \includegraphics[width=\textwidth]{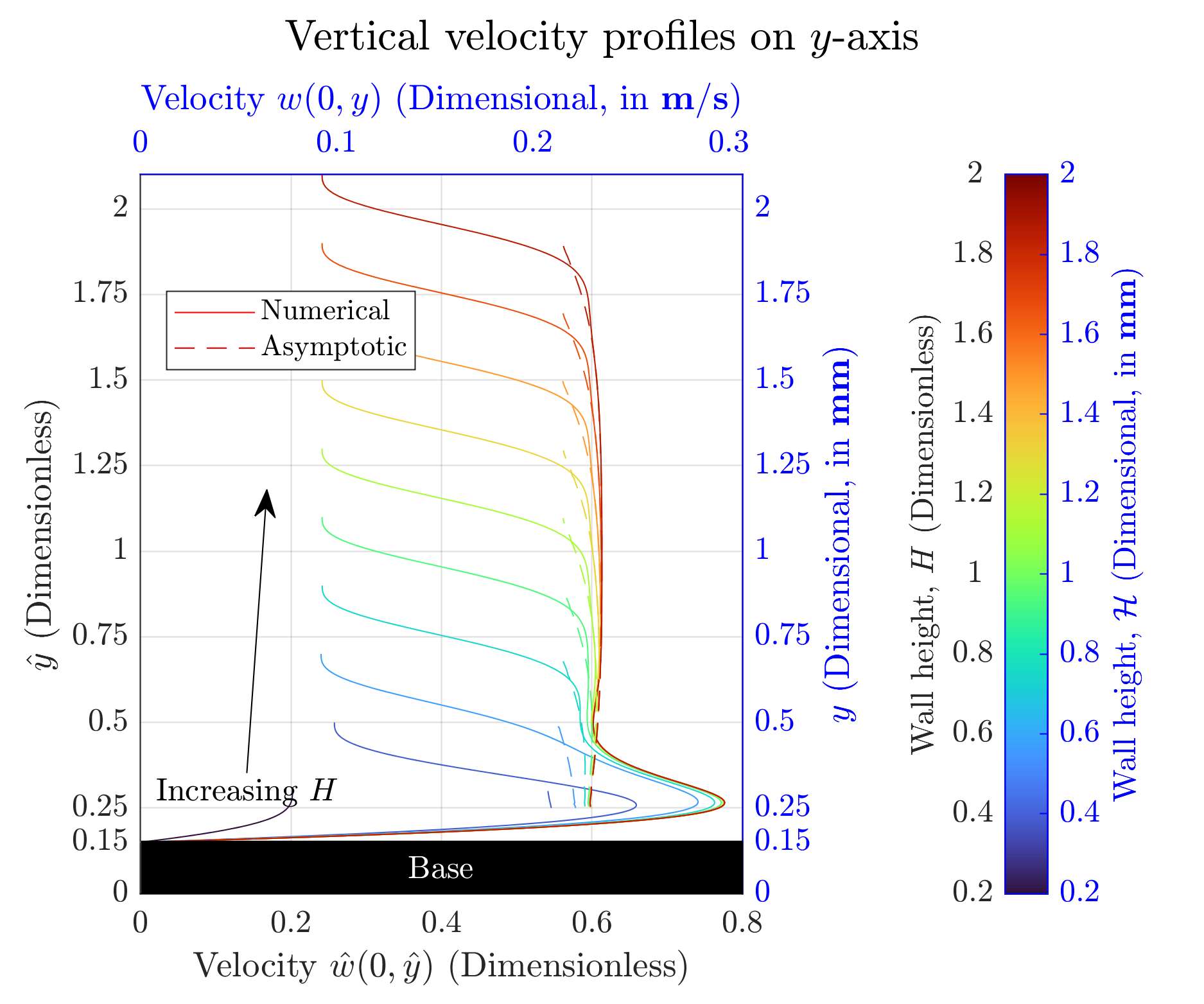}\\
    \textbf{(f)}\\
    \includegraphics[width=\textwidth]{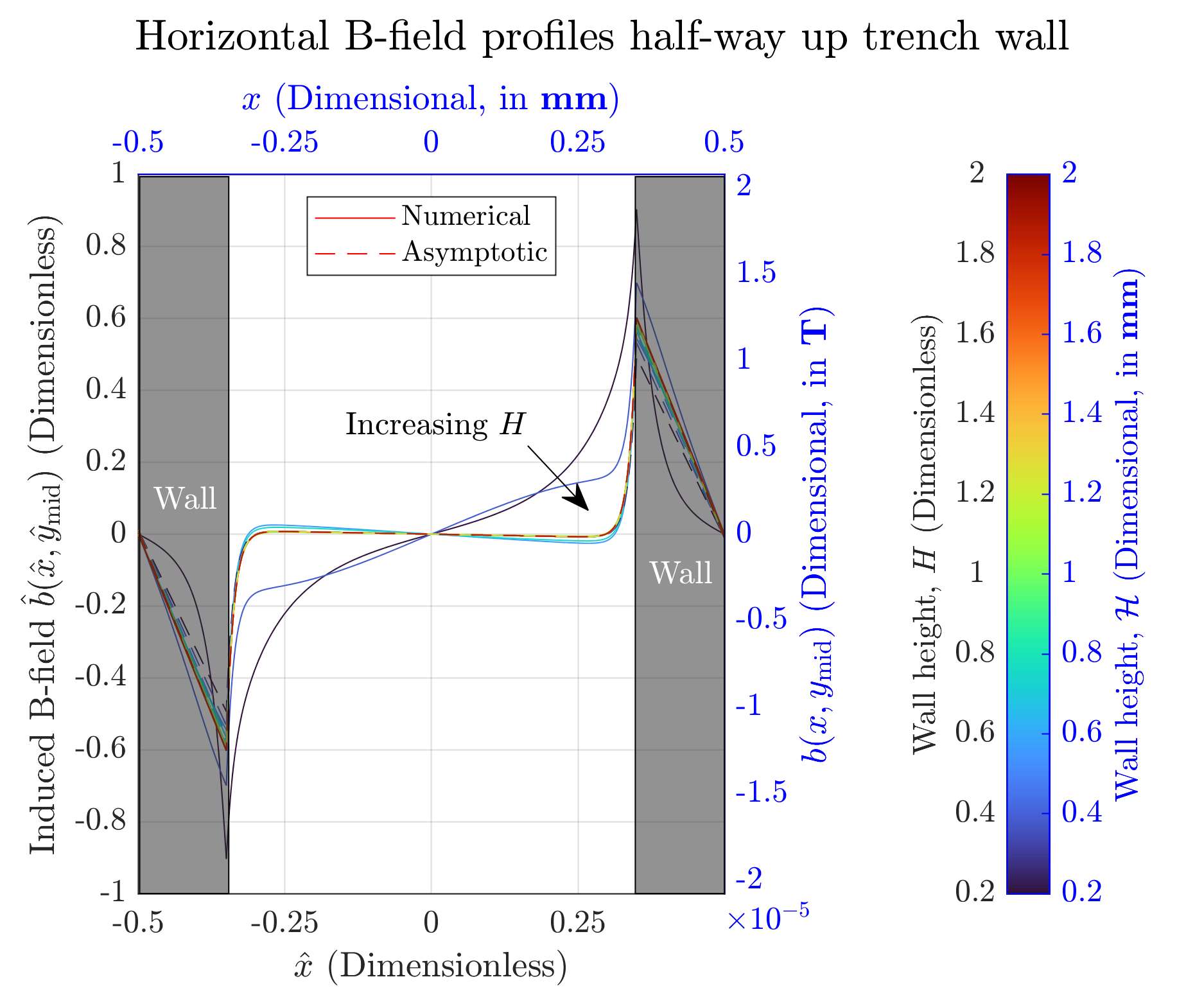}\\
    \end{minipage}
    \caption{Numerical plots of \textbf{(a,d)} horizontal velocity profiles half-way up the trench wall, \textbf{(b,e)} vertical velocity profiles along the $\dimensionless{y}$-axis, and \textbf{(c,f)} horizontal magnetic field profiles half-way up the trench wall, for dimensionless base thickness $\varsigma \in \{ 0.01, 0.1, 0.2,\ldots,0.9,0.99 \}$ (in \textbf{(a,b,c)}) and dimensionless wall height $H \in \{0.2,0.4,\ldots,1.8,2 \}$ (in \textbf{(d,e,f)}).}
    \label{fig:increasing_sigma_and_H}
\end{figure}

\subsection{Changing the vertical dimensions \texorpdfstring{$\varsigma$}{sigma} and \texorpdfstring{$H$}{H}}

The effects of changing the base thickness $\varsigma$ and wall height $H$ are demonstrated in figure~\ref{fig:increasing_sigma_and_H}, assuming that the free surface thickness $f$ is kept constant. The horizontal velocity plots in~\textbf{(a)} and~\textbf{(d)} show that the bulk velocity approaches a limiting value as either $\varsigma\to0$ or $H\to\infty$, and in either case the asymptotic and numerical predictions are in generally good agreement. However, as either $\varsigma\to H$ or $H\to \varsigma$, the velocity profile features local maxima near the trench walls,
caused by
the conducting-base jet intersecting the central horizontal line $\dimensionless{y}=\frac{1}{2}(H+\varsigma)$, thus revealing a velocity profile characteristic of conducting-wall MHD duct flow \citep{Hunt1967}. The asymptotic approximation \eqref{eq:2dTrenchThinWallProblem-ApproxHorizontalProfile-NonThin-Velocity} breaks down in either of these limits where the base becomes close to the top of the walls, and the velocity ultimately collapses to zero as the TEMHD effect vanishes along with the side walls. Moreover, the boundary layers near the bottom and top of the trench merge and annihilate one another.

In plot~\textbf{(b)}, we see that making the base thicker shifts the velocity profile upwards and slightly strengthens the conducting-base jet.
Increasing $H$ has a similar shifting effect on the velocity profile near the top as shown in plot~\textbf{(e)}.
In both cases, the bulk velocity profile demonstrates a uniform shape away from the jet and the top of the wall, and in this region the asymptotics and numerics show good agreement.

In plots~\textbf{(c)} and \textbf{(f)}, we see that the numerical solutions agree well with
the asymptotic prediction that the induced magnetic field is close to zero outside the Hartmann layers, except when the base and top of the walls are brought very close together and the corresponding boundary layers start to affect the outer flow.

\begin{figure}
    \begin{minipage}{0.5\textwidth}
    \centering
    \textbf{(a)}\\
    \includegraphics[width=0.8\textwidth]{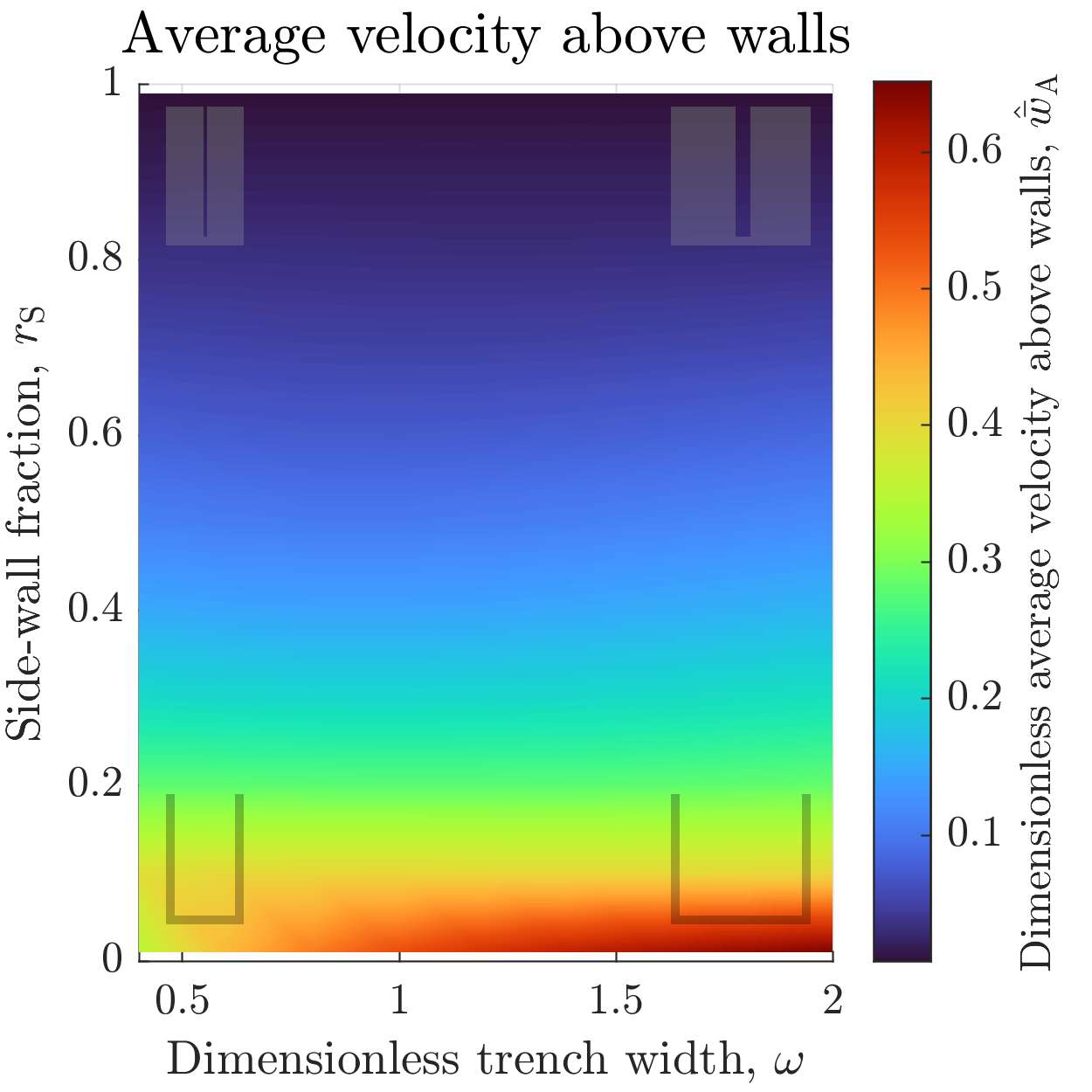}\\
    \textbf{(b)  }\\
    \includegraphics[width=0.8\textwidth]{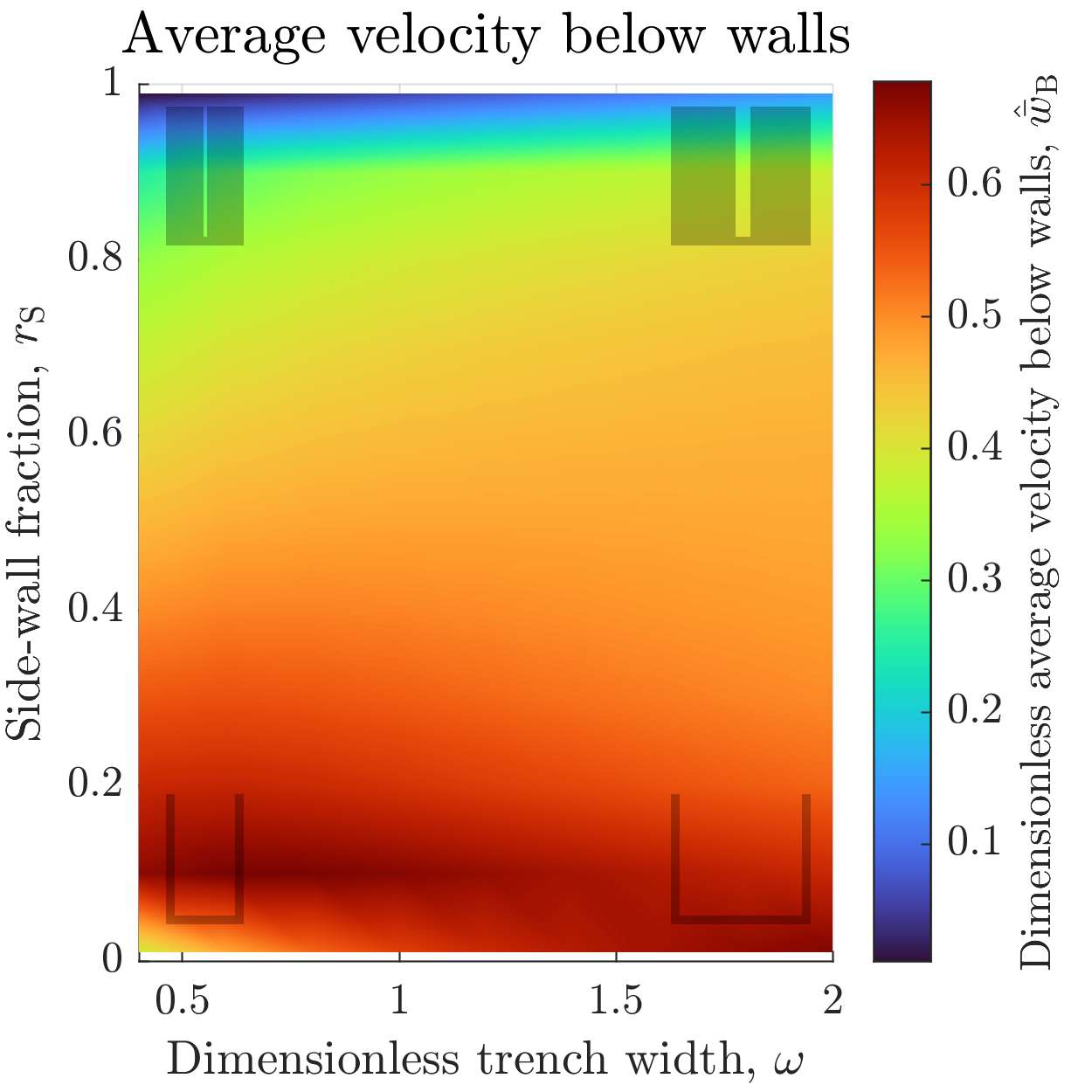}\\
    \textbf{(c)  }\\
    \includegraphics[width=0.8\textwidth]{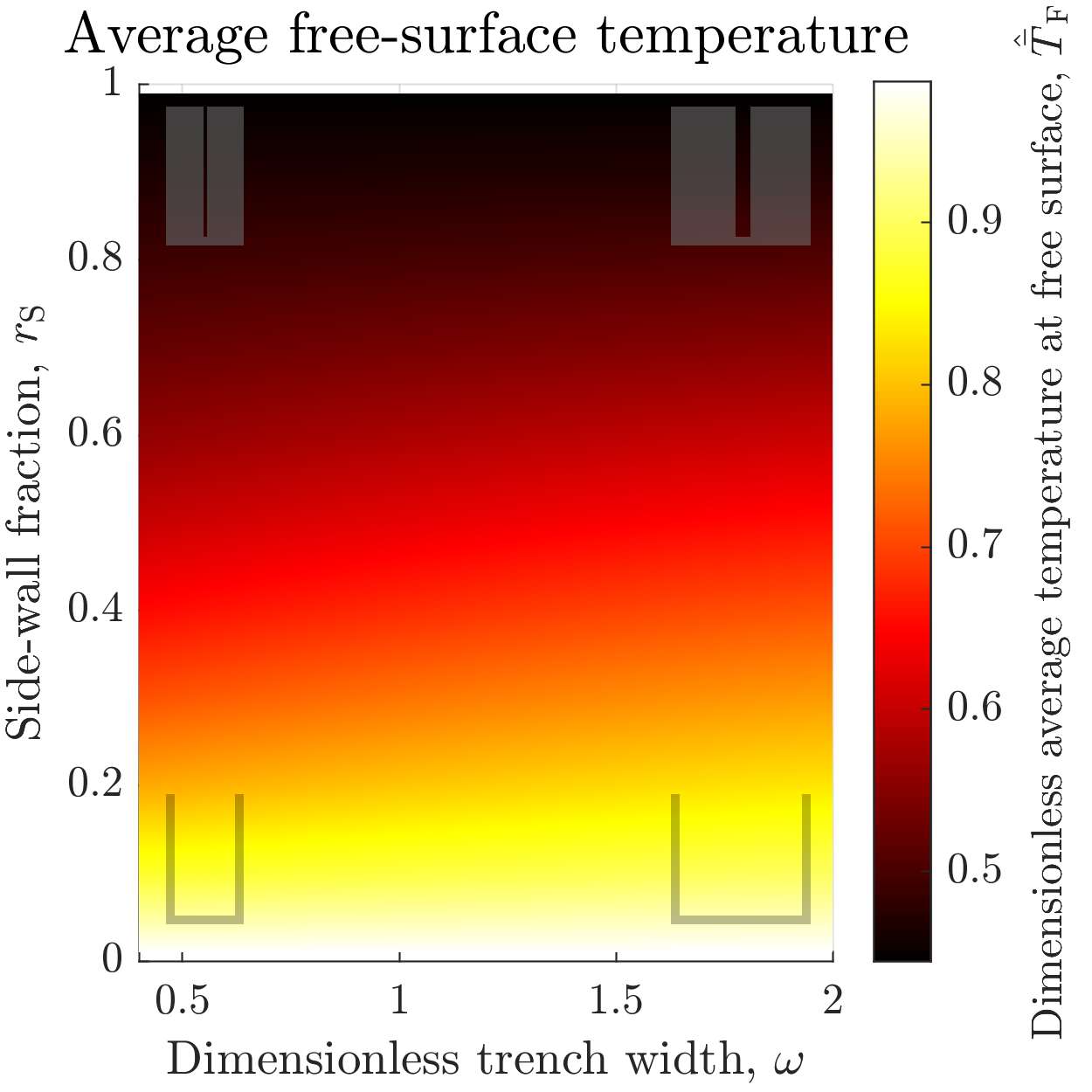}
    \end{minipage}
    \begin{minipage}{0.5\textwidth}
    \centering
    \textbf{(d)  }\\
    \includegraphics[width=0.8\textwidth]{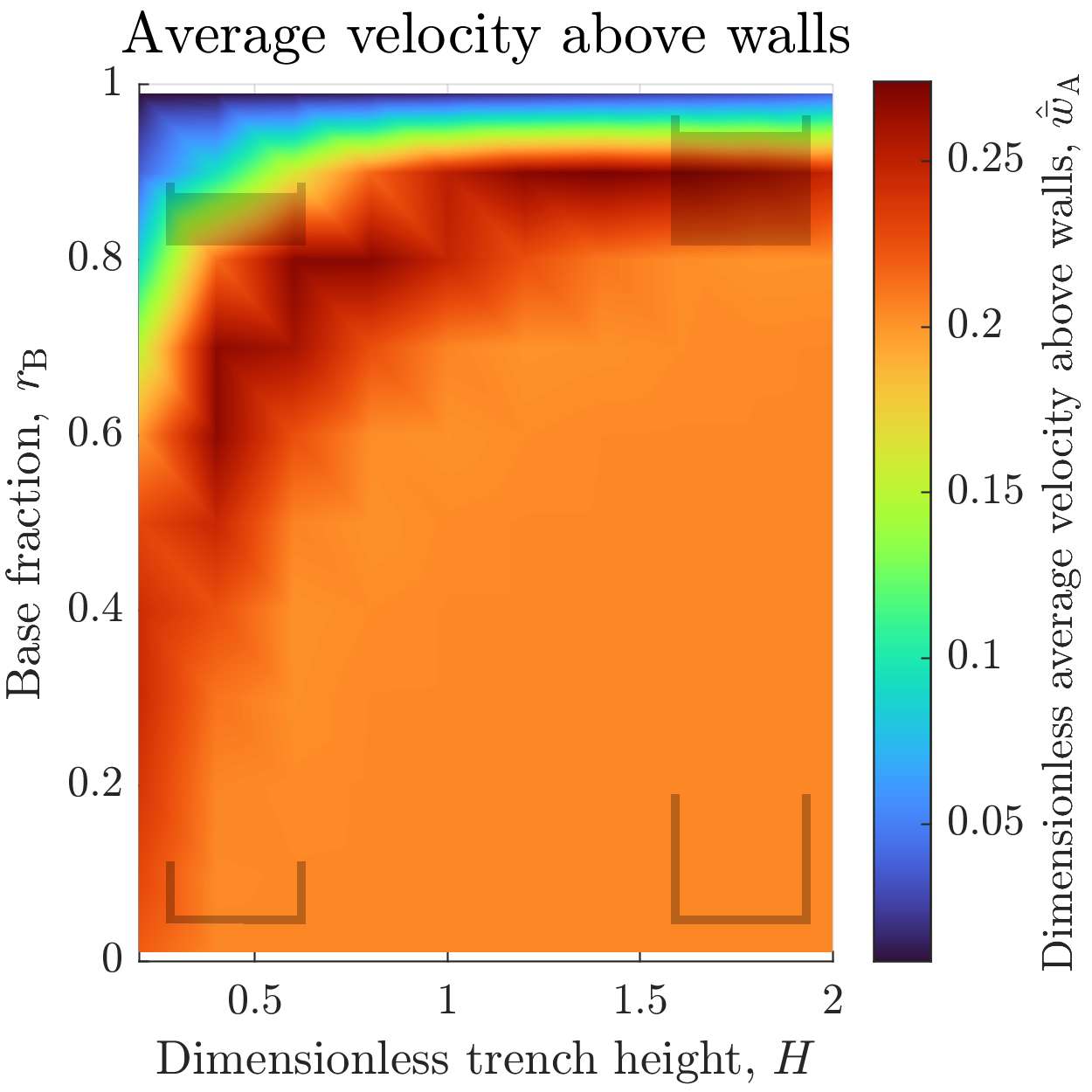}\\
    \textbf{(e)  }\\
    \includegraphics[width=0.8\textwidth]{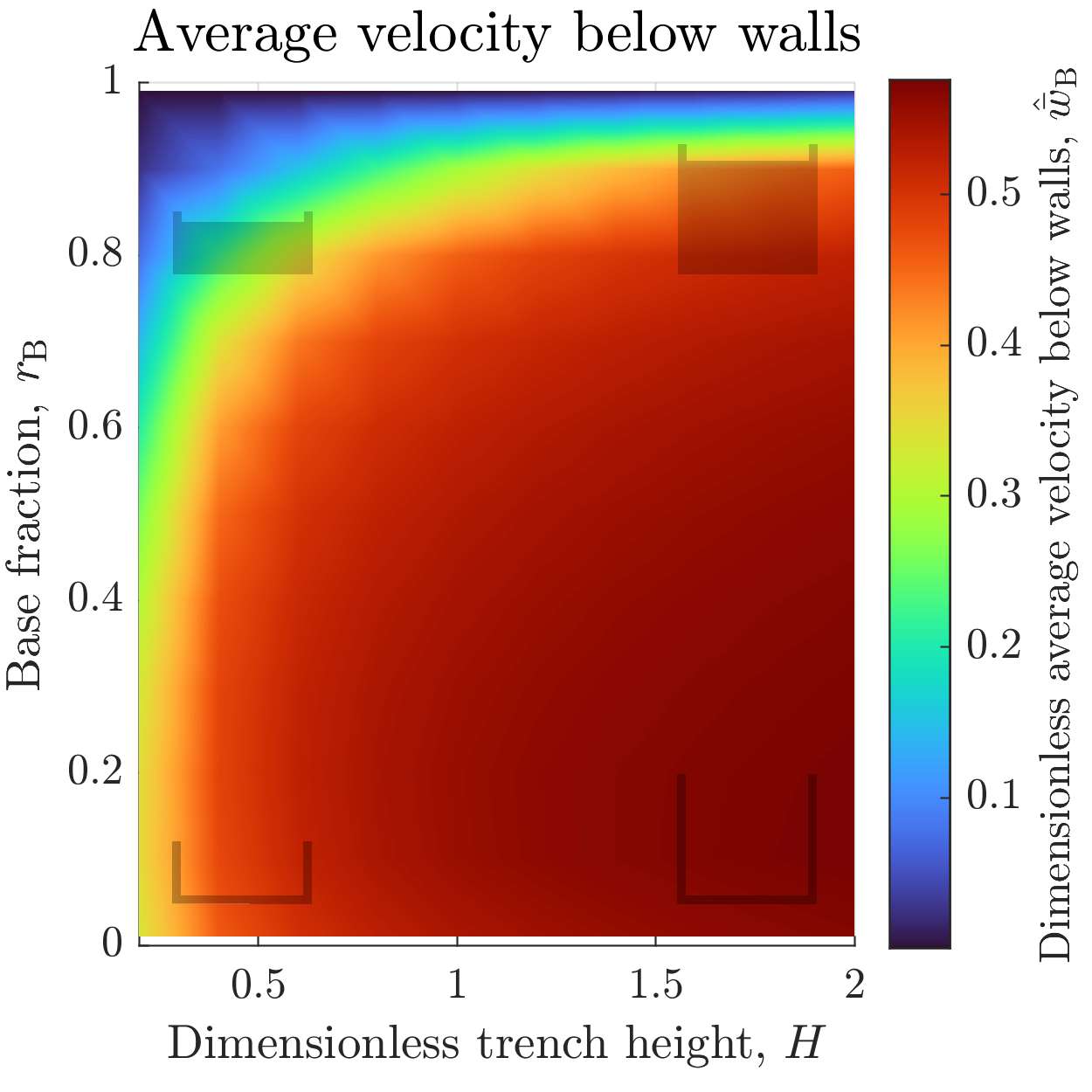}\\
    \textbf{(f)  }\\
    \includegraphics[width=0.8\textwidth]{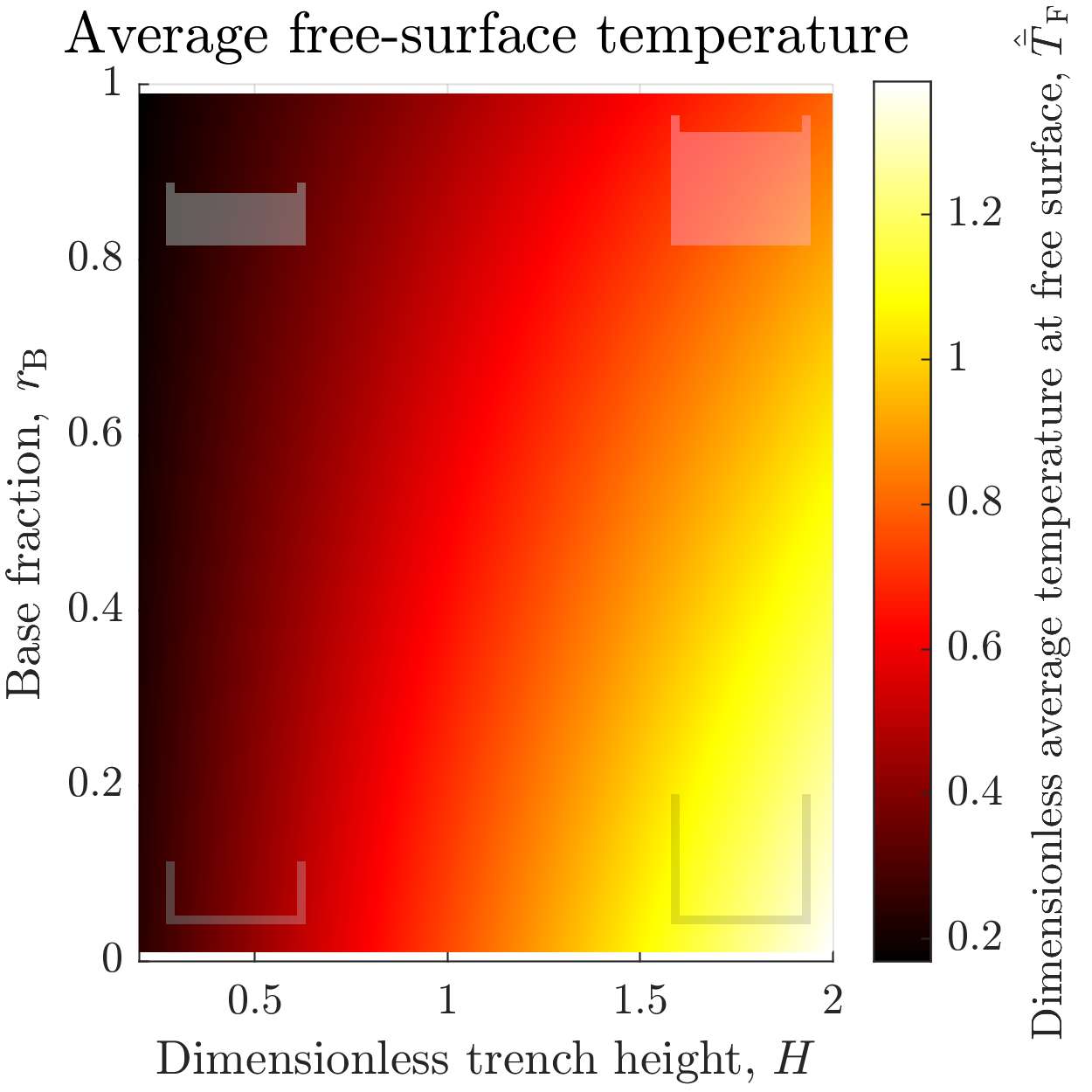}\\
    \end{minipage}
    \caption{The dependence of the average velocity above and inside the trench (\textbf{(a,d)} and \textbf{(b,c)}, respectively) and the average free-surface temperature \textbf{(c,f)} on the trench width $\omega$ and side-wall fraction $r_{\rm S}$ \textbf{(a-c)} and on the trench height $H$ and base fraction $r_{\rm B}$ \textbf{(d-f)}, for $\omega\in[0.4,2]$, $H\in\{0.2,2\}$, and $r_{\rm S},r_{\rm B}\in[0.01,0.99]$.
    Schematics of the trench geometry are shown in the corners of each plot.
    }
    \label{fig:increasing_omega_fS_H_fB}
\end{figure}

\subsection{Geometrical parameter space}\label{ss:gps}

We now examine the effects on the global flow properties of varying the trench geometry. We consider three averaged quantities that describe the global flow characteristics, namely the average flow rates above and below the trench walls, defined by
\begin{subequations}
\begin{align}
\hat{\bar{w}}_{{\rm A}} & =\dfrac{1}{2\omega f}\int_{H}^{H+f}\int_{-\omega}^{\omega}\hat{w}(\hat{x},\hat{y}){\rm d}\hat{x}{\rm d}\hat{y},\label{eq:averageVelocityAboveWalls}\\
\hat{\bar{w}}_{{\rm B}}& =\dfrac{1}{2(\omega-\tau)(H-\varsigma)}\int_{\varsigma}^{H}\int_{-\omega+\tau}^{\omega-\tau}\hat{w}(\hat{x},\hat{y}){\rm d}\hat{x}{\rm d}\hat{y},\label{eq:averageVelocityBetweenWalls}
\end{align} 
respectively, and the average surface temperature, defined by
\begin{equation}
\hat{\bar{T}}_{{\rm F}}=\int_{-\omega}^{\omega}\hat{T}^{\mathfrak{L}}(\hat{x},H+f){\rm d}\hat{x}.\label{eq:averageSurfaceTemperature}
\end{equation}
\end{subequations}
For the LiMIT concept to be successful, reasonably high flow rates both above and within the trench should be maintained, whilst the surface temperature should not get too high.

We start by varying the  horizontal wall dimensions $\omega$ and $r_{\rm S}=\tau/\omega$ in Figure~\ref{fig:increasing_omega_fS_H_fB}\textbf{(a,b,c)}. Plot~\textbf{(a)} demonstrates that the average velocity above the walls is largely unaffected by the trench width, but increasing the trench wall thickness, and thus the effective conductivity, significantly reduces the flow rate.
The velocity inside the trench is maximised for a relatively thin trench with a side-wall fraction of around 10\%,
as shown in plot \textbf{(b)}.
When the side wall fraction is less than approximately 5\%, the average velocities above and below the trench walls take similar values. Otherwise, however, the two average velocities can differ markedly, so that the flow speed inside the trench cannot readily be inferred by observation of the free surface. As plot \textbf{(c)} shows, making the trench wider does not greatly affect the average surface temperature. However, making the walls thinner does result in a hotter free surface, because there is a smaller volume of solid wall through which heat can conduct (recall that the solid thermal conductivity is roughly three times that of the liquid). On the other hand, making the trench walls too thick results in very little flow either above or below the walls.

Next we turn our attention to plots~\textbf{(d,e,f)} in figure~\ref{fig:increasing_omega_fS_H_fB}, where the effects of varying the wall height $H$ and the base fraction $r_{\rm B}$ are examined. The velocities above and below the walls appear to show similar distributions. However, plot~\textbf{(d)} reveals a curved band of optimal values for the flow above the trench, where there is only relatively moderate flow inside the trench. On the left of this curved region, we see that very little flow is generated by a trench with a very high base and walls resembling short stumps. On the other hand, an optimal value of the flow below the walls in plot~\textbf{(e)} is achieved when the trench is tall, but with a relatively thin base. 

This latter regime is also where the free surface is hottest, as shown in plot~\textbf{(f)}, due to there being less conducting wall for heat flux to pass through. In practice,
we suggest that trench dimensions along the optimal band in figure~\ref{fig:increasing_omega_fS_H_fB}\textbf{(d)} represent a good trade-off between increasing the average velocities above and below the walls, and reducing the temperature of the free surface. This region of parameter space corresponds to 
a
configuration
with
a thick divertor plate and only slightly higher trench walls.

\begin{figure}
    \centering
    \begin{minipage}{0.49\textwidth}
    \centering
    \textbf{(a)}\\
    \includegraphics[width=\textwidth]{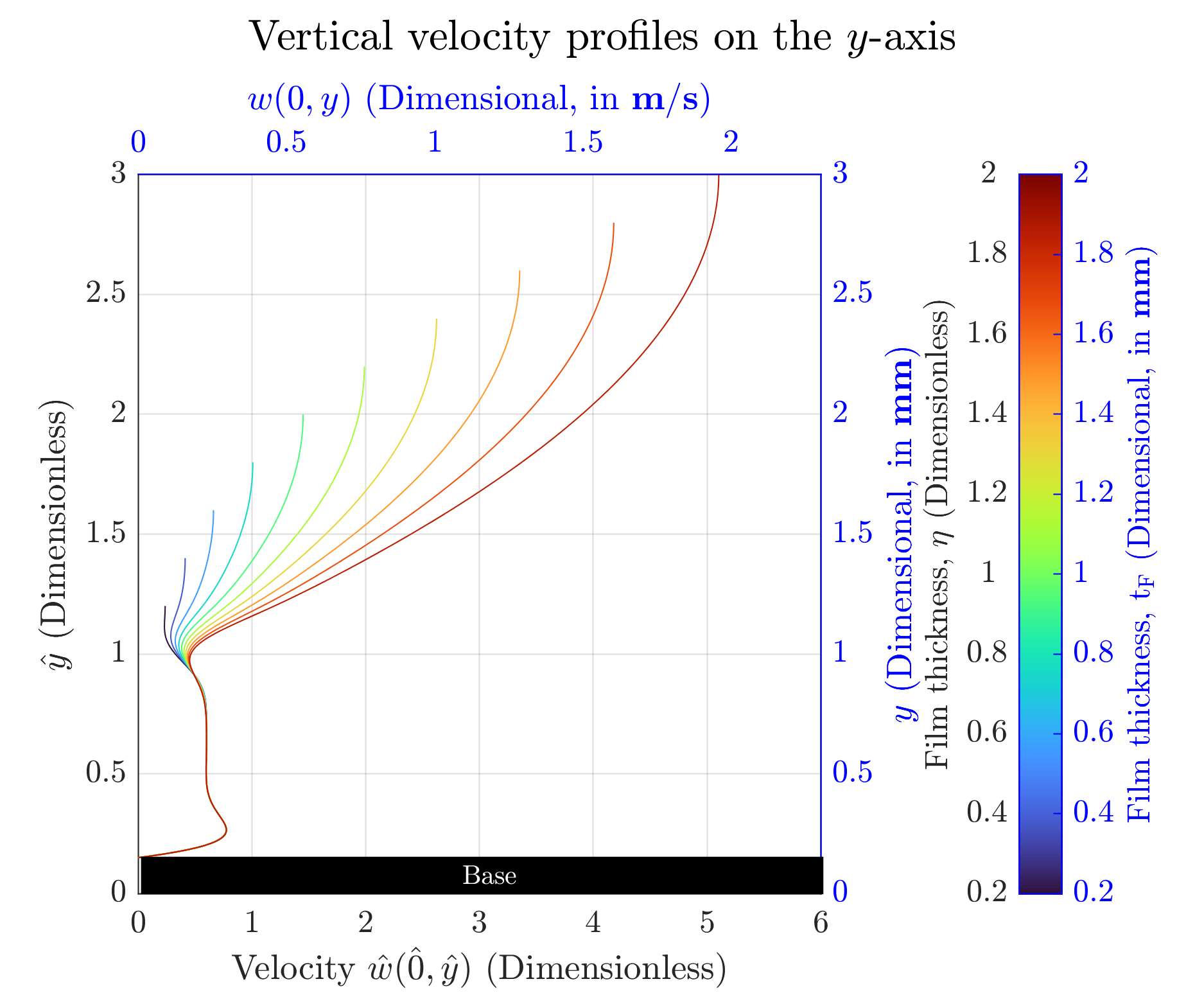}\\
    \end{minipage}
    \begin{minipage}{0.49\textwidth}
    \centering
    \textbf{(b)}\\
    \includegraphics[width=\textwidth]{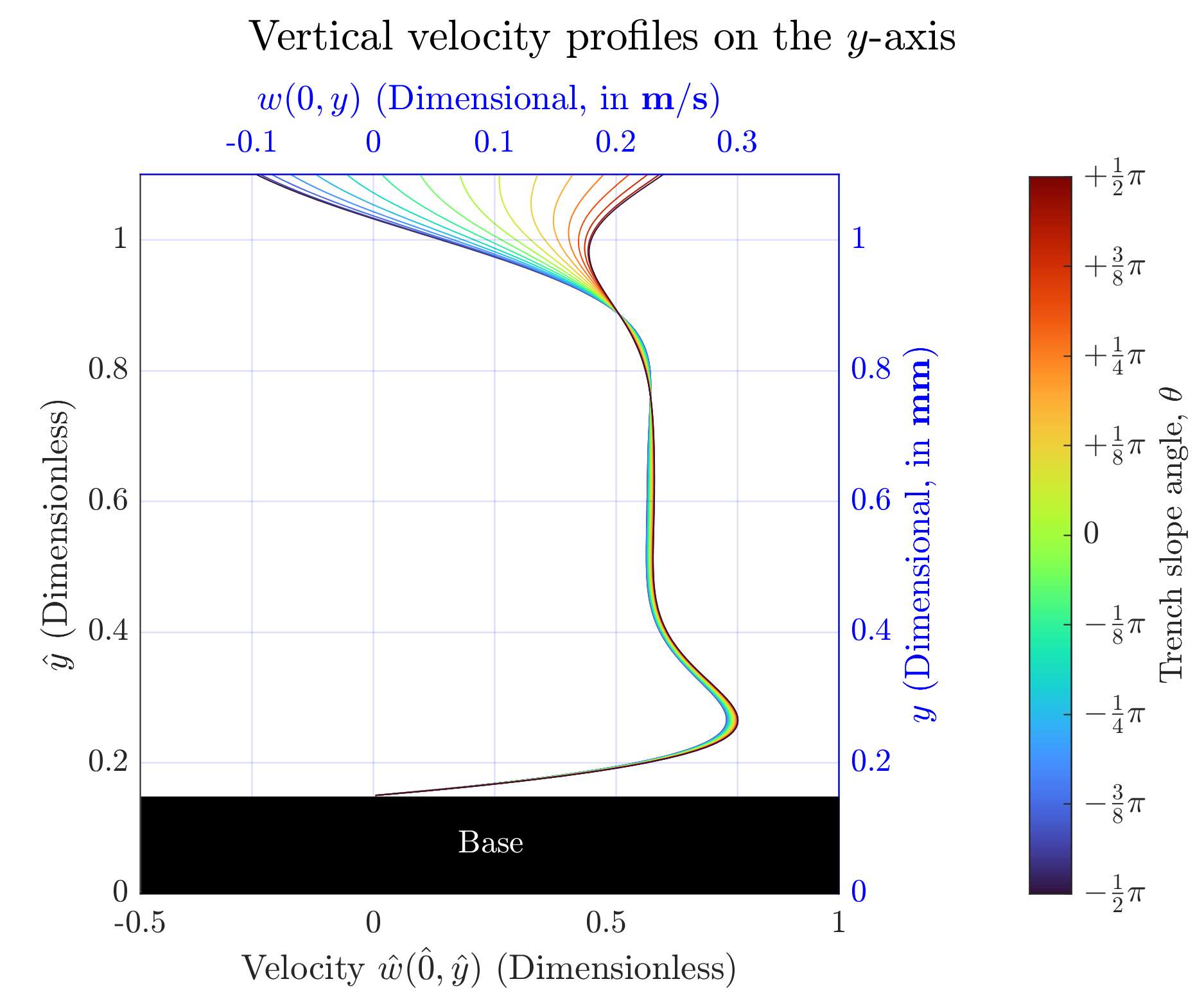}
    \end{minipage}
    \caption{Plots showing the effect of varying \textbf{(a)} the film thickness $f$, \textbf{(b)} the trench slope angle $\theta$, on the vertical velocity profile.}
    \label{fig:Increasing_eta_and_theta}
\end{figure}

\subsection{Changing the film thickness \texorpdfstring{$f$}{f} and the trench angle \texorpdfstring{$\theta$}{theta}}

The effects of varying the film thickness $f$ and the trench slope angle $\theta$ are shown in figure~\ref{fig:Increasing_eta_and_theta}.
Recall that the trench slope angle $\theta$ enters the problem through the gravitational parameter $\varGamma$ defined by \eqref{eq:ndparams}.

In plot~\textbf{(a)} we see that overfilling the trench has a marked impact on the nature of the flow above the walls without significantly affecting the flow inside the trench. The velocity above the trench may be much larger than that inside, and approximately quadratic in $\dimensionless{y}$, with its maximum at the free surface, in qualitative agreement with the asymptotic prediction \eqref{eq:2dTrenchThinWallProblemNavierStokesEquationAboveTrenchRescaled-AE-LO-Sol}. 
Recalling that $\theta = 0.1$ in these solutions, we infer that even moderate trench slope angles can result in large flow rates above the walls. In practice, the intention is to use thermoelectric effects as the principal driver of the flow rather than relying on gravity, not least because similar divertor plates may be used at the top of the tokamak, where gravity acts in the opposite direction.

As figure~\ref{fig:Increasing_eta_and_theta} \textbf{(b)} shows, the velocity profile between the walls is mostly unaffected by the slope angle $\theta$, because TEMHD is the dominant driving mechanism there. On the other hand, increasing $\theta$ increases the flow speed above the trench walls, and in the case where $f = 0.1$ as plotted,
while making
$\theta < 0$ slows down the flow above the trench and can lead to backflows. The fluid above the walls experience both the drag from the TEMHD-driven flow inside the trench and the gravitational forcing, and when $\theta<0$ these effects act in opposite directions.

\subsection{Varying the applied field \texorpdfstring{$\mathcal{B}^{\rm a}$}{Ba}}

\begin{figure}
    \centering
    \begin{minipage}{0.49\textwidth}
    \centering
    \textbf{(a)}\\
    \includegraphics[width=0.9\textwidth]{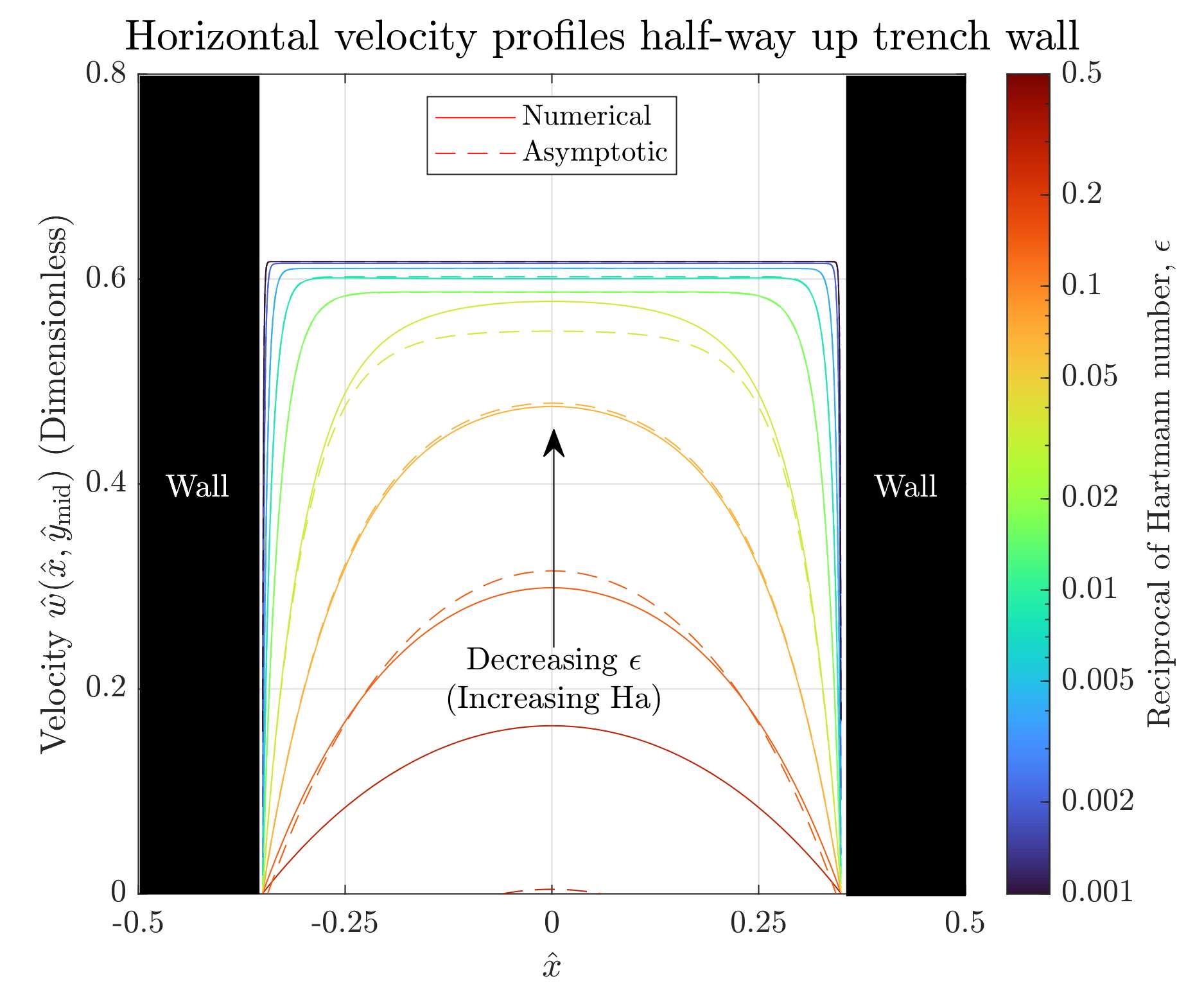}\\
    \textbf{(b)}\\
    \includegraphics[width=0.9\textwidth]{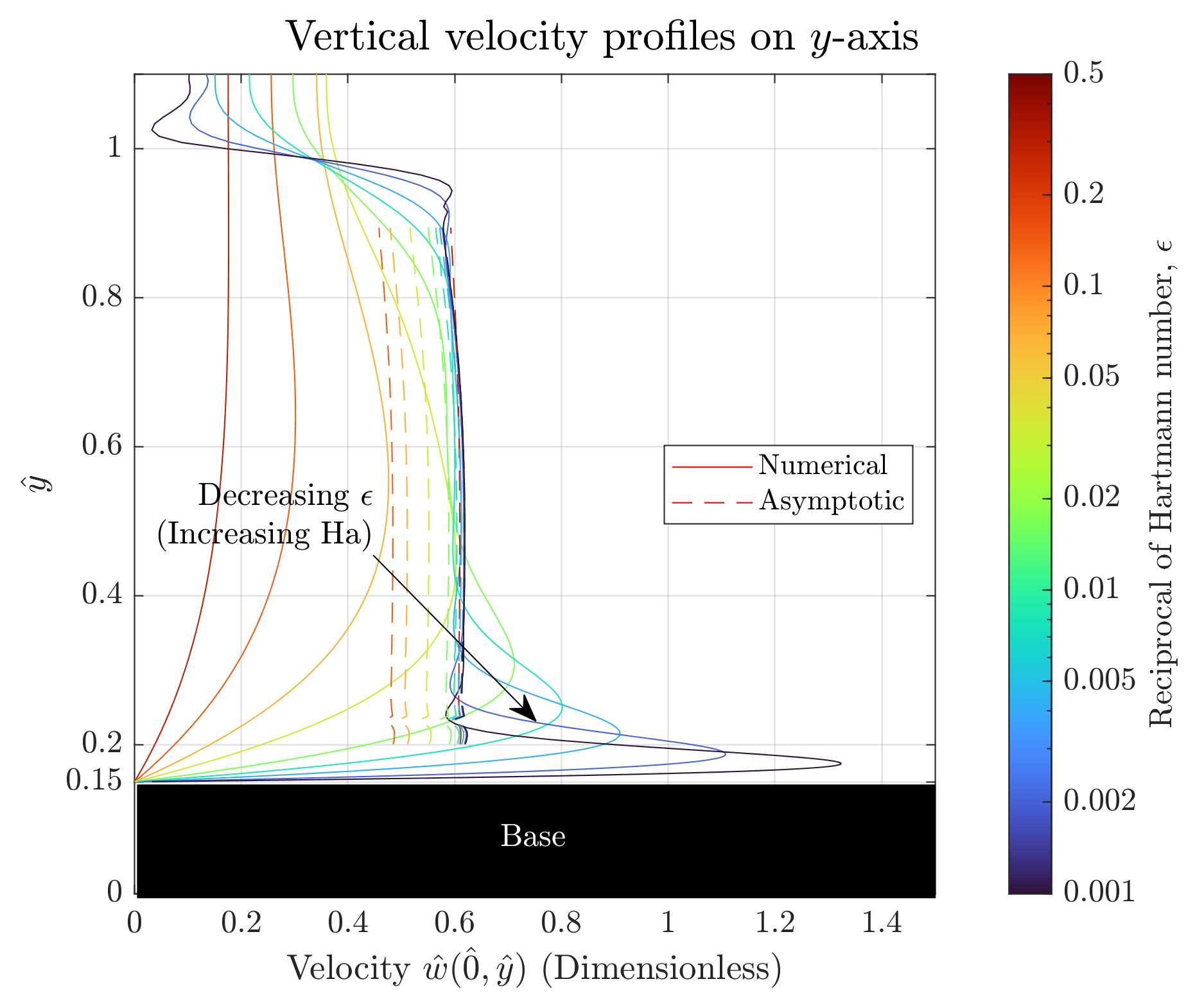}\\
    \textbf{(c)}\\
    \includegraphics[width=0.9\textwidth]{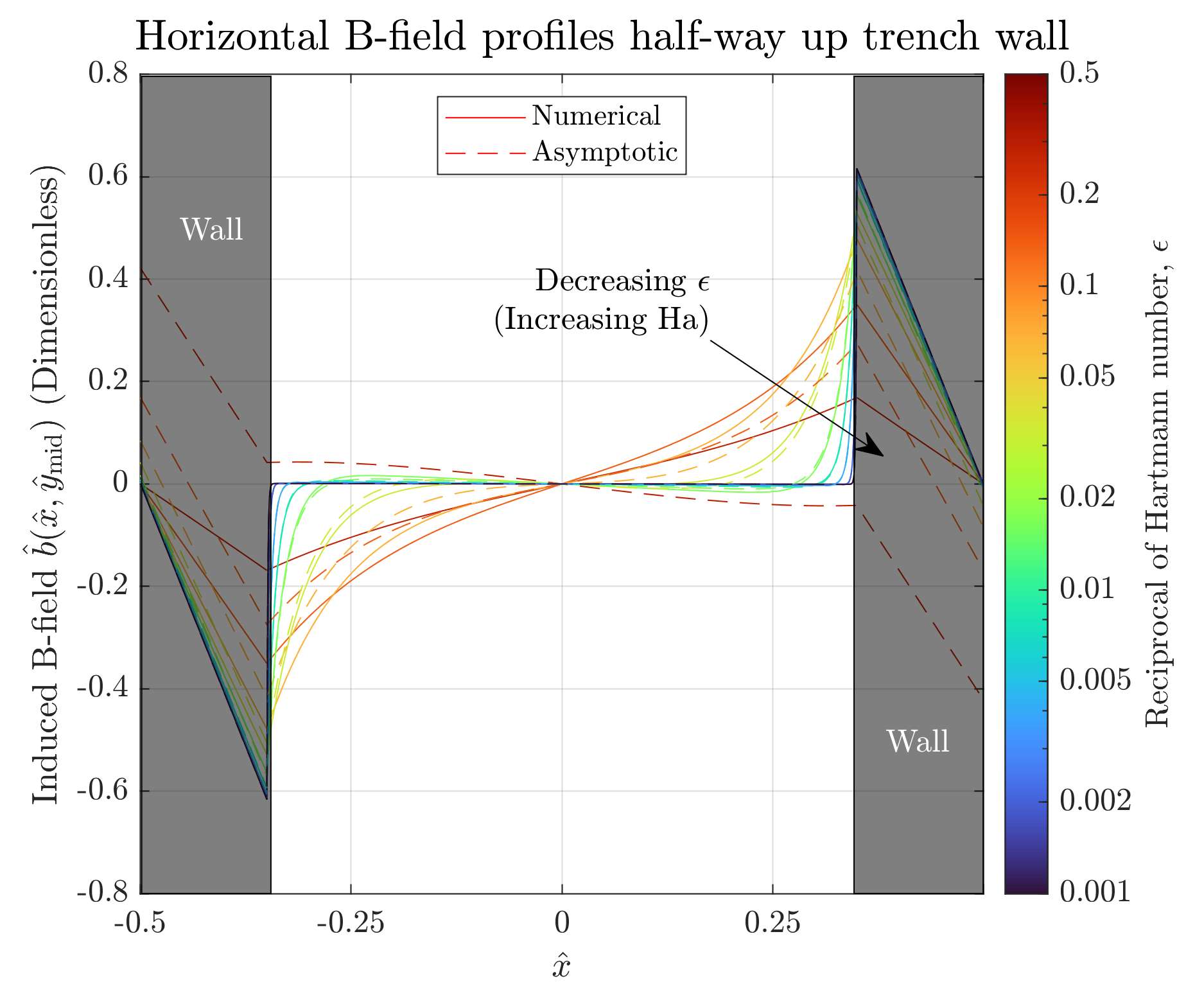}
    \end{minipage}
    \begin{minipage}{0.49\textwidth}
    \centering
    \textbf{(d)}\\
    \includegraphics[width=0.9\textwidth]{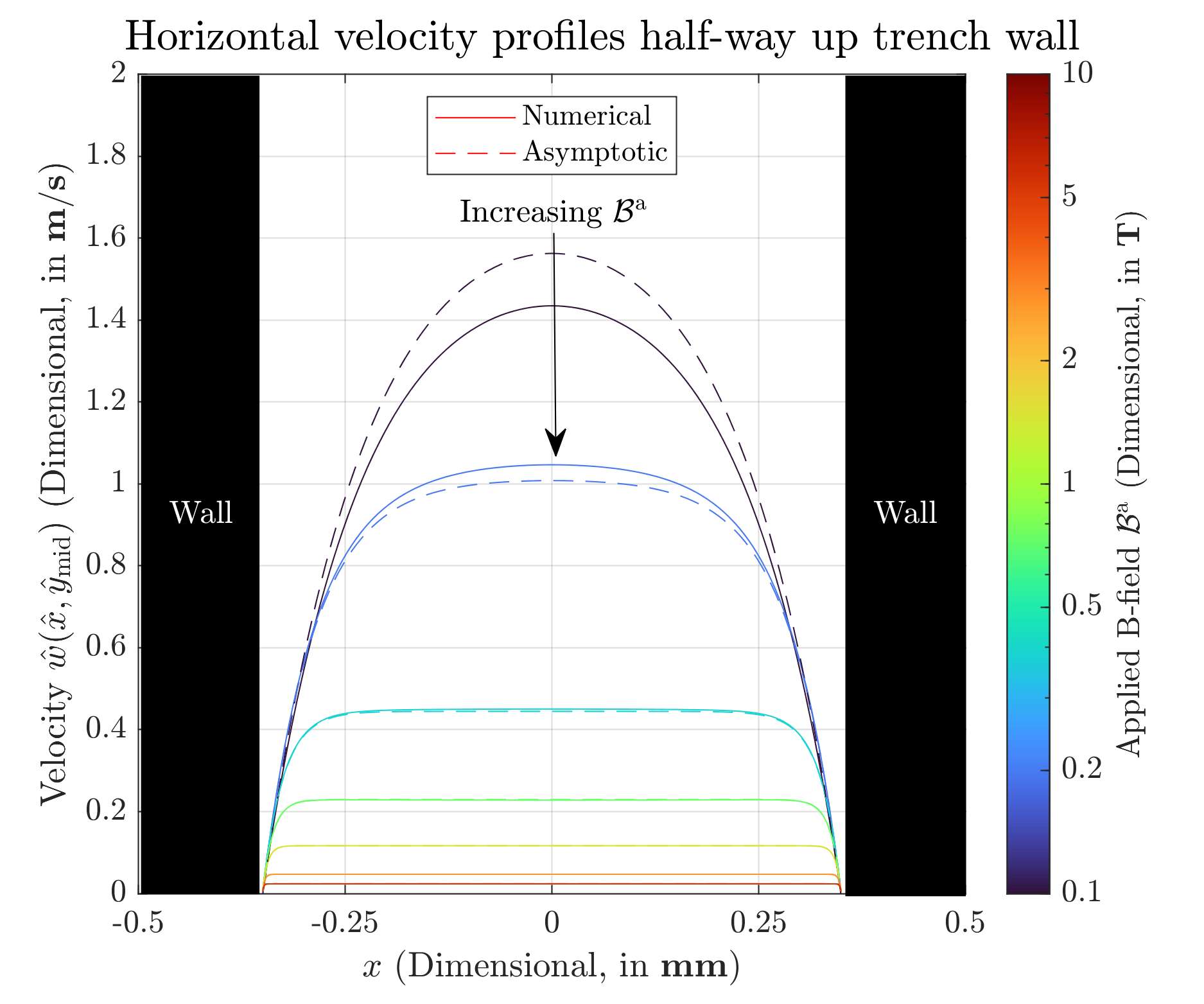}\\
    \textbf{(e)}\\
    \includegraphics[width=0.9\textwidth]{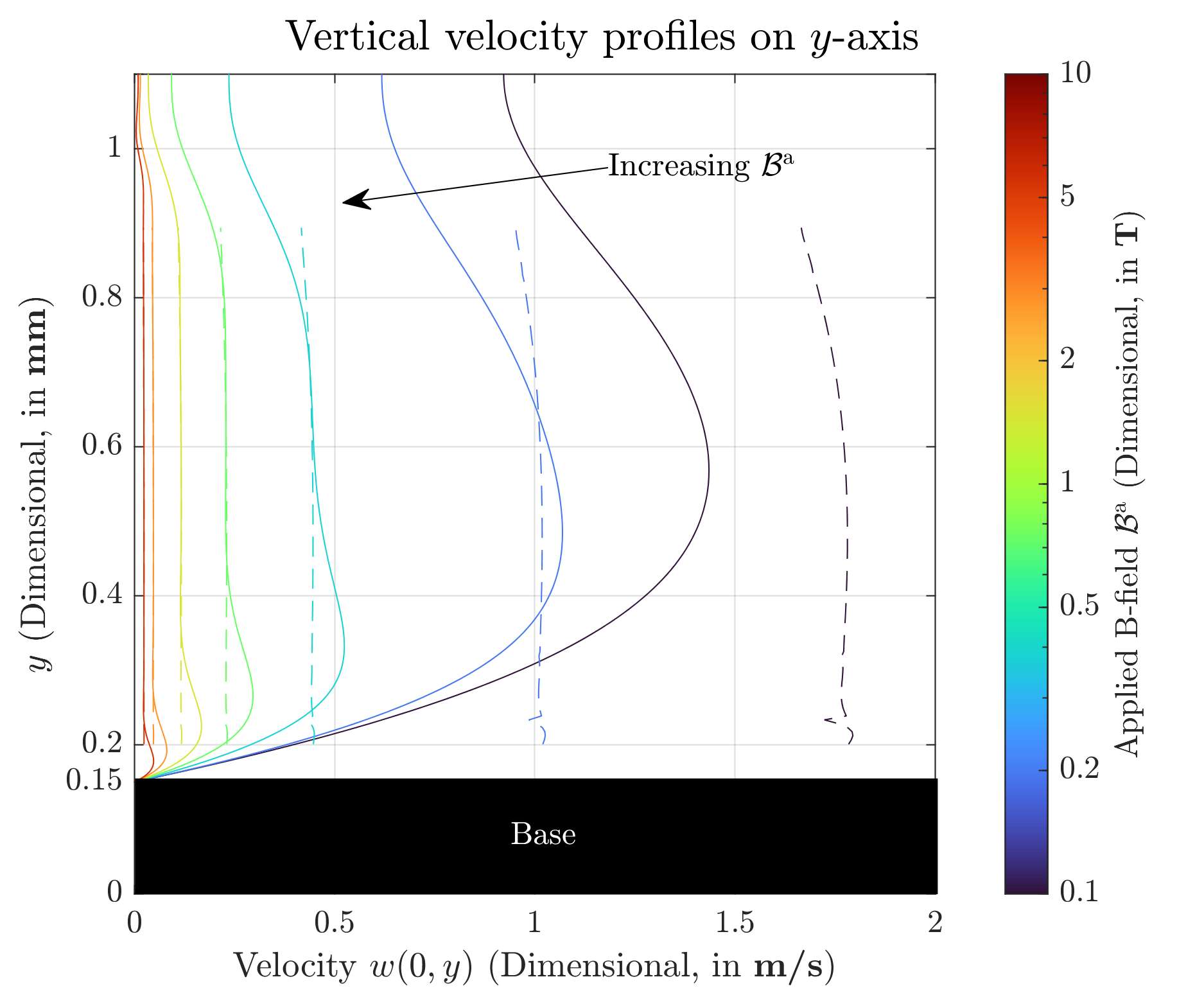}\\
    \textbf{(f)}\\
    \includegraphics[width=0.9\textwidth]{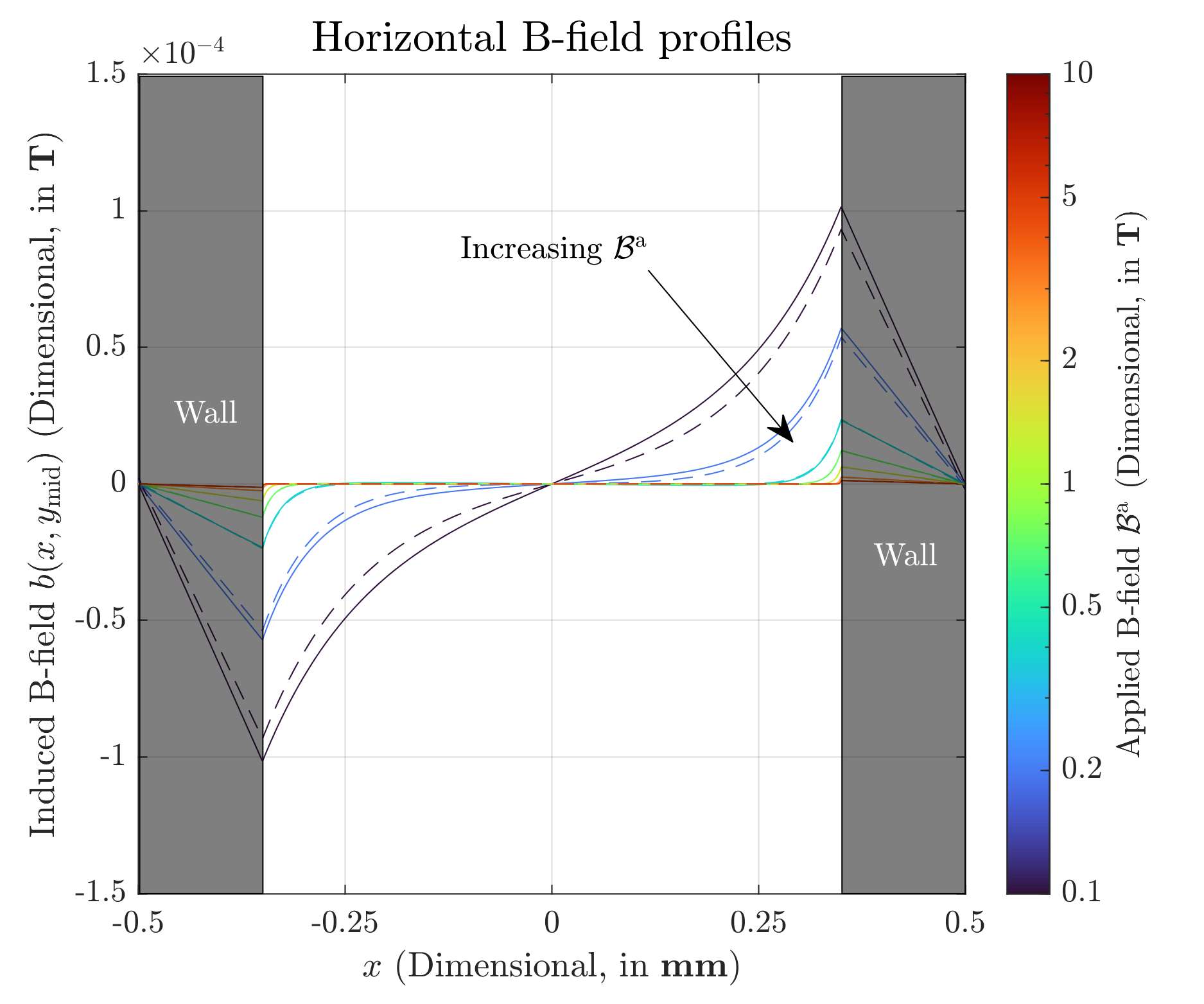}
    \end{minipage}
    \caption{Plots illustrating the behaviour of \textit{dimensionless} solutions in \textbf{(a,b,c)} for $\epsilon\in\{0.5,0.2,0.1,0.05,0.02,0.01,0.005,0.002,0.001\}$, and the \textit{dimensional} solutions in \textbf{(d,e,f)} for ${\cal B}^{\rm a}\in\{0.1,0.2,0.5,1,2,5,10\}$T.}
    \label{fig:increasing_Ha_and_Ba}
\end{figure}

In figure~\ref{fig:increasing_Ha_and_Ba}, we explore the effects of varying the reciprocal Hartmann number $\epsilon$. Recall that $1/\epsilon$ is proportional to the applied magnetic field strength ${\cal B}^{\rm a}$, while the gravitational parameter $\varGamma$ is independent of ${\cal B}^{\rm a}$. Provided the material and geometrical properties are kept constant, varying $\epsilon$ is thus mathematically equivalent to varying ${\cal B}^{\rm a}$. However, when considering the 
practical implications of the results,
we must recall that the velocity scaling \eqref{eq:nondimensionalisationVelocity-Steady} contains a factor of $1/{\cal B}^{\rm a}$.
We therefore also plot the results using dimensional variables to determine what would be observed in practice as
the magnetic field strength is increased.

In figure~\ref{fig:increasing_Ha_and_Ba}\textbf{(a)}, we clearly observe the singular behaviour of the horizontal velocity profiles as $\epsilon\to0$, where the velocity approaches a constant value, determined by the vertical temperature gradient, outside Hartmann layers whose width is of $\order(\epsilon)$. In plot~\textbf{(b)}, we observe that the conducting-wall jet at the base becomes narrower and stronger as $\epsilon$ decreases. Meanwhile, the
velocity above the trench walls decreases with decreasing $\epsilon$. We also see in plot~\textbf{(c)} that, as $\epsilon$ decreases, the induced magnetic field approaches a piecewise linear function of $\dimensionless{x}$, negligible inside the fluid but with a finite gradient inside the walls. In all cases, the asymptotic predictions (dashed curves) become increasingly accurate with decreasing $\epsilon$, as expected.

However, when considering dimensional quantities, increasing the magnetic field strength ${\cal B}^{\rm a}$ has a retarding effect not apparent in plots~\textbf{(a-c)}. In plot~\textbf{(d)}, we see that increasing the magnetic field strength does create boundary layers near the walls, but simultaneously decreases the maximum velocity. This effect is also captured in plot~\textbf{(e)} where, even though the velocity jet near the trench base still appears, it is much less prominent.
%
Plot~\textbf{(f)} shows that increasing the applied magnetic field also has a suppressing influence on the induced magnetic field.
This perhaps counterintuitive behaviour is consistent with our scaling (\ref{eq:NDTwb}\textit{c}) for the induced magnetic field, which is proportional to 
$1/{\cal B}^{\rm a}$.

\subsection{Changing \texorpdfstring{$\psi$}{psi}, the applied field inclination}

\begin{figure}             
    \centering
\includegraphics[width=0.95\textheight,angle=270]{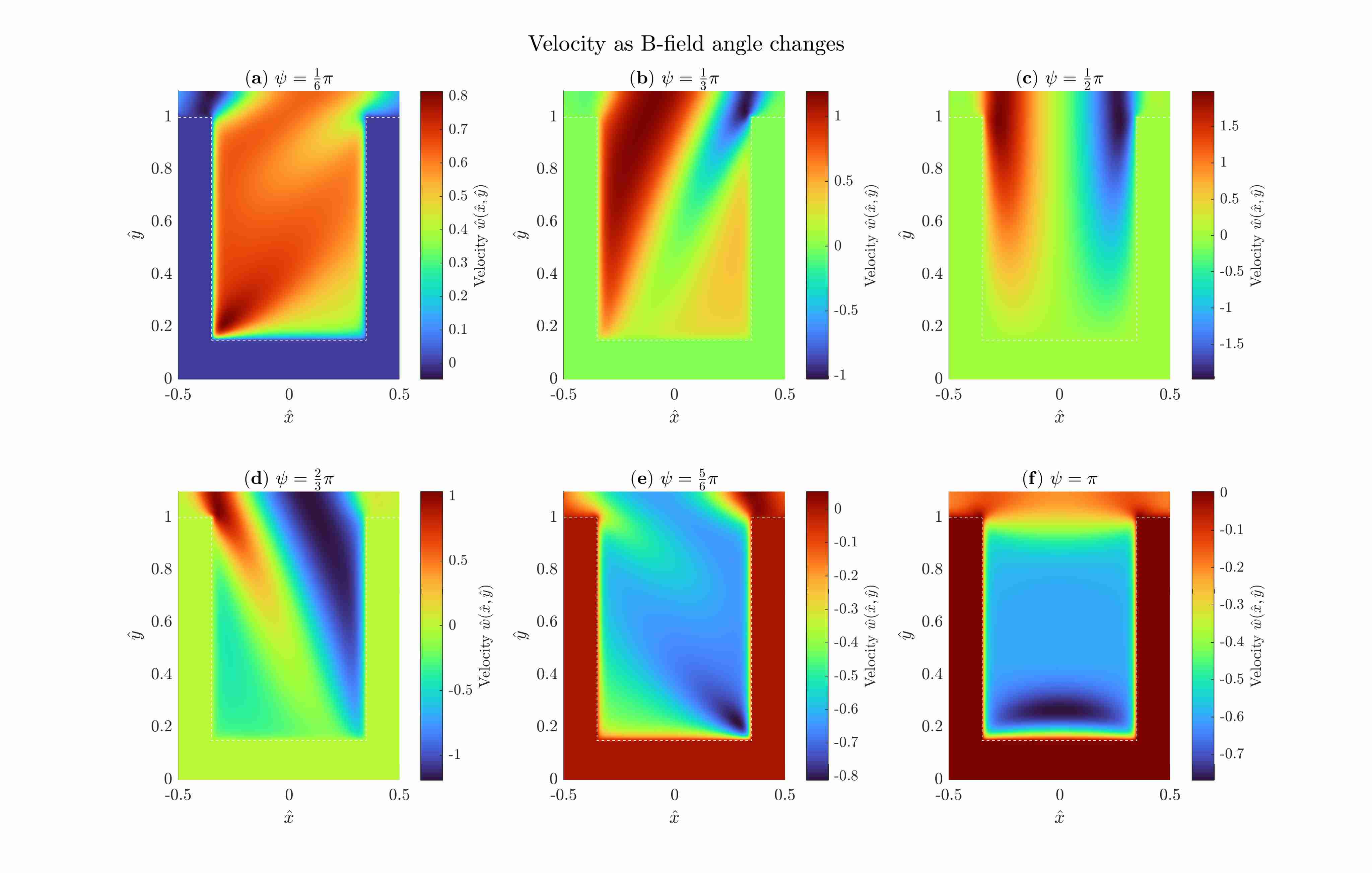}
    \caption{Surface plots illustrating the effect of changing the applied magnetic field angle $\psi$ on the dimensionless velocity field.}
    \label{fig:increasing_psi_velocity}
\end{figure}
\begin{figure}
    \centering
\includegraphics[width=0.95\textheight,angle=270]{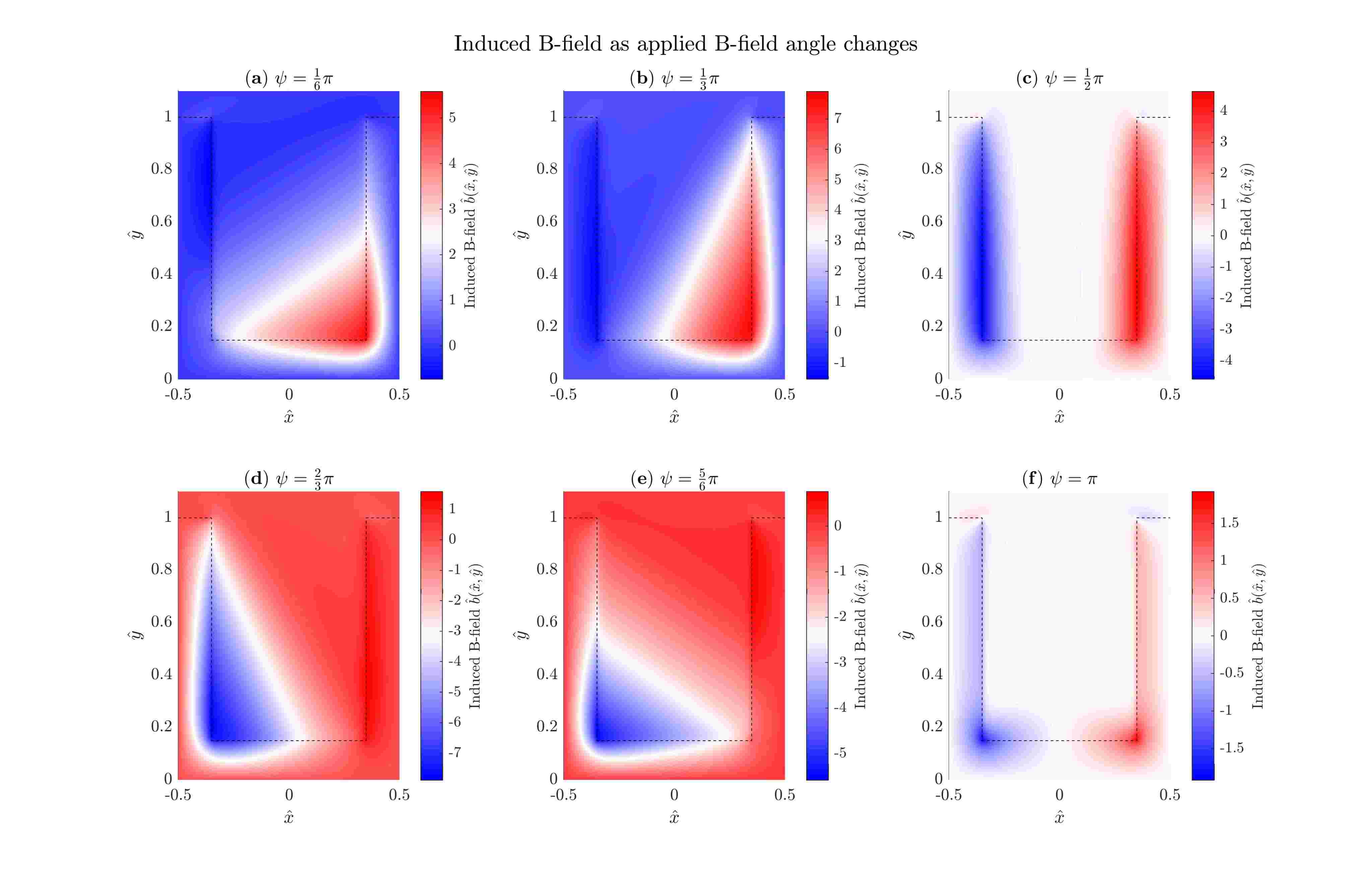}
    \caption{Surface plots illustrating the effect of changing the applied magnetic field angle $\psi$ on the dimensionless induced magnetic field.}
\label{fig:increasing_psi_magnetic_field}
\end{figure}

Although the applied magnetic field is likely to be horizontal in practice, due to the toroidal symmetry of
a tokamak, for completeness we illustrate the effect of the applied field angle $\psi$ on the velocity and the induced magnetic field in figures~\ref{fig:increasing_psi_velocity} and~\ref{fig:increasing_psi_magnetic_field}, respectively.
The most obvious observation from plot~\textbf{(f)} in both figures is that the flow and magnetic field are reversed, as expected, when $\psi=\pi$ and the applied magnetic field is pointing from right to left, rather than left to right.

In figure~\ref{fig:increasing_psi_velocity}\textbf{(a)} with $\psi=\pi/6$, the typical Hunt velocity jet shifts over to the bottom-left corner, and the thickness of the boundary layer near the base decreases. We notice that this conducting jet appears to be stretched out in the direction of the magnetic field, and a small stagnant region also emanates from the top right corner. It appears that the flow profiles are being distorted along dividing subcharacteristics, since the leading-order TEMHD trench problem (see figure~\ref{fig:leading_order}) becomes hyperbolic in the limit $\epsilon\to 0$. This phenomenon, where the velocity gradient appears to become discontinuous across dividing characteristics passing through the corners, was described by \citet{Alty1971,Morley1996}.

In plot~\textbf{(b)}, with increasing $\psi$, the velocity becomes close to zero near the subcharacteristic passing through the top-right corner, in a neighbourhood of which the flow begins to reverse. In plot~\textbf{(c)}, with $\psi=\pi/2$, there is a perfectly symmetrical velocity dipole with the liquid in the right half of the trench flowing at an equal and opposite speed to that in the left, and the average velocity is equal to zero. The same behaviour as in plots~\textbf{(a)} and~\textbf{(b)} is demonstrated in plots~\textbf{(d)} and~\textbf{(e)}, albeit with reversed direction. In any case, it appears that the net velocity in the trench is maximised when the applied magnetic field is orthogonal to the temperature gradient, as expected.

Turning our attention to figure~\ref{fig:increasing_psi_magnetic_field}, we notice that the induced field is close to zero outside a triangular region near the bottom-right corner in plots \textbf{(a)} and \textbf{(b)}. The dividing line follows a subcharacteristic inclined at an angle $\psi$ to the horizontal, which encloses a region of high induced field strength, counteracting the dipole formed by the rest of the magnetic field profile. In plot~\textbf{(c)}, the applied field is parallel to the trench walls, where the induced magnetic field exhibits side layers of thickness of $\order(\epsilon^{1/2})$. These layers a stronger electrical current through the walls, and hence a stronger Lorentz force which drives the fluid at a higher speed in both the left- and right-hand sides of the trench, although in opposite directions.

\section{Discussion}\label{sec:discussion}

In this paper, we performed a hybrid numerical/asymptotic analysis of steady unidirectional TEMHD flow inside a trench akin to those seen in LiMIT. We obtained asymptotic solutions for the velocity, magnetic field and temperature in the limit of large Hartmann number and thin trench walls, demonstrating how temperature gradients drive fluid flow. We obtained numerical solutions to the problem in different parameter cases corresponding to both laboratory and fusion-relevant applications, and these demonstrated very good agreement with the asymptotic predictions.


Upon performing extensive parameter sweeps, we came to the following general conclusions. Firstly, making the side walls thicker can have a retarding effect on the fluid flow above the trench, whilst making the trenches wider can increase the flow speed up to a point. Increasing the base thickness can also increase the flow speed, provided that it is less than approximately 90\% of the wall height.
Increasing the wall height was found to reinforce the velocity jet near the conducting base, thus increasing the speed of the flow inside the trench, but only up to a point. The flow speed above the trench can be optimised by using either a trench that is very wide with short walls, or one whose base is very thick compared to the wall height. The latter scenario represents a good trade-off between optimising the flow and making sure that the free-surface temperature is relatively low.

If the trench is overfilled too much, then the liquid above the trench can flow several times faster than that inside the trench.
We hypothesise that such a situation may be susceptible to free-surface instability and thus to be avoided in practice.
From the numerical simulations and asymptotic estimates, 
one can in principle determine how much the trench should be filled for a given slope angle, so that velocities above and below the trench remain approximately equal.


We confirmed our asymptotic predictions by taking $\epsilon\ll 1$ in the numerics.
However, we emphasise that intuition may be lost when considering the dimensionless solutions, because the velocity scaling is inversely proportional to the applied magnetic field. When this scaling is taken into account, our results agree qualitatively with the findings by \cite{Xu2014} that increasing the magnetic field strength can have a retarding effect on the flow velocity. Furthermore, increasing the applied magnetic field strength paradoxically has an attenuating effect on the induced magnetic field.

Changing the inclination of the applied magnetic field revealed some interesting behaviour which lines up with classically observed phenomena \citep{Morley1996}, such as dividing sub-characteristics in the solutions. We confirmed that reversing the orientation of the magnetic field reverses the direction of the flow, as expected, and that making the applied magnetic field vertical introduces regions of flow reversal, resulting in zero average velocity.


There are several interesting avenues that have arisen as a result of the modelling work we have presented.
One could attempt to model the cooling tubes beneath the trench in more detail, e.g., by replacing the Dirichlet boundary condition on the temperature on the base with a Robin boundary condition.
We considered only steady solutions in this paper, but unsteady effects may be important in practice, for example during start-up of a tokamak, when 
the applied magnetic field is ramped up from zero to a several Teslas over a number of minutes.
Also of practical interest is the effect on the flow in the trench of plasma instabilities such as edge-localised modes, which can cause large localised transient spikes in heat flux and/or applied field.

We have focused on rectangular trenches in this modelling work, because because they are the most straightforward to design and manufacture. However, 3D printing capabilities can in principle produce novel trench shapes, including cross-sections that vary along the trench. It would be interesting to pose a shape optimisation problem, for example,
to maximise the average flow speed inside the trench, given engineering constraints on the wall and base properties.

We have assumed that the physically imposed properties, including the impinging heat flux and applied magnetic field, are all uniform in the flow direction along the trench. In practice, the impinging heat flux is likely to be concentrated around a strip near the centre of the trench (described by a Gaussian in \cite{Xu2014}), and the applied magnetic field is also likely to vary slightly with distance along the trench. The next step is to relax the assumption of unidirectional flow and take into account variations along the trench. Since the trench can be assumed to be much longer than it is wide, a slowly-varying version of the model can be derived, which varies parametrically with distance $z$ along the trench. In principle the resulting model could be solved by considering a sequence of two-dimensional slices, each of which resembles the purely two-dimensional problem analysed in this paper.

\section{Acknowledgements}

We would like to thank Dr P.\ F.\ Buxton from Tokamak Energy for his continued support and insight into confined nuclear fusion, and for facilitating contact with others who have helped to guide this work.
OGB would like to thank C.~Moynihan and S.~Stemmley from University of Illinois at Urbana-Champaign for providing software assistance and giving insight into the engineering aspects of LiMIT. Prof.~J.~R.~Ockendon, Prof.~J.~C.~R.~Hunt and Dr~D.~J.~Allwright contributed to helpful team discussions. This publication is based on work supported by the EPSRC Centre For Doctoral Training in Industrially Focused Mathematical Modelling (EP/L015803/1) in collaboration with Tokamak Energy.

\section{Declaration of interests}
The authors report no conflict of interest.

 
\bibliographystyle{jfm}
\bibliography{jfm-instructions}

\end{document}